\documentclass[twocolumn]{svjour3}

\usepackage{fix-cm}
\usepackage{graphicx}%
\usepackage{multirow}%
\usepackage{bm,braket,amsmath,amssymb,amsfonts,hyperref}%
\usepackage{mathrsfs}%
\usepackage[title]{appendix}%
\usepackage{xcolor}%
\usepackage{textcomp}%
\usepackage{manyfoot}%
\usepackage{booktabs}%
\usepackage{algpseudocode}%
\usepackage{listings}%
\usepackage{url,longtable,tabularx}

\newcommand\+{\dagger}
\newcommand\jr{j_{\rho}}
\newcommand\mr{m_{\rho}}

\newcommand\pesdft{E_{\rm SCMF}(\beta,\gamma)}
\newcommand\pesibm{E_{\rm IBM}(\beta,\gamma)}
\newcommand\pesdftoct{E_{\rm SCMF}(\beta_2,\beta_3)}
\newcommand\pesibmoct{E_{\rm IBM}(\beta_2,\beta_3)}
\newcommand\pesdfthex{E_{\rm SCMF}(\beta_2,\beta_4)}
\newcommand\pesibmhex{E_{\rm IBM}(\beta_2,\beta_4)}
\newcommand\bh{{\hat H}_{\rm IBM}}
\newcommand\hamffm{{\hat H}_{\rm IBFFM}}
\newcommand\hf{\hat{H}_{\mathrm{F}}}
\newcommand\hbf{\hat{V}_{\mathrm{BF}}}

\newcommand\hff{\hat{V}_{\nu\pi}}
\newcommand\vd{v_{\mathrm{d}}}
\newcommand\vssd{v_{\mathrm{ssd}}}
\newcommand\vsss{v_{\mathrm{ss}}}
\newcommand\vt{v_{\mathrm{t}}}

\newcommand\ft{\log{}_{10}ft}
\newcommand\mgt{M_{\mathrm{GT}}}
\newcommand\mf{M_{\mathrm{F}}}
\newcommand\ga{g_{\mathrm{A}}}
\newcommand\gv{g_{\mathrm{V}}}
\newcommand\db{\beta\beta}
\newcommand\tnbb{2\nu\beta\beta}
\newcommand\znbb{0\nu\beta\beta}

\begin{document}

\title{Mapped interacting boson model for nuclear structure studies}
\author{Kosuke Nomura}

\institute{K. Nomura \at
           Department of Physics\\
	   Hokkaido University\\
	   Sapporo 060-0810, Japan \and
           Nuclear Reaction Data Center\\
	   Hokkaido University\\
	   Sapporo 060-0810, Japan\\
	   \email{nomura@sci.hokudai.ac.jp}
}

\date{Received: date / Accepted: date}

\maketitle

\begin{abstract}
The present status of the mapped interacting 
boson model studies on nuclear structure 
is reviewed. 
With the assumption that the nuclear surface 
deformation induced by the multi-nucleon dynamics 
is simulated by bosonic degrees of 
freedom, the interacting-boson Hamiltonian 
that provides energy spectra and wave functions 
is determined by mapping the 
potential energy surface that is obtained  
from self-consistent mean-field 
calculations based on the energy density functional 
onto the corresponding energy surface of the 
boson system. 
This procedure has been shown to be valid 
in general cases of the quadrupole 
collective states, and has allowed for 
systematic studies on spectroscopic properties 
of medium-heavy and heavy nuclei, 
including those that are far from the 
line of $\beta$ stability. 
The method has been extended to study 
nuclear structure phenomena that 
include shape phase transitions and coexistence, 
octupole deformation and collectivity, 
and the coupling of the single-particle 
to collective degrees of freedom, which is 
crucial to describe structures of odd nuclei, 
and $\beta$ and $\beta\beta$ decays. 
\end{abstract}

\section{Introduction\label{sec:intro}}

The atomic nucleus is a quantal system of 
strongly correlated neutrons and protons. 
A prominent feature of the nucleus is that 
it has a shell structure arising from the 
motions of independent particles 
in an average field \cite{SM}. 
When the nucleus takes 
specific nucleon (magic) numbers 
2, 8, 20, 28, 50, 82, 126, etc., 
the system acquires a notable stability, 
and is spherically symmetric in shape. 
In addition to these features resulting 
from the single-particle motions, 
the nucleus exhibits stunning 
collective properties arising 
from the deformation of the nuclear shape. 
Dominant collective modes of excitations 
are of quadrupole type, 
which include the oscillations of the nuclear surface, 
leading to the vibrational spectra, 
the rotational motion of an axially symmetric 
deformed nucleus, resulting in the 
rotational bands \cite{BM}, and the motions 
of an axially asymmetric ($\gamma$-soft) 
deformed nucleus. 
Correlations between the single-particle and collective 
degrees of freedom play a significant role in determining 
the low-energy nuclear structure and, in particular, 
the microscopic description of the nuclear collective motions 
in terms of the nucleonic degrees of freedom has been 
a central problem in the field of low-energy nuclear physics.

The nuclear energy density functional (EDF) 
theory has been successful in 
describing accurately 
intrinsic properties of finite nuclei 
including binding energies, radii, deformations, 
properties of infinite nuclear matter, 
and collective excitations 
in a broad range of the nuclear chart 
\cite{RS,bender2003,vretenar2005,niksic2011,robledo2019}. 
Representative classes of the nuclear EDFs 
are those based on the zero-range, Skyrme \cite{Skyrme,VB}, 
and finite range, Gogny \cite{Gogny}, 
interactions, which are formulated in the 
nonrelativistic framework, and those interactions 
derived from the relativistic mean-field Lagrangian. 
The nuclear EDFs are usually implemented in 
self-consistent mean-field (SCMF) methods, 
and the SCMF calculations that employ a nonrelativistic 
or relativistic EDF have been extensively made 
for describing ground-state properties all over
the nuclear chart.

The self-consistent solutions, including the 
potential energy surface (PES) in terms of 
the collective coordinates, are characterized by 
the symmetry breaking, e.g., of
rotational, particle number, parity, etc.
To study quantitatively 
excitation energies and transition rates, 
it is necessary to extend the EDF framework 
to include the dynamical correlations that 
are associated with the restoration of broken 
symmetries in the mean-field approximation, 
and with the fluctuations around the mean-field 
solution \cite{RS}. Both types of the correlations are  
taken into account by the projections onto 
states with good symmetry quantum numbers and 
by the configuration mixing of mean-field solutions 
in a straightforward manner 
by means of the generator coordinate method (GCM)
(see Refs.~\cite{bender2003,niksic2011,robledo2019}
for reviews).
It provides a fully microscopic and consistent
description of the ground and excited states
of nuclei including all relevant many-body correlations.
The GCM-type calculations with the 
symmetry conservations and configuration mixing 
have been performed extensively in 
nuclear structure studies, which also utilize 
massive computational power. 
An alternative method is to derive the 
collective Hamiltonian microscopically, 
in such a way that the deformation-dependent 
collective mass parameters and moments of inertia 
are determined by the constrained SCMF 
solutions with a given EDF
(see, e.g., Refs.~\cite{libert1999,prochniak2009,niksic2009,delaroche2010}
for earlier studies and, e.g., 
Refs.~\cite{delaroche2024,washiyama2024,suzuki2024}
for most recent ones, which considered
the five-dimensional collective Hamiltonian
for quadrupole collective states).

Another promising method to  
calculate low-energy collective excitations 
in heavy nuclei consists in the fermion-to-boson mapping 
of the mean-field solutions onto the 
interacting boson model (IBM) \cite{nomura2008}. 
Since its advent in the 1970s, the IBM 
has been remarkably successful in phenomenological 
studies of low-energy collective states 
in medium-heavy and heavy nuclei \cite{IBM}. 
The basic assumption of the model is that 
correlated monopole, $S$ 
(with spin and parity $J^\pi=0^+$), 
and quadrupole, $D$ (with $J^\pi=2^+$), 
pairs of nucleons in a given valence space 
are represented by monopole $s$ and quadrupole 
$d$ bosons, respectively. 
A prominent feature of the IBM is that 
if the model Hamiltonian is expressed in 
terms of Casimir operators it is exactly solvable, 
resulting in the three dynamical symmetries 
U(5), SU(3), and O(6), which correspond to 
the energy spectra for the anharmonic vibrator, 
axially-deformed rotor, and $\gamma$-unstable rotor, 
respectively. 
In the majority of cases, however, 
exact dynamical symmetries rarely occur, and 
the eigenvalue problem should be solved numerically. 
Also because the parameters of the IBM were determined 
by fitting directly to the experimental energy spectra, 
the model had been applied only to those nuclei 
for which experimental data exist.

Microscopic foundation of the IBM has also been 
studied, that is, an attempt to derive 
the bosonic Hamiltonian from nucleonic degrees 
of freedom. 
A conventional way of deriving the IBM Hamiltonian 
was based on the nucleon-to-boson mapping within the 
nuclear shell model, which is 
carried out in such a way that 
the full Hilbert space is truncated to a 
subspace consisting of the $S$ and $D$ pairs using 
the seniority scheme, and the matrix element of 
a shell-model Hamiltonian in the 
$SD$-pair states is mapped onto the 
corresponding matrix element of the IBM 
Hamiltonian in the $sd$-boson states \cite{OAIT,OAI}. 
This approach was shown to be valid for those 
nuclear systems that are characterized by 
moderate deformations, that is, nearly 
spherical and $\gamma$-soft nuclei \cite{OAI,mizusaki1997}.

The SCMF solutions resulting from a 
given universal EDF have immediate relevance 
to the nuclear deformations and shapes, 
and are used as a reasonable starting 
point for constructing collective models. 
The nuclear EDF, in general, 
is calibrated to reproduce accurately 
the intrinsic properties of nuclei 
including the deformation 
of the ground-state shape, and appears 
to give an appropriate microscopic 
input to construct the IBM Hamiltonian 
for general cases that include strongly
deformed nuclei. 
A method of deriving 
the IBM Hamiltonian by using the framework 
of the nuclear EDF theory 
was proposed in Ref.~\cite{nomura2008}. 
Under the assumption that multi-nucleon dynamics 
of the nuclear surface deformation is simulated by 
bosonic degrees of freedom,
the PES that is computed by the constrained SCMF method 
within the EDF framework is mapped onto the 
expectation value of the IBM Hamiltonian 
in the boson intrinsic state.
Strength parameters of the boson Hamiltonian used
for describing eve-even nuclear systems are
derived by this mapping procedure.
Energy eigenvalues and wave functions with 
good symmetry quantum numbers in the laboratory frame 
are obtained from the numerical
diagonalization of the mapped IBM Hamiltonian. 
The method was shown to work in the three 
limiting cases of the quadrupole collective states, 
namely, nearly spherical vibrational 
\cite{nomura2008,nomura2010}, 
axially-deformed rotational \cite{nomura2011rot}, 
and $\gamma$-unstable \cite{nomura2010} states, 
and in the transitional regions. 
An improvement over the conventional IBM 
is that it has become possible to determine 
the parameters of IBM without having to fit to 
the experimental data, and to predict  
spectroscopic properties for those nuclei that 
are in thus far unknown regions of the mass chart.

The framework of the IBM mapping has been extended 
to study various nuclear structure phenomena: 
(i) Contributions of intruder states 
and configuration mixing effects have been introduced 
in the IBM, and shown to play a significant role 
in describing phenomena of shape coexistence and mixing, 
and in interpreting the nature of the low-lying 
excited $0^+$ states \cite{nomura2012sc,nomura2016sc}. 
(ii) In order to describe 
both positive and negative parity 
states, in addition to $s$ and $d$ bosons, octupole 
$f$ bosons, with spin and parity $J^{\pi}=3^-$, 
have been introduced in the IBM mapping procedure 
\cite{nomura2013oct,nomura2014}, 
in which the Hamiltonian of the $sdf$-boson model 
is determined by the quadrupole and octupole 
constrained SCMF method. 
The mapped $sdf$-IBM calculations 
have been extensively performed 
to investigate the nuclear structural evolution 
in terms of both the quadrupole and octupole 
shape degrees of freedom. 
(iii) Coupling of the single-particle to 
collective motions has been incorporated,
which has allowed to calculate the
low-lying structure of
those nuclei with odd nucleon numbers \cite{nomura2016odd}. 
The calculations for the odd nuclei, i.e., 
odd-mass and odd-odd nuclei, 
are also essential to model nuclear decay 
processes such as single ($\beta$) 
and double-$\beta$ ($\beta\beta$) decays.

This contribution provides an extensive review on the 
nuclear structure studies within 
the framework of the EDF-mapped IBM. 
The basic notions, formulations, and proof of the method 
are illustrated in simple examples of
quadrupole collective states in Sec.~\ref{sec:model}.
Extensions of the method 
to include intruder states and 
configuration mixing, and 
several illustrative applications to study 
shape coexistence and shape transition 
are discussed in Sec.~\ref{sec:co}. 
In Sec.~\ref{sec:ho}, 
higher-order deformation effects including
those of octupole and hexadecapole types on 
spectroscopic properties are discussed. 
Section~\ref{sec:odd} gives the formulation to 
include the coupling of the collective 
to single-particle degrees of freedom 
in odd nuclei, and 
the relevant results of the calculations 
on low-energy spectroscopic properties, 
and on $\beta$ and $\beta\beta$ decays 
in the odd-mass and odd-odd systems. 
Section~\ref{sec:summary} gives a summary and 
perspectives for future developments.

\section{Basic notions and proof of the method\label{sec:model}}

\subsection{Mean-field framework\label{sec:basic}}

As the starting point, 
the constrained SCMF calculations 
using a given nuclear EDF and a pairing interaction 
are performed to yield the PES, that is, 
total mean-field energy as a function 
of quadrupole deformations. 
The self-consistent calculations are carried 
out in a standard way \cite{RS,bender2003} 
that employs, e.g., 
the Hartree-Fock-Bogoliubov (HFB), and 
Hartree-Fock plus BCS (HF+BCS) frameworks. 
The constraints imposed on the SCMF calculations 
are on the expectation values 
of multipole moments of rank $\lambda$, 
$Q_{\lambda\mu}$, with $\mu=-\lambda,\ldots,+\lambda$. 
The quadrupole ($\lambda=2$), octupole ($\lambda=3$), 
hexadecapole ($\lambda=4$), $\ldots$ moments 
are associated with the intrinsic 
deformations $\beta_{\lambda\mu}$. 
For the quadrupole collectivity, in particular, 
the relevant collective coordinates are 
the axial deformation
\begin{align}
\label{eq:beta2}
 \beta_2
=\frac{\sqrt{5\pi}}{3AR_0^2}
\sqrt{\braket{\hat Q_{20}}^2 + 2\braket{\hat Q_{22}}^2 }
\end{align}
and the triaxiality
\begin{align}
\label{eq:gamma}
 \gamma = \tan^{-1}{\left(\sqrt{2}
\frac{\braket{\hat Q_{22}}}{\braket{\hat Q_{20}}}\right)}
\; ,
\end{align}
where $R_0=1.2 A^{1/3}$ fm. 
In the following, the variable $\beta_2$ is replaced 
simply with $\beta$ 
when discussing the quadrupole collective states. 
In Eqs.~(\ref{eq:beta2}) 
and (\ref{eq:gamma}) 
the $\beta$ takes zero or positive values 
$\beta \geqslant 0$, and 
the $\gamma$ deformation takes values within 
the range $0^{\circ} \leqslant \gamma \leqslant 60^{\circ}$.

A set of the constrained calculations for a given 
even-even nucleus yields the PES in terms 
of the quadrupole triaxial deformations $(\beta,\gamma)$, 
which is denoted by $E_{\rm SCMF}(\beta,\gamma)$. 
Figure~\ref{fig:pes-smba} shows the SCMF PESs 
for the even-even $^{148}$Sm, $^{154}$Sm, and $^{134}$Ba nuclei 
computed by the constrained 
HF+BCS method \cite{ev8} 
using the Skyrme SkM* EDF \cite{skms} 
and density-dependent zero-range 
pairing interaction with the strength 1250 MeV fm$^3$. 
These nuclei are representatives of 
the nearly spherical anharmonic 
vibrator ($^{148}$Sm), prolate deformed rotor ($^{154}$Sm), 
and $\gamma$-unstable rotor ($^{134}$Ba). 
The PES for $^{148}$Sm indeed exhibits 
a weakly deformed minimum at $\beta\approx 0.15$. 
For the $^{154}$Sm nucleus an well developed 
prolate minimum at $\beta\approx 0.35$ is obtained. 
As regards $^{134}$Ba, the PES exhibits a pronounced 
$\gamma$ softness.

\subsection{IBM mapping\label{sec:ibm}}

In this review, 
the neutron-proton IBM (IBM-2) \cite{IBM,OAIT,OAI}, 
in which the neutron
and proton degrees of freedom are distinguished, 
is mainly considered, since it is  
more realistic than the original simpler version 
of the IBM, which is referred to as IBM-1. 
The building blocks of the IBM-2 are neutron $s_\nu$ and 
$d_{\nu}$ bosons, and proton $s_\pi$ and 
$d_{\pi}$ bosons. The neutron (proton) 
$s_\nu$ ($s_{\pi}$) and $d_\nu$ ($d_{\pi}$) bosons 
reflect, from a microscopic point of view, 
the collective monopole $S_\nu$ ($S_\pi$) and 
quadrupole $D_\nu$ ($D_\pi$) pairs of valence neutrons (protons), 
respectively. 
The number of neutron (proton) bosons, 
denoted by $N_\nu$ ($N_\pi$), is equal to that of 
the valence neutron (proton) pairs. 
Here $N_{\nu}$ satisfies $N_\nu = N_{s_{\nu}} + N_{d_{\nu}}$, 
with $N_{s_{\nu}}$ and $N_{d_{\nu}}$ being 
the numbers of neutron $s_{\nu}$ and $d_{\nu}$ 
bosons, respectively, and similarly  
the proton boson number 
$N_\pi = N_{s_{\pi}} + N_{d_{\pi}}$. 
For a given nucleus, the total number 
of bosons $N_{\rm B} = N_\nu + N_\pi$ 
is conserved.

The form of the most general IBM-2 
Hamiltonian with up to two-body terms 
is too complicated to be implemented in the 
mapping procedure. 
However, the following simplified form is often 
employed in the literature, and 
is shown to be adequate for most of the realistic 
nuclear structure studies \cite{IBM}.
\begin{align}
\label{eq:ham1}
 \hat H_{\rm IBM}
= \epsilon_d \hat n_d + \kappa \hat Q_{\nu} \cdot \hat Q_{\pi} \; .
\end{align}
Here the first term represents the number operator 
of $d$ bosons, 
$\hat n_d = \hat n_{d_{\nu}} + \hat n_{d_{\pi}}$, 
with $\hat n_{d_{\rho}} = d^\+_{\rho} \cdot \tilde d_{\rho}$ 
($\rho=\nu$ or $\pi$). 
$\epsilon_d$ is the single $d$ boson energy 
relative to the $s$-boson one. 
Note that the notation 
$\tilde d_{\rho\mu} = (-1)^{\mu} d_{\rho-\mu}$. 
The second term stands for the quadrupole-quadrupole 
interaction with strength parameter $\kappa$ between 
neutron and proton bosons. 
The quadrupole operator $\hat Q_{\rho}$ is defined by
\begin{align}
\label{eq:bquad}
 \hat Q_{\rho} = 
[s_{\rho}^\+ \times \tilde d_{\rho}
+ d_{\rho}^\+ \times \tilde s_{\rho} ]^{(2)}
+ \chi_{\rho}
[ d_{\rho}^\+ \times \tilde d_{\rho} ]^{(2)} \; ,
\end{align}
where $\chi_\rho$ are dimensionless parameters 
and determine whether the neutron and proton boson 
systems are prolate 
or oblate deformed, 
depending on their sign: 
$\chi_\rho<0$ for prolate, 
$\chi_\rho>0$ for oblate, and $\chi_\rho\approx 0$ 
for $\gamma$-unstable deformations. 
The use of the Hamiltonian of the form (\ref{eq:ham1}) is also 
justified by the microscopic consideration \cite{OAI,IBM}
that the pairing-type interaction 
between identical nucleons and the quadrupole-quadrupole 
interaction between non-identical nucleons in 
the residual nucleon-nucleon interaction 
in the nuclear shell model 
make dominant contributions 
to the low-lying quadrupole 
collective states.

A connection of the IBM-2  
to the geometrical collective model is made 
by calculating the expectation value of the Hamiltonian 
$\braket{\Phi|\bh|\Phi}$  
in the boson intrinsic wave function 
or coherent state $\ket{\Phi}$, which is given as
\cite{ginocchio1980,dieperink1980,bohr1980}
\begin{align}
 \ket{\Phi} = \frac{1}{\sqrt{N_{\nu}!N_{\pi}!}}
(\lambda_\nu^\+)^{N_\nu}
(\lambda_\pi^\+)^{N_\pi}
\ket{0} \; .
\end{align}
The ket $\ket{0}$ stands for the inert core, 
and $\lambda^\+_\rho$ is defined as
\begin{align}
 \lambda_\rho^\+ = 
\frac{1}{\sqrt{1+\sum_{\mu=-2}^{+2} \alpha_{\rho\mu}^2}}
\left(
s^\+_\rho + 
\sum_{\mu=-2}^{+2} \alpha_{\rho\mu} d^\+_{\rho\mu}
\right) \; ,
\end{align}
with $\alpha_{\rho\mu}$ being amplitudes,
which are expressed as
\begin{align}
&\alpha_{\rho 0}=\beta_{{\rm B}\rho} \cos\gamma_{{\rm B}\rho}\\
&\alpha_{\rho\pm1}=0\\
&\alpha_{\rho\pm2}=\frac{1}{\sqrt{2}}\beta_{{\rm B}\rho} \sin\gamma_{{\rm B}\rho} \; .
\end{align}
The variables $\beta_{{\rm B}\rho}$ 
and $\gamma_{{\rm B}\rho}$ represent the 
boson analogs of quadrupole axial
and triaxial deformations, respectively. 
It is assumed that the neutron 
and proton boson systems have the 
same deformations, i.e., 
$\beta_{{\rm B}\nu}=\beta_{{\rm B}\pi} \equiv \beta_{{\rm B}}$ 
and 
$\gamma_{{\rm B}\nu}=\gamma_{{\rm B}\pi} \equiv \gamma_{{\rm B}}$. 
In addition, the intrinsic variables 
$\beta_{\rm B}$ and $\gamma_{\rm B}$ can be related 
to their fermionic counterparts \cite{ginocchio1980} 
so that $\beta_{\rm B}$ is proportional to $\beta$ 
and that $\gamma_{\rm B}$ is identical to $\gamma$: 
\begin{align}
\label{eq:cbeta}
\beta_{\rm B} = C_{\beta} \beta
\quad ,
\quad
\gamma_{\rm B} = \gamma
\end{align}
with $C_{\beta}$ being a constant of proportionality. 
The constant $C_{\beta}$ 
usually takes a value $C_{\beta}\approx 5$, 
indicating that the bosonic $\beta$ deformation 
is larger than the fermionic one. 
The difference between the fermionic and bosonic 
$\beta$ deformations reflects the fact that 
the boson system is comprised of only 
valence nucleons, whereas in realistic 
nuclei all composite nucleons should be taken into account.

The energy surface for the boson system, 
$\pesibm=\braket{\Phi|\bh|\Phi}$ has  
an analytical form 
\begin{align}
\label{eq:pesibm2}
&\pesibm
=\frac{\epsilon(N_{\nu}+N_{\pi})
 \beta_{\rm B}^2}{1+\beta_{\rm B}^2}
+ \frac{\kappa N_{\nu}N_{\pi}\beta_{\rm B}^2}{(1+\beta_{\rm B}^2)^2}
\nonumber\\
&\times 
\left[
4-2\sqrt{\frac{2}{7}}(\chi_{\nu}+\chi_{\pi})\beta_{\rm B}\cos{3\gamma_{\rm B}}+\frac{2}{7}\chi_{\nu}\chi_{\pi}\beta_{\rm B}^2
\right] \; .
\end{align}
The above formula represents an energy surface in terms of 
the $\beta_{\rm B}$ and $\gamma_{\rm B}$ 
deformations.
Depending on the values of the Hamiltonian parameters, 
the bosonic energy surface gives 
an equilibrium minimum at a spherical,
a prolate (if $\chi_{\nu}+\chi_{\pi}<0$), or
an oblate (if $\chi_{\nu}+\chi_{\pi}>0$) configuration. 
If $\chi_{\nu}+\chi_{\pi}=0$, 
the energy surface (\ref{eq:pesibm2}) is independent of 
and completely flat along the $\gamma_{\rm B}$ deformation, 
that is, it represents 
an ideal $\gamma$-unstable limit. 

%-------- Sm-Ba PES --------------
\begin{figure}[h]
\centering
\includegraphics[width=\linewidth]{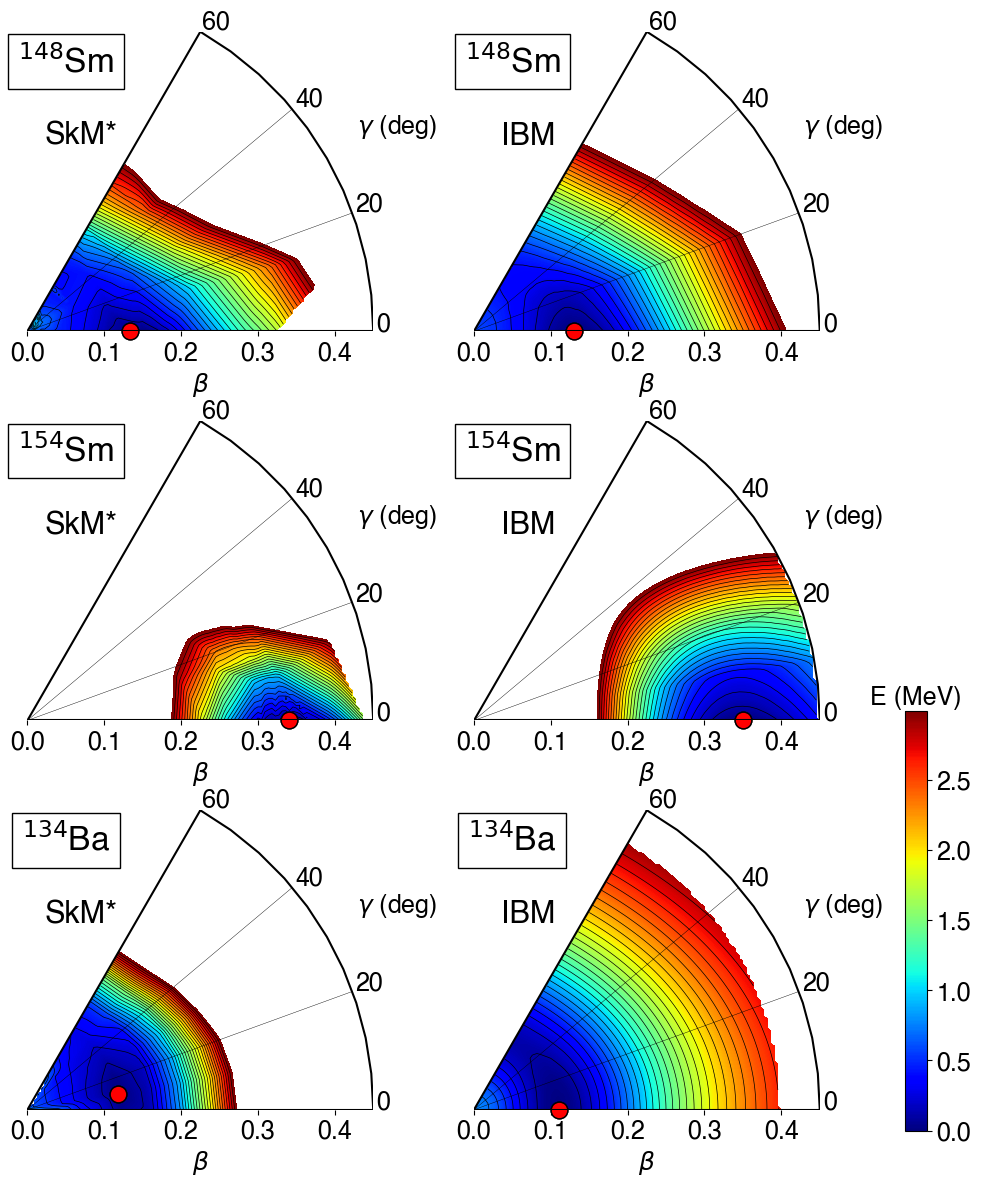}
\caption{(Left) 
Triaxial quadrupole $(\beta,\gamma)$
potential energy surfaces
for $^{148}$Sm, $^{154}$Sm, and $^{134}$Ba 
obtained from the constrained SCMF
calculations using the
Hartree-Fock plus BCS
method with the Skyrme SkM* EDF and 
density-dependent delta-type 
pairing interaction with the 
strength of 1250 MeV fm$^{3}$. 
(Right) Corresponding mapped IBM-2 
energy surfaces. The global minimum is 
identified by the solid circle.}
\label{fig:pes-smba}
\end{figure}
%---------------------------------

The strength parameters of the IBM-2 Hamiltonian (\ref{eq:ham1}), 
$\epsilon_d$, $\kappa$, $\chi_\nu$ and $\chi_\pi$, 
and the proportionality coefficient $C_\beta$ (\ref{eq:cbeta}) 
are determined by mapping the SCMF PES $\pesdft$ 
onto the corresponding boson energy surface $\pesibm$.
In other words, these parameters are determined 
so that the IBM PES should be similar in topology to 
the SCMF PES, satisfying the relation
\begin{align}
\label{eq:mapping}
 \pesibm \approx \pesdft \; .
\end{align}
An approximate equality symbol 
($\approx$) is used in the above expression, 
because the mapping should be considered in 
the vicinity of the global minimum, 
which corresponds to the SCMF PES 
up to $\approx$ 2-3 MeV above the minimum. 
Optimal values of the IBM parameters 
are determined that reproduce 
basic characteristics of the SCMF PES 
within this energy range, such as 
the curvatures in $\beta$ and $\gamma$ deformations, 
and the location and depth of the minimum. 
The shape of the PES is supposed to reflect 
essential features of many-nucleon systems, 
such as the Pauli principle, antisymmetrization, 
and nuclear force. 
In particular, the 
mean-field configurations that are in the 
neighborhood of the global minimum
are most relevant to low-energy 
collective nuclear structure, 
as in the case of the beyond-mean-field 
calculations. 
By reproducing the basic features of 
the fermionic PES that is calculated by the 
microscopic framework of the nuclear EDF, 
the important correlations in nuclear 
many-body system are supposed to be 
incorporated in the IBM system. 
However, one should not try to reproduce those  
regions of the PES that correspond to 
very large deformations compared with that 
which gives the global minimum, since 
in these configurations with large deformations 
quasiparticle degrees of freedom come to play a 
role, which are, by construction, 
not included in the IBM-2 model space.

The mapped IBM-2 PESs for $^{148}$Sm, $^{154}$Sm, 
and $^{134}$Ba are depicted in Fig.~\ref{fig:pes-smba}. 
One can see that the SCMF PESs for these nuclei 
are well reproduced by the IBM-2 ones near 
the global minimum. 
For large $\beta$ deformations, the mapped PESs 
for $^{154}$Sm and $^{134}$Ba 
are flatter than the SCMF PESs. 
These differences mainly arise from the 
fact that the IBM-2 consists of valence nucleons only, 
which may not be sufficient to reproduce 
accurately the SCMF PES.

The determination of the IBM-2 parameters
is made in a standard way, e.g.,
by means of the $\chi$-square fit of the 
formula (\ref{eq:pesibm2}) to the SCMF PES. 
In Ref.~\cite{nomura2010}, a method of finding 
optimal IBM-2 parameters in the mapping procedure 
was proposed, that exploited the technique 
of wavelet transform.
The wavelet transform was specifically used 
for the mapping, partly in order to 
address the uniqueness of the derived 
values of the IBM-2 parameters. 
A major conclusion of the analysis
in Ref.~\cite{nomura2010} is that 
the method that employs the wavelet transform 
allowed to determine the
IBM Hamiltonian parameters for Sm isotopes
in a more unambiguous way
than that using the straightforward $\chi$-square fit. 
In fact, in some cases, the $\chi$-square fit 
results in a set of parameters 
that perfectly reproduces the original SCMF PES, 
but that is also unphysical, e.g., positive values of the 
strength $\kappa$. 
In most of the realistic nuclear structure 
calculations within the mapped IBM framework, 
however, a standard parameter search 
including the $\chi$-square fit appears 
to be just adequate, since there is little difference 
between the derived values of the IBM parameters 
from the $\chi$-square fits and those from the 
wavelet transform.

%-------- Sm parameters ----------
\begin{figure}[h]
\centering
\includegraphics[width=.8\linewidth]{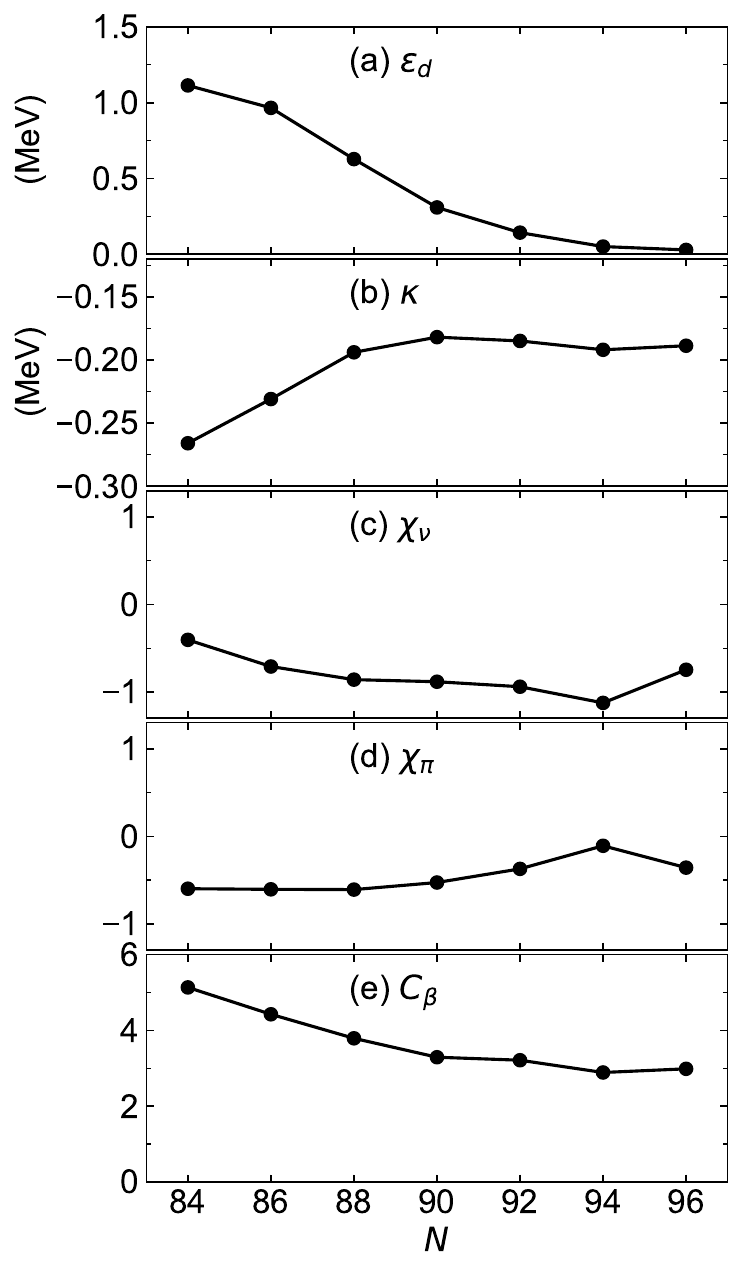}
\caption{Derived IBM-2 parameters for the 
even-even nuclei $^{146-158}$Sm plotted as functions of 
neutron number $N$. The Skyrme SkM* interaction is used.}
\label{fig:sm-para}
\end{figure}
%---------------------------------

The derived values of the IBM-2 parameters, 
$\epsilon_d$, $\kappa$, 
$\chi_{\nu}$, $\chi_{\pi}$ and $C_{\beta}$, 
are given in Fig.~\ref{fig:sm-para}. 
The systematic behaviors of these parameters 
as functions of the 
neutron number $N$ reflect the 
shape phase transition between 
nearly spherical and deformed regions, 
which is suggested by the SCMF PES. 
The single-$d$ boson energy $\epsilon_d$ keeps 
decreasing with $N$ 
[Fig.~\ref{fig:sm-para}(a)],
indicating that the quadrupole 
collectivity becomes stronger for larger $N$. 
The quadrupole-quadrupole strength $\kappa$, 
shown in Fig.~\ref{fig:sm-para}(b), 
exhibits a significant decrease in magnitude 
in the region $84 \leqslant N \leqslant 90$, 
and is approximately constant for $N\geqslant 90$. 
Both the $\chi_{\nu}$ and $\chi_{\pi}$ parameters 
[Fig.~\ref{fig:sm-para}(c) and \ref{fig:sm-para}(d)]
take negative values. 
This reflects the fact that 
the corresponding PESs for Sm nuclei 
exhibit axially symmetric prolate 
deformations (Fig.~\ref{fig:pes-smba}). 
The $\chi_{\nu}$ shows a stronger 
dependence on $N$ than $\chi_{\pi}$, and is closer 
to the SU(3) limit of the IBM, 
$-\sqrt{7}/2\approx -1.32$, in deformed 
region with $N>90$. 
The parameter $C_{\beta}$ shows a gradual 
decrease with $N$ [Fig.~\ref{fig:sm-para}(e)], 
as the deformation becomes stronger.

The $N$-dependencies of the derived 
IBM-2 parameters $\epsilon_d$, 
$\kappa$, $\chi_{\nu}$, and $\chi_{\pi}$ 
for Sm isotopes were also shown to be 
qualitatively similar to those used in the 
conventional IBM-2 calculations
(e.g., Ref.~\cite{scholten1980phd}), and 
to those derived in earlier microscopic IBM 
studies employing the mapping method 
of Otsuka, Arima, and Iachello (OAI) \cite{OAI}. 
A significant difference between the EDF-mapped 
and empirical IBM-2 parameters 
arises in the quadrupole-quadrupole 
interaction strength $\kappa$: 
the derived $\kappa$ value is generally
larger in magnitude than the phenomenological ones
by approximately a factor of 2-3.
This deviation appears to be a general 
feature of the EDF-mapped IBM-2, 
and is commonly encountered in describing,
in particular, deformed nuclei.
In the mapping procedure, 
the $\kappa$ value is mainly determined 
so as to reproduce the depth and steepness 
of the potential valley, and these features 
depend largely on the underlying EDF. 
In the earlier microscopic IBM-2 studies, 
including the one that employed the OAI mapping, 
the unexpectedly large negative values of the 
strength $\kappa$ were also obtained 
(see, e.g., Ref.~\cite{mizusaki1997}).

%-------- Sm spectra -------------
\begin{figure}[h]
\centering
\includegraphics[width=\linewidth]{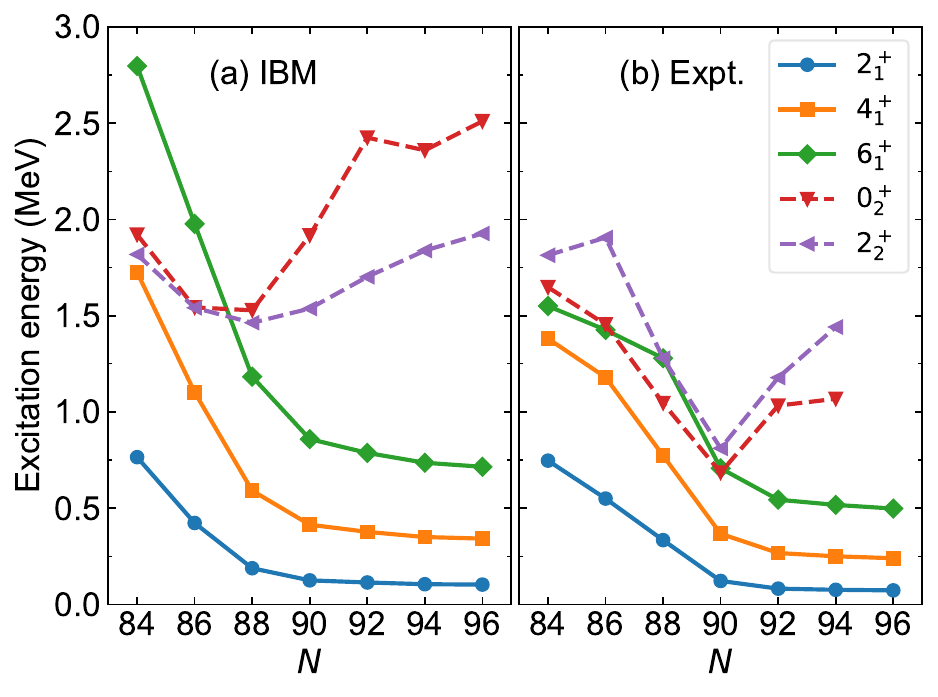}
\caption{
Calculated and experimental \cite{data} energy spectra 
for the low-lying states in the even-even 
$^{146-158}$Sm isotopes.}
\label{fig:sm-level}
\end{figure}
%---------------------------------

%-------- 134Ba spectra ----------
\begin{figure}[h]
\centering
\includegraphics[width=\linewidth]{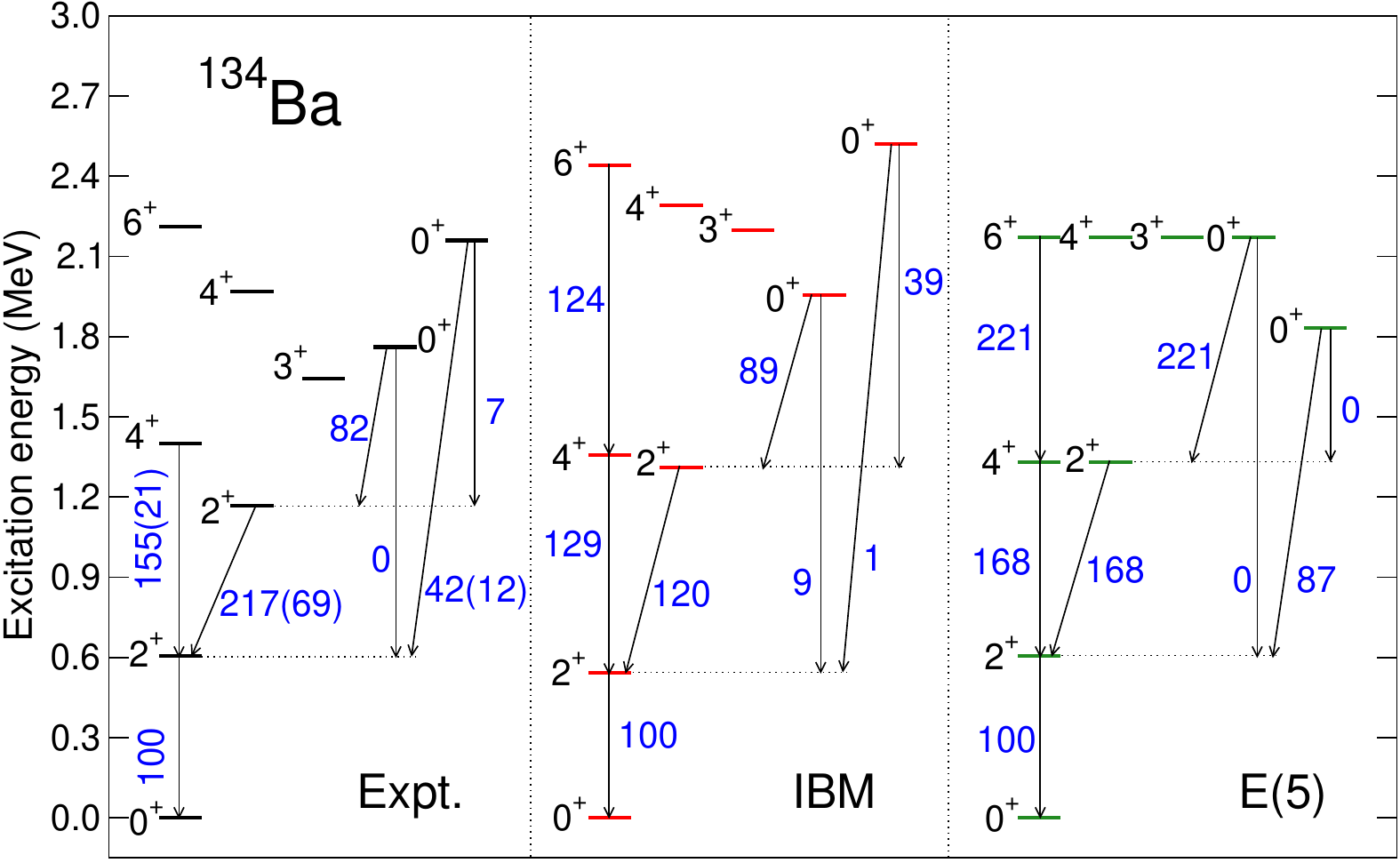}
\caption{Calculated, experimental \cite{data}, 
and E(5) energy spectra 
and ratios of $B(E2)$ transitions 
for $^{134}$Ba. The Skyrme SkM* EDF is used 
for the IBM-2 mapping calculation. 
The $2^+_1$ energy level for the E(5) model 
is normalized to the experimental one, 605 keV. 
}
\label{fig:ba134-level}
\end{figure}
%---------------------------------

Calculated low-lying levels for the Sm isotopes,
obtained from the diagonalization of 
the mapped IBM-2 Hamiltonian, 
are given in Fig.~\ref{fig:sm-level}(a), 
and are compared with the experimental data \cite{data}, 
shown in Fig.~\ref{fig:sm-level}(b).
The calculated energy levels near $^{148}$Sm 
resemble vibrational spectra of the 
anharmonic vibrator, 
with the predicted energy ratio $R_{4/2}$ of the $4^+_1$ 
to $2^+_1$ states being $R_{4/2} \approx 2.61$. 
In heavier Sm nuclei with $N\geqslant 92$, 
the calculation produces rotational spectra, 
with the ratio $R_{4/2} \approx 3.3$. 
The intermediate nuclei near $N \approx 90$ 
correspond to the transitional region and, 
in particular, $^{152}$Sm was suggested to be 
a candidate for the X(5) critical point symmetry
\cite{iachello2001X5,casten2001X5} for
the U(5)-SU(3) first-order
quantum phase transitions (QPTs)
in nuclear shapes \cite{cejnar2010}. 
Overall behaviors of the calculated levels 
as functions of $N$ 
are similar to those of the experimental ones. 
However, the mapped IBM-2 substantially overestimates 
the energy levels of the non-yrast 
states $0^+_2$ and $2^+_2$.
These discrepancies indicate that certain 
improvements of the method are necessary, 
e.g., by inclusions of additional 
terms in the IBM-2 Hamiltonian, and/or 
boson degrees of freedom.

With the IBM-2 wave functions of given states, 
electromagnetic transition probabilities, including the 
electric quadrupole ($E2$) and magnetic dipole ($M1$) 
transition rates and moments, are calculated. 
The corresponding operators for the $E2$ and $M1$
transitions in the IBM-2 are given by
\begin{align}
\label{eq:e2op}
& \hat T^{(E2)} = 
e_{\nu}^{\rm B} \hat Q_{\nu}
+ e_{\pi}^{\rm B} \hat Q_{\pi}
\\
\label{eq:m1op}
& \hat T^{(M1)} = 
g_{\nu}^{\rm B} \hat L_{\nu}
+ g_{\pi}^{\rm B} \hat L_{\pi} \; ,
\end{align}
respectively. 
In Eq.~(\ref{eq:e2op}), $e_{\rho}^{\rm B}$ stands for 
the boson effective charge, and $\hat Q_{\rho}$ 
is the quadrupole operator defined in (\ref{eq:bquad}). 
As conventionally made,
the same value of the parameter 
$\chi_\rho$ as that used in the Hamiltonian (\ref{eq:ham1}) 
is employed \cite{warner1983}.
In Eq.~(\ref{eq:m1op})
$g_{\rho}^{\rm B}$ stands 
for the effective $g$-factor for neutron or proton bosons, 
and $\hat L_\rho$ represents the angular momentum 
operator in the boson system, defined by
\begin{align}
\label{eq:bang}
 \hat L_\rho = \sqrt{10}
[d_{\rho}^\+ \times \tilde d_{\rho}]^{(1)} \; .
\end{align}

Figure~\ref{fig:ba134-level} shows 
the calculated and 
experimental energy spectra and the $B(E2)$ ratios 
for the nucleus $^{134}$Ba. 
One can see that the overall feature of the experimental 
energy level scheme is reproduced by the IBM-2. 
The $^{134}$Ba was also suggested \cite{casten2000E5} 
to be empirical evidence for 
the E(5) critical point symmetry \cite{iachello2000E5} 
of the second-order shape QPT between the U(5) 
and O(6) limits. 
The corresponding energy spectrum predicted by 
the E(5) symmetry is also shown in 
Fig.~\ref{fig:ba134-level}.

The results shown in Figs.~\ref{fig:sm-level} 
and \ref{fig:ba134-level} demonstrate that 
the IBM-2 mapping method 
works reasonably well in all three limits of the 
quadrupole collective states, i.e., 
nearly spherical U(5), deformed rotational SU(3), 
and $\gamma$-unstable O(6) limits of the IBM.

%-------- W-Os spectra -------------
\begin{figure}[h]
\centering
\includegraphics[width=\linewidth]{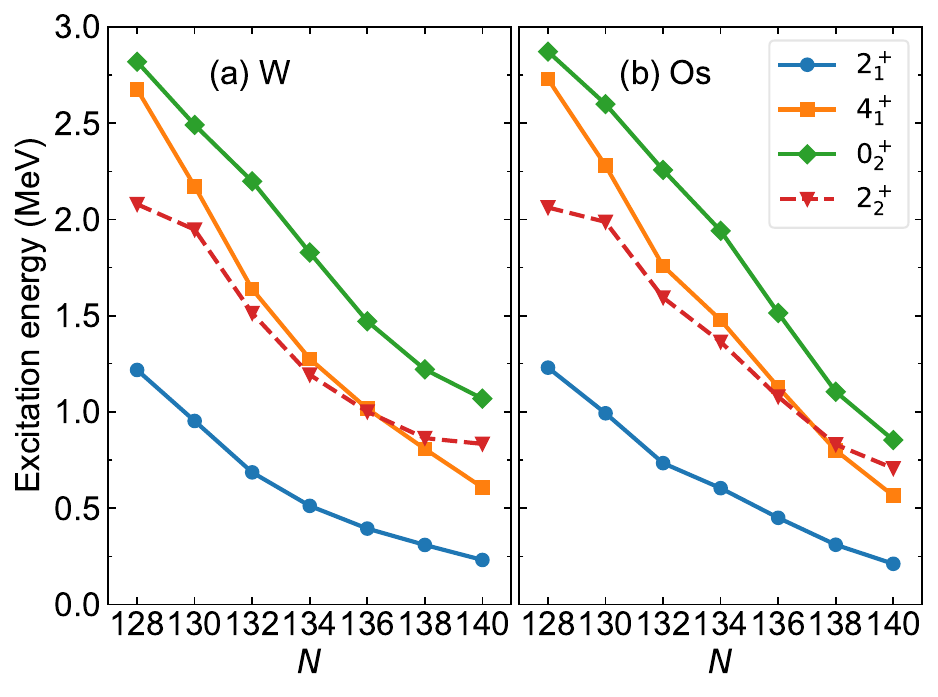}
\caption{
Predicted low-energy spectra for exotic isotopes 
$^{202-214}$W and $^{204-216}$Os.
Skyrme SkM* EDF is used.}
\label{fig:wos-level}
\end{figure}
%---------------------------------

In addition to those nuclei for which experimental 
data exist, the IBM-2 mapping procedure 
was applied to heavy exotic nuclei that 
are experimentally unknown. 
The ``south-east'' of $^{208}$Pb, that is, 
the region corresponding to $N>126$ and $Z<82$ 
is among the challenging regions from an 
experimental point of view. 
Figure~\ref{fig:wos-level} shows 
low-energy excitation spectra 
for the exotic even-even $^{202-214}$W 
and $^{204-216}$Os nuclei 
predicted within the mapped IBM-2 that is 
based on the HF+BCS calculations using 
the Skyrme SkM* interaction \cite{nomura2008,nomura2010}. 
The Skyrme SCMF PESs for many of these nuclei exhibit
a pronounced $\gamma$ softness \cite{nomura2010}. 
One can see in Fig.~\ref{fig:wos-level} that, 
while the levels are gradually lowered 
with $N$ as the quadrupole collectivity increases, 
many of these W and Os nuclei 
exhibit level structures that 
resemble that of O(6) or E(5) symmetry, that is, 
the $2^+_2$ and $4^+_1$ levels are nearly equal 
in energy for $N=130-138$. 
The emergence of a number of the E(5)/O(6) 
structure has not been observed in stable 
nuclei, and may be a characteristic of 
the exotic nuclei in this mass region. 
The $2^+_2$ levels appear to go up for $N>138$. 
This evolution of the calculated energy levels 
indicates a transition from $\gamma$-soft to 
axially deformed shapes in the vicinity 
of $N=140$.

The framework of the EDF-mapped IBM-2 
was further applied to study various 
types of shape phase transitions 
in different mass regions. 
In Ref.~\cite{nomura2010}, 
transitions from a nearly spherical to weakly 
prolate deformed, to $\gamma$-soft, and to spherical 
shapes in the even-even Ru, Pd, Xe and Ba isotopes 
within the range $50<N<82$ were analyzed using the 
Skyrme SLy4 \cite{sly} and SkM* \cite{skms} functionals. 
Prolate-to-oblate shape transitions and 
related spectroscopic properties were studied 
using the Gogny-EDF HFB constrained calculations 
for the even-even nuclei in the 
Yb, Hf, W, Os, and Pt isotopes in the 
$82<N<126$ and $Z<82$ region
\cite{nomura2011pt,nomura2011wos,nomura2011sys}. 
In particular, the nuclei with $N=116$ 
were suggested, both theoretically 
\cite{nomura2011wos,nomura2011sys} and 
experimentally \cite{sahin2024}, 
to exhibit a pronounced $\gamma$ softness, 
and were interpreted as the transition point 
between the prolate and oblate shapes.

Furthermore, an application was made to neutron-rich 
exotic Kr isotopes with mass $A\approx 100$ and 
neutron numbers $N=50-60$. 
The Gogny-D1M-to-IBM-2 mapping was performed 
to obtain the spectroscopic properties of 
these nuclei, and produced 
a smooth evolution of the low-lying levels 
consistently with the experimental finding 
\cite{albers2012,albers2013}.

From a computational point of view,
the mapped IBM calculations for low-energy
collective states are comparatively feasible,
and are suitable for systematic studies.
The computer resources the
mapped IBM calculations require are
mainly for the constrained
SCMF calculations of the PES, and for the numerical
diagonalization of the mapped IBM Hamiltonian.
The constrained calculations are performed
at each set of the relevant deformations,
which can be done in parallel,
and the CPU time and computational resources
required for the SCMF calculations with
EDFs are of the same order of magnitude
as in other EDF-based models.
See, e.g., Ref.~\cite{hfodd249}, in which
computing time and accuracy of the SCMF calculations
using the standard DFT solver HFODD are described
in detail.
The diagonalization of the standard
IBM-2 Hamiltonian takes only a few CPU seconds
even with personal computers
to yield excitation energies and wave functions
for a number of states for a given nucleus,
and the computational
time for the diagonalization increases with
the number of bosons, with additional
boson degrees of freedom being introduced,
and/or within the configuration mixing model
described in Sec.~\ref{sec:co}.
The diagonalization of the interacting boson-fermion
and boson-fermion-fermion model Hamiltonians
for odd nuclear systems, discussed in more detail
in Sec.~\ref{sec:odd}, generally
requires more computational
time than in the case of the even-even nuclei,
and amounts to a few CPU minutes
to obtain low-lying levels and
electromagnetic transition properties for
a given odd nucleus in open shell regions
corresponding to the boson numbers
$N_\nu$ and $N_\pi$ larger than 5.
The most time consuming case among
those discussed in this review is the diagonalization
of the quadrupole-octupole boson-fermion Hamiltonian,
which in some cases takes a few CPU hours to
obtain relevant spectroscopic properties for
an odd-mass nucleus.
By using the simpler IBM-1 model, the computational times
for the numerical diagonalization are
significantly reduced.

\subsection{Rotational response\label{sec:LL}}

The mapping procedure described in Sec.~\ref{sec:model} 
is shown to be valid in describing low-energy 
quadrupole collective spectra. 
As one can see from Fig.~\ref{fig:sm-level}, 
however, while the IBM-2 mapping 
reproduces overall features 
of the rotational bands in strongly 
deformed nuclei $^{154}$Sm, $^{156}$Sm and $^{158}$Sm, 
these calculated energy levels are much higher 
than the observed ones, 
that is, the moments of inertia of the 
rotational bands are significantly 
underestimated.

These underestimates 
illustrate a deficiency of the IBM 
for rotational nuclei in particular, when it is formulated 
within a microscopic nuclear structure theory. 
The problem of not being able to reproduce the 
moments of inertia of the rotational band was 
indeed encountered in earlier microscopic IBM 
studies \cite{otsuka1979phd,otsuka1981prl}, 
and was attributed to the limited degrees 
of freedom of $s$ and $d$ bosons in the IBM. 
It was also pointed out in Ref.~\cite{bohr1980}
that on the basis of the analysis using 
the Nilsson plus BCS approach, 
the truncation to the $SD$ subspace 
might not be sufficient to 
reproduce properties of deformed rotational nuclei. 
This critique shed light upon the microscopic 
justification of the IBM for rotating nuclear systems, 
and invoked a number of related studies 
(see Ref.~\cite{nomura2011rot}, and references are therein). 
A straightforward remedy was to introduce new  
degrees of freedom, e.g., 
the hexadecapole $G$ ($J^\pi=4^+$) pairs, 
in addition to the $S$ and $D$ pairs,  
to renormalize the $G$-pair effects into the $sd$-boson 
sector through a fermion-to-boson mapping, 
or to explicitly consider the corresponding 
boson image, i.e., a $g$ boson. 
It was also suggested that 
the $SD$-pair dominance in the intrinsic states 
holds to a good extent for 
rotating nuclei \cite{otsuka1981,OAY,bes1982}. 
However, there had not been 
a conclusive mapping procedure that covers 
strongly deformed rotational nuclei.

In Ref.~\cite{nomura2011rot} it was proposed that 
the response of the boson system 
to the rotational cranking with 
a finite angular frequency $\omega\neq0$ 
was substantially 
different from that of the nucleonic system, and that 
this difference accounted for the deviation of the 
calculated rotational band levels from the 
observed ones. 
In addition to the 
mapping of the PES, which is a quantity 
in the intrinsic state at rest ($\omega=0$), 
one should go one step further 
for deformed rotational systems so that 
the rotational response of the boson system 
with non-zero $\omega$ values should match that of 
the corresponding fermionic system. 
To incorporate 
the rotational response in the boson system 
while keeping the result of the PES mapping unchanged, 
an additional term $\hat L \cdot \hat L$ 
should be introduced in 
the Hamiltonian of Eq.~(\ref{eq:ham1}):
\begin{align}
\label{eq:ham1-LL}
 \bh' = \bh + \alpha \hat L \cdot \hat L \; ,
\end{align}
where $\hat L = \hat L_\nu + \hat L_\pi$ with 
$\hat L_\rho$ being defined in (\ref{eq:bang}), 
and $\alpha$ is the strength parameter to be determined. 
Since the $\hat L \cdot \hat L$ term is diagonal 
with eigenvalue $I(I+1)$, 
with $I$ being the total angular momentum, 
its effect is either to compress or stretch 
the rotational bands without altering 
the wave functions. 
This term also does not make a 
unique contribution to the energy surface (\ref{eq:pesibm2}) 
and cannot be determined by the PES mapping, 
because its expectation value in the coherent state
has the same analytical form as that of the one-body $d$-boson term.
The parameter $\alpha$ can be, therefore, simply 
renormalized into the $d$-boson energy $\epsilon_d$ 
so that the mapped PES should remain unaltered.

The first step to determine the full 
Hamiltonian $\bh'$ (\ref{eq:ham1-LL}) 
is to derive the five parameters, 
$\epsilon_d$, $\kappa$, $\chi_\nu$, $\chi_\pi$ and $C_\beta$, 
from the energy-surface mapping. 
Having fixed these parameter values, 
the $\alpha$ value is determined 
so that the cranking moment of inertia 
in the boson system, calculated in the coherent 
state \cite{schaaser1986} at a specific deformation, 
e.g, that corresponds to the global minimum, 
is equated to the corresponding 
moment of inertia computed by the SCMF method 
using standard formulas such as those of 
Inglis and Belyaev \cite{inglis1956,belyaev1961} 
and of Thouless and Valatin \cite{TV}.

\begin{figure}[h]
\centering
\includegraphics[width=\linewidth]{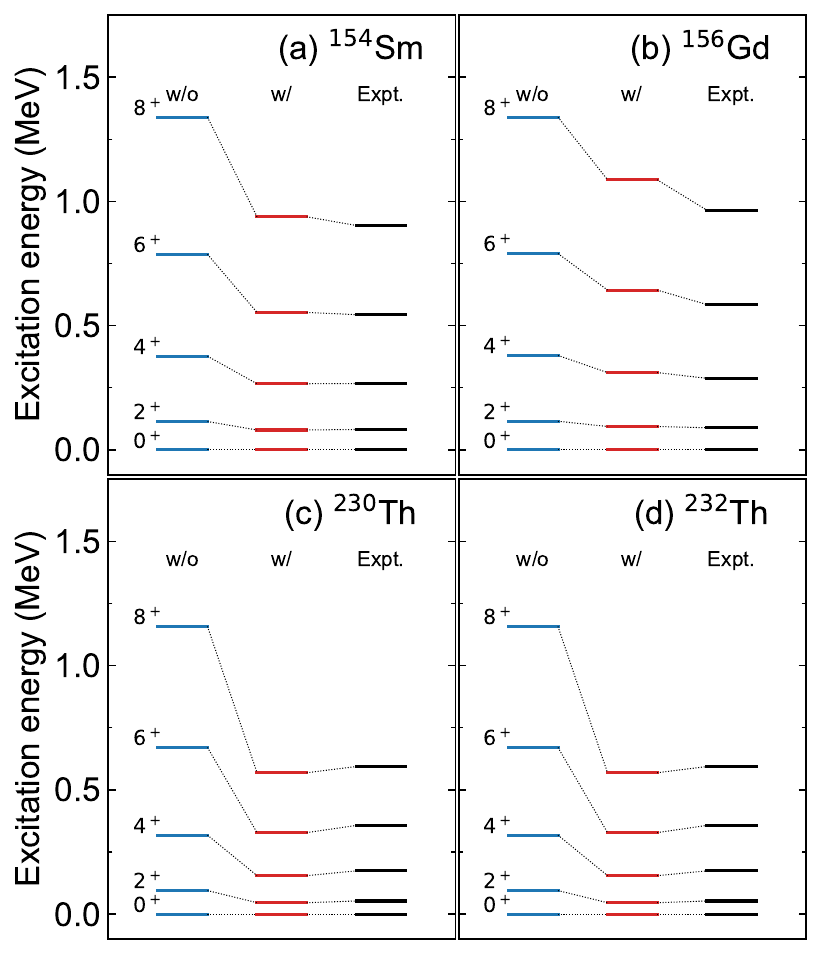}
\caption{Rotational bands of
the axially deformed nuclei $^{154}$Sm,
$^{156}$Gd, $^{230}$Th, and $^{232}$U
calculated by the Skyrme-SkM* mapped IBM-2
with (``w/'') and without (``w/o'')
the $\hat L\cdot \hat L$ term in the Hamiltonian.
The experimental data are taken from Ref.~\cite{data}.}
\label{fig:LL}
\end{figure}

Figure~\ref{fig:LL} depicts the calculated 
ground-state rotational bands 
of strongly deformed nuclei 
$^{154}$Sm, $^{156}$Gd, $^{230}$Th and $^{232}$U. 
The energy spectra computed by the IBM-2 
in which the rotational response is taken 
into account by the inclusion of the 
$\hat L \cdot \hat L$ term 
(denoted by ``w/'' in the figure), are compared 
with those in which this effect 
is not introduced (``w/o''). 
As is evident from Fig.~\ref{fig:LL}, 
the mapped IBM-2 description of the rotational bands 
is significantly improved by incorporating
in the model the rotational response
through the $\hat L \cdot \hat L$ term.

This finding and those described 
in Sec.~\ref{sec:ibm}, 
indicate that the microscopic formulation of the 
IBM in terms of the mean-field framework 
appears to be valid for deformed nuclei: 
the low-energy spectra 
of nearly spherical and weakly deformed nuclei 
are described sufficiently 
well by the PES mapping procedure; 
for the strongly deformed axially 
symmetric nuclei, in addition to the PES 
the inclusion of the rotational response 
allows for an accurate reproduction 
of rotational bands.

\begin{figure}[h]
\centering
\includegraphics[width=\linewidth]{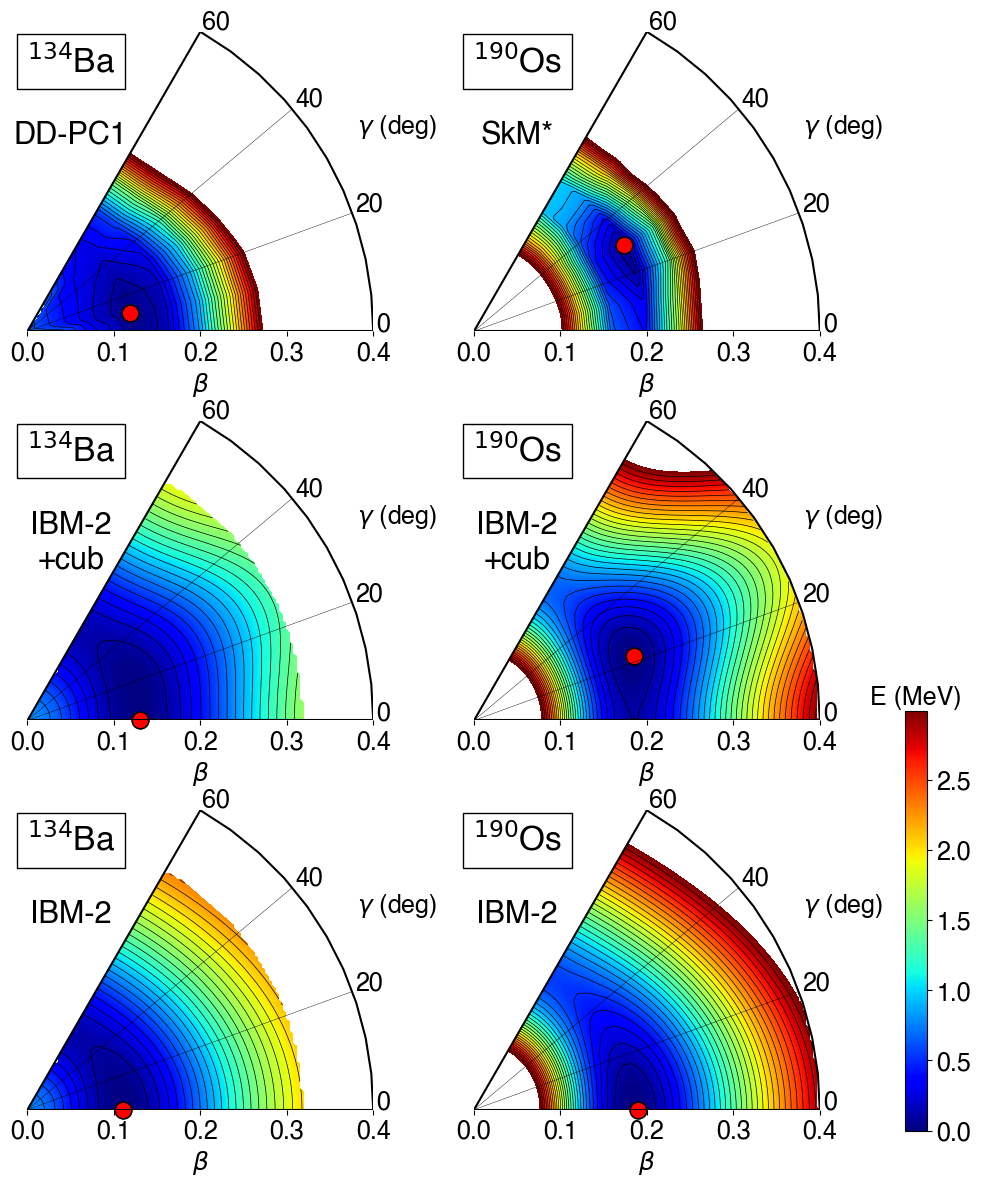}
\caption{
Triaxial quadrupole PESs for the $\gamma$-soft 
nuclei $^{134}$Ba (left) and $^{190}$Os (right). 
In the upper, middle, and bottom rows plotted 
are the SCMF PESs, mapped IBM-2 PESs with (``IBM-2+cub'')
and without (``IBM-2'') the cubic term, respectively. 
The relativistic functional DD-PC1 \cite{DDPC1} 
and the Skyrme SkM* EDF \cite{skms} are used for $^{134}$Ba
and $^{190}$Os, respectively. 
The global minimum is indicated by the 
solid circle.}
\label{fig:3b}
\end{figure}

\subsection{Cubic term and triaxiality in the IBM\label{sec:cubic}}

The IBM-2 Hamiltonian 
in Eq.~(\ref{eq:ham1}) or (\ref{eq:ham1-LL}) 
consists of one- and two-body boson interactions. 
A limitation of using 
the two-body IBM-2 Hamiltonian is that it is 
not able to produce 
a triaxial minimum in the energy surface, 
while the SCMF calculations suggest 
the triaxial equilibrium minima in a number 
of heavy nuclei. 
Within the IBM, the triaxial minimum can be produced 
by the inclusion of three-body (cubic) 
terms \cite{IBM,vanisacker1981,heyde1984}. 
In Ref.~\cite{nomura2012tri}, the cubic boson 
interaction of the form
\begin{align}
\label{eq:3b}
 \hat V_{\rm 3B} = \sum_{\rho\neq\rho'}
\sum_{L}
\theta_{\rho}^{(L)}
[d^\+_{\rho} \times d^\+_{\rho} \times d^\+_{\rho'}]^{(L)}
\cdot
[\tilde d_{\rho'} \times \tilde d_{\rho} \times \tilde d_{\rho}]^{(L)} \; ,
\end{align}
with $\theta_{\rho}^{(L)}$ being strength parameters, 
was introduced in the mapped IBM-2 framework. 
This was also the first to implement the cubic terms 
in the IBM-2, while these terms had been considered
only in the IBM-1.
The cubic term $\hat V_{\rm 3B}$ 
(\ref{eq:3b}) contains 
a number of independent terms corresponding to 
different angular momenta $L$, but it has been shown 
in the phenomenological IBM-1 calculations 
that the term with $L=3$ is most effective to 
give rise to a minimum at $\gamma=30^{\circ}$ 
\cite{vanisacker1981,heyde1984}.
In analogy to the IBM-1, the IBM-2 mapping 
procedure developed in Ref.~\cite{nomura2012tri} 
adopted the $L=3$ term of $\hat V_{\rm 3B}$. 
The expectation value of these terms 
in the coherent state of the IBM-2 
takes the form
\begin{align}
\label{eq:3b-coherent}
 -\frac{1}{7}
\theta^{(3)}
N_{\nu}N_{\pi}
(N_{\nu}+N_{\pi}-2)
\frac{\beta_{\rm B}^3}{(1+\beta_{\rm B})^3}
\sin^2{3\gamma_{\rm B}} \; ,
\end{align}
where $\theta_{\nu}^{(3)}=\theta_{\pi}^{(3)}\equiv \theta^{(3)}$ 
is assumed. 
The formula (\ref{eq:3b-coherent}) 
indeed gives rise to a minimum 
at $\gamma=30^{\circ}$ if a positive 
value $\theta^{(3)}>0$ is chosen.

Figure~\ref{fig:3b} gives the SCMF PESs 
for $\gamma$-soft nuclei $^{134}$Ba and $^{190}$Os. 
The PES for $^{134}$Ba was computed by the relativistic 
Hartree-Bogoliubov (RHB) method \cite{vretenar2005,niksic2011}
employing the density-dependent 
point-coupling (DD-PC1) \cite{DDPC1}
EDF and separable pairing force 
of finite range of Ref.~\cite{tian2009}. 
The PES for $^{190}$Os was obtained from the 
constrained HF+BCS method with the Skyrme 
SkM* interaction and density-dependent $\delta$-type 
pairing force with the strength 1250 MeV fm$^3$. 
The Skyrme HF+BCS PES exhibits for $^{190}$Os 
a triaxial minimum at $\gamma\approx 30^{\circ}$. 
The RHB-SCMF PES for $^{134}$Ba 
shows a shallow triaxial minimum at $\gamma\approx 10^{\circ}$, 
but is considerably soft 
in the $\gamma$ deformation. 
The corresponding mapped IBM-2 PES for $^{190}$Os 
that is obtained by using the Hamiltonian 
including the cubic boson term of 
the type (\ref{eq:3b}) shows a global minimum 
at $\gamma\approx 30^{\circ}$, 
as in the case of the Skyrme-SkM* PES. 
The IBM-2 PESs computed by the IBM-2 
that consists only of two-body boson terms 
do not exhibit any triaxial global minimum 
for both nuclei. 
For $^{134}$Ba, 
the IBM-2 PES with the cubic term 
does not show a triaxial minimum. 
However, the triaxial minimum predicted 
in the DD-PC1 PES is also so shallow that 
it is not of crucial importance to try to 
reproduce very precisely the location of the minimum. 
One can indeed observe differences between the IBM-2 
PESs with and without the inclusion 
of the cubic term for $^{134}$Ba: 
the former looks more 
extended in the $\gamma\approx 30^{\circ}$ 
direction than the latter and is consistent with 
the DD-PC1 PES.

%-------- 190Os spectra ----------
\begin{figure}[h]
\centering
\includegraphics[width=\linewidth]{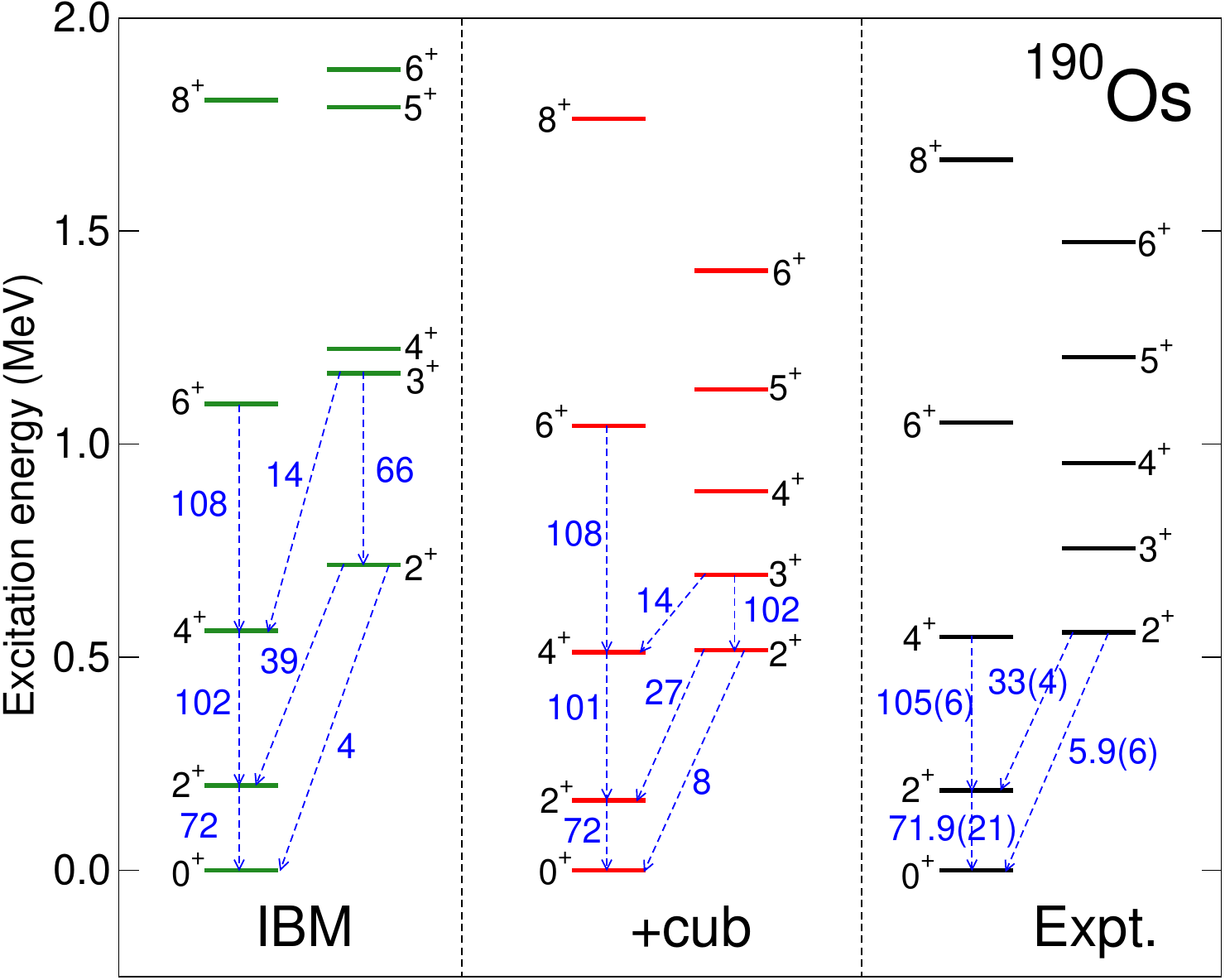}
\caption{
Low-energy spectra and $B(E2)$ transition 
strengths in Weisskopf units (W.u.) 
for the $^{190}$Os nucleus 
obtained from the mapped IBM-2 with (``+cub'') 
and without (``IBM'') including the 
cubic term. The experimental data are taken 
from NNDC \cite{data}. 
The Skyrme SkM* EDF is used. 
Effective charge for the $E2$ operator is 
fitted to reproduce the experimental
$B(E2;2^+_1 \to 0^+_1)$ value.}
\label{fig:os190-3b}
\end{figure}
%---------------------------------

As an example illustrating effects of
the cubic term on the spectroscopic 
properties, Fig.~\ref{fig:os190-3b} 
gives the low-energy levels for $^{190}$Os, 
obtained from the IBM-2 Hamiltonians with up to 
three-body and two-body boson terms. 
The diagonalization of the Hamiltonians 
was made in the boson 
$m$-scheme basis \cite{nomura2012phd}.
The two calculated results are noticeably different:
the mapped IBM-2 calculation 
without the cubic term produces the band built on 
the $2^+_2$ state that 
exhibits an odd-even-spin staggering 
of energy levels  
$2^+,(3^+,4^+),(5^+,6^+),\ldots$, 
whereas in the calculated results obtained with 
the cubic term 
the $2^+_2$-band levels including the odd-spin 
members are significantly lowered in energy, 
and form a harmonic structure that is 
characterized by almost equal energy-level spacing.
The predicted $2^+_2$ band 
with the cubic term are consistent with the observed 
$\gamma$-vibrational band. 
The calculated $B(E2)$ transition 
probabilities (the numbers shown along 
allows in the figure) in the mapped IBM-2 
are in the same order of magnitude as 
the experimental values \cite{data}.

An important consequence of introducing the cubic 
boson term in the IBM-2 concerns 
the interpretation of the level structures 
of $\gamma$-soft nuclei. 
There are two geometrical limits to describe structures of 
$\gamma$-soft nuclei: the $\gamma$-unstable rotor model 
of Wilets and Jean (W-J) \cite{wilets1956}, which 
is also equivalent to the 
O(6) limit of the IBM, and the rigid-triaxial rotor 
model of Davydov and Filippov (D-F) \cite{davydov1958}. 
In the W-J limit, the $\gamma$-band exhibits a 
staggering pattern 
$2^+,(3^+,4^+),(5^+,6^+),\ldots$, whereas in the 
D-F limit a different staggering pattern 
$(2^+,3^+),(4^+,5^+),\ldots$ appears. 
In most of the realistic nuclei in medium-mass 
and heavy regions, however, these two geometrical 
limits are rarely realized. The observed 
$\gamma$-vibrational bands show features that 
are rather in between those obtained in 
the two limits: 
the $2^+$, $3^+$, $4^+$, $5^+$, $6^+$, 
$\ldots$ levels appear with equal energy spacing, 
as shown in Fig.~\ref{fig:os190-3b}.  
Within the IBM-2 framework 
these observed energy level patterns are reproduced 
naturally only by the inclusion of 
the cubic boson term.

On the basis of the results given in this section,
it appears that for the description of the 
low-energy level structures of $\gamma$-soft nuclei 
it is necessary to include three-body interactions 
in the IBM-2 framework. 
This conclusion is justified by the fact 
that the cubic term is introduced 
so as to reproduce the triaxial minimum predicted 
by the EDF-SCMF calculations, and suggests the 
optimal IBM-2 description of $\gamma$-soft nuclei.

\section{Configuration mixing and shape coexistence\label{sec:co}}

The present section outlines the extension 
of the IBM mapping procedure 
to study those nuclei for which more than one 
mean-field minimum appear in the energy surfaces. 
Most of the formulations given below, 
and applications to the neutron-deficient 
Pb and Hg nuclei are discussed in 
a previous review article of Ref.~\cite{nomura2016sc}, 
which is focused on the mapped IBM study 
on shape coexistence. 
In this section,  
the shape evolution and coexistence in the neutron-rich 
isotopes with mass $A\approx 100$ are mainly discussed, 
while the corresponding formulations and applications 
in the Pb-Hg regions are described briefly.

\subsection{Formalism}

In the standard IBM the boson model space 
corresponds to a given valence space. 
As in the case of the large-scale nuclear shell model, 
it is in some cases insufficient 
to consider only one major oscillator shell, and 
certain core polarization effects have to be 
incorporated by taking a larger model space. 
Such an extension is important 
especially to compute the low-lying 
excited $0^+$ states that are 
interpreted as the intruder states. 
The very low-lying $0^+$ levels near the 
$0^+_1$ ground state have been observed in 
a number of nuclei, and are attributed to a 
signature of shape coexistence, a phenomenon 
in which multiple shapes appear simultaneously 
in the vicinity of the ground state 
\cite{wood1992,heyde2011,garrett2022,leoni2024}. 
Empirical evidence for the shape coexistence 
in heavy-mass regions has been found, 
e.g., in the neutron-deficient Pb and Hg 
nuclei near the middle of the 
neutron major shell $N=82-126$, 
and neutron-rich isotopes with $Z \approx 40$ 
and $N\approx 60$.

A method to incorporate the deformed intruder 
configurations and to mix the normal and 
intruder states in the IBM-2 was introduced in 
\cite{duval1981,duval1982}. 
The intruder states are considered to
be shell-model-like states arising from 
two-particle-two-hole ($2p-2h$), 
four-particle-four-hole ($4p-4h$), $\ldots$ 
excitations from a next major oscillator shell. 
The normal or $0p-0h$ configuration 
comprises $N_{\rm B} = N_{\nu} + N_{\pi}$ bosons 
in a given valence space, while, assuming 
that the particle and hole states 
are not distinguished, the intruder 
$2mp-2mh$ ($m \geqslant 1$) configurations 
correspond to the boson spaces consisting of 
$N_{\rm B}+2m$ bosons. 
In the following, 
the particle-hole excitations of protons 
are considered, since the empirical evidence 
for the shape coexistence is found mainly 
in those nuclei (e.g., Pb and Hg) that are close to the 
proton major shell closure, and the proton cross-shell 
excitations are more likely to occur. 
The entire boson space that includes 
proton $2mp-2mh$ ($m \geqslant 0$) 
configurations is expressed as a direct 
sum of subspaces that differ in boson number by 2:
\begin{align}
\label{eq:cmspace}
 [N_{\nu}\otimes N_{\pi}] 
\oplus [N_{\nu}\otimes(N_\pi+2)] 
\oplus [N_{\nu}\otimes(N_\pi+4)]
\oplus \cdots \; .
\end{align}
The corresponding IBM-2 Hamiltonian
with configuration mixing (IBM-CM) is written as
\begin{align}
\label{eq:ibmcm}
 \hat H_{\rm IBM}^{\rm (CM)} = 
& \sum_{m=0}
\left\{
\hat P_{m} ( \bh^{(m)} + \Delta_{m} ) \hat P_{m}
\right.
\nonumber\\
&
\quad
\quad
\quad
\left.
+\hat P_{m} \hat V_{\rm mix}^{(m,m+1)} \hat P_{m+1} 
+ ({\rm H.c.})
\right\} \; .
\end{align}
$\bh^{(m)}$ and $\hat P_m$ 
represent the IBM-2 Hamiltonian for and the projection 
operator onto the $[N_{\rm B}+2m]$ space, respectively.
The unperturbed Hamiltonian $\bh^{(m)}$
is of the form given in Eq.~(\ref{eq:ham1}).
$\Delta_m$ is the energy required to promote 
$2m$ particles from the next major shell. 
For the normal configuration, $\Delta_0=0$. 
The term $\hat V_{\rm mix}^{(m,m+1)}$ stands for the 
interaction that mixes the configurations 
$[N_{\rm B}+2m]$ and $[N_{\rm B}+2(m+1)]$, and 
takes the form
\begin{align}
\label{eq:mix}
\hat V_{\rm mix}^{(m,m+1)}
= & \omega_{s}^{(m,m+1)} s^\+_{\pi} \cdot s^\+_{\pi}
\nonumber \\
  & 
\quad\quad
+ \omega_{d}^{(m,m+1)} d^\+_{\pi} \cdot d^\+_{\pi}
+ ({\rm H.c.}) \; ,
\end{align}
where $\omega_{s}^{(m,m+1)}$ and $\omega_{d}^{(m,m+1)}$ 
are mixing strength parameters for 
$s$ and $d$ boson terms, respectively, 
and are assumed to be equal,  
$\omega_{s}^{(m,m+1)}= \omega_{d}^{(m,m+1)} \equiv \omega^{(m,m+1)}$.
Note that only the configuration mixing between 
$2mp-2mh$ and $2(m+1)p-2(m+1)h$ spaces is 
considered, since there is no two-body interaction 
that couples states with nucleon numbers differing 
by more than 2. 
The transition operators 
are also extended in a similar way 
to the Hamiltonian. 
The $E2$ operator, for instance, is given as
\begin{align}
 \hat T^{(E2)} 
= \sum_{m}
\hat P_{m}
\left(
e^{{\rm B}}_{\nu,m} \hat Q_{\nu,m} + 
e^{{\rm B}}_{\pi,m} \hat Q_{\pi,m}
\right)
\hat P_{m} \; ,
\end{align}
where $\hat Q_{\rho,m}$ is of the form (\ref{eq:bquad}), 
and $e^{{\rm B}}_{\rho,m}$ is the effective 
charge for the $2mp-2mh$ configuration.

%-------- pb186 PES ----------
\begin{figure}[h]
\centering
\includegraphics[width=\linewidth]{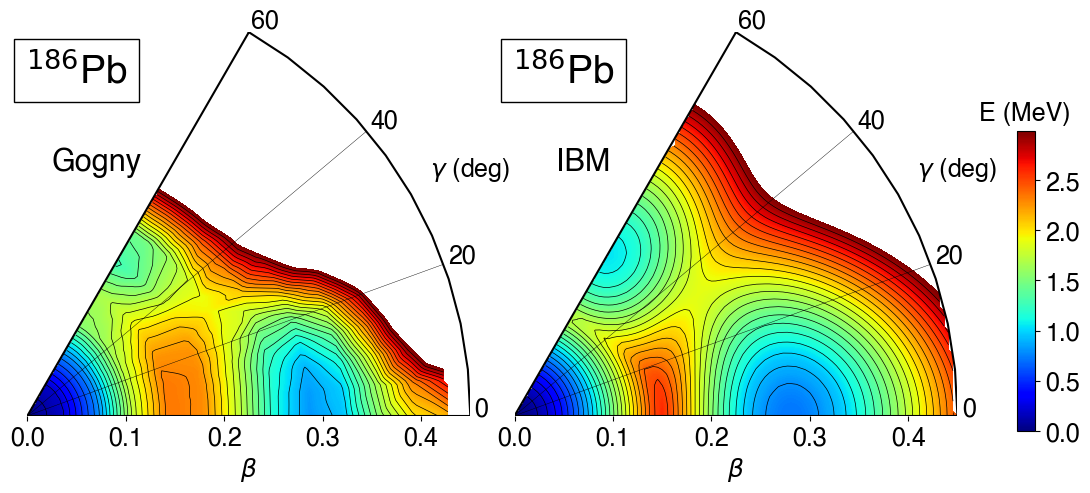}
\caption{SCMF and mapped IBM-CM PESs for $^{186}$Pb.
The Gogny-D1M interaction is used.}
\label{fig:pb186-pes}
\end{figure}
%-----------------------------

The method developed in \cite{nomura2012sc} 
provides a way of determining 
a large number of parameters for the 
IBM-CM Hamiltonian (\ref{eq:ibmcm}), that is, 
the strength parameters for each of the 
unperturbed Hamiltonians associated with 
$2mp$-$2mh$ configurations, 
mixing strengths $\omega$, and 
energy offsets $\Delta$. 
Consider, for example, the nucleus $^{186}$Pb, 
in which a spectacular triple 
shape coexistence of spherical, 
oblate, and prolate shapes was suggested 
to emerge \cite{andreyev2000}. 
The corresponding PES, computed by the 
HFB method employing the Gogny D1M EDF \cite{D1M}, 
shown in Fig.~\ref{fig:pb186-pes}, exhibits 
the spherical global minimum, and two local 
minima on the oblate and prolate sides. 
To account for the three mean-field minima, 
the IBM-CM space should contain up to three 
IBM-2 subspaces. 
The procedure to determine the IBM-CM parameters 
is summarized as follows. 
\begin{enumerate}
 \item 
The parameters for the unperturbed 
$0p-0h$, $2p-2h$, and $4p-4h$ Hamiltonians 
are determined so as to reproduce 
the topology of the PES in the neighborhoods 
of the spherical global, 
oblate local, and prolate local minima, respectively. 
The association of the unperturbed Hamiltonians 
with the different mean-field minima 
is based on the assumption that 
higher particle-hole excitations correspond to 
those mean-field minima with 
larger intrinsic deformations \cite{nazarewicz1993}. 

\item
The energy offsets $\Delta_1$ and $\Delta_2$ are fixed 
so that the energy differences between the 
global and local mean-field minima, 
that is, between the spherical and oblate minima, 
and between the spherical and prolate minima, 
respectively, should be reproduced. 

\item
The final step is to introduce 
the mixing interaction (\ref{eq:mix}) and 
fix the parameters $\omega^{(0,1)}$ 
and $\omega^{(1,2)}$. 
The expectation value of the IBM-CM Hamiltonian 
in the coherent state is given 
as an eigenvalue of the following 
$3\times 3$ matrix \cite{frank2004}. 
\begin{align}
\label{eq:pescm}
&{\cal E}(\beta,\gamma)
\nonumber\\
&=\left(
\begin{array}{ccc}
E_{0}(\beta,\gamma) & \Omega_{0,1}(\beta,\gamma) & 0 \\
\Omega_{1,0}(\beta,\gamma) & E_1(\beta,\gamma)+\Delta_{1} & \Omega_{1,2}(\beta,\gamma)
 \\
0 & \Omega_{2,1}(\beta,\gamma) & E_2(\beta,\gamma)+\Delta_{2} \\
\end{array}
\right) \; ,
\end{align}
where 
$E_m(\beta,\gamma)$ stands for the 
expectation value in the coherent state 
of $(N_\nu,N_\pi+2m)$ bosons, and the non-diagonal 
element $\Omega_{m,m+1}(\beta,\gamma)=\Omega_{m+1,m}(\beta,\gamma)$, 
is the expectation value of $\hat V_{\rm mix}^{m,m+1}$
in the coherent states for the 
$[N_\pi+2m]$ and
$[N_\pi+2(m+1)]$ spaces. 
The $\omega^{(0,1)}$ 
and $\omega^{(1,2)}$ values are then 
fixed so that the lowest eigenvalue of (\ref{eq:pescm}) 
reproduces the barriers between the 
spherical and oblate, and between the 
oblate and prolate minima, respectively.
In this step, the $\Delta$ values are also 
readjusted so that the energy differences 
between neighboring minima remain unchanged. 
\end{enumerate}
The mapped IBM-CM PES is shown 
in Fig.~\ref{fig:pb186-pes}, exhibiting three minima 
consistently with the Gogny-HFB PES. 

%-------- pb186 spectra ----------
\begin{figure}[ht]
\centering
\includegraphics[width=\linewidth]{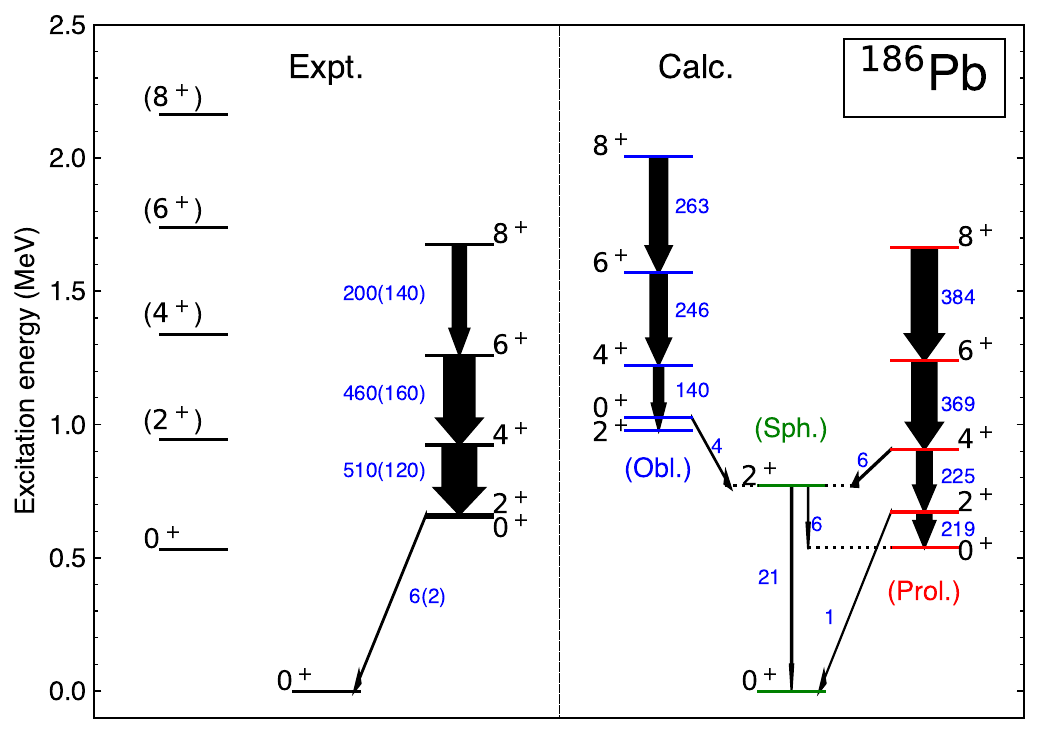}
\caption{Calculated and experimental \cite{data} 
low-energy spectra and $B(E2)$ values in W.u. 
(numbers along arrows) for $^{186}$Pb.}
\label{fig:pb186-spec}
\end{figure}
%---------------------------------

The mapped IBM-CM Hamiltonian is diagonalized 
in the extended boson space defined in (\ref{eq:cmspace}). 
The resultant low-energy spectra are shown 
in Fig.~\ref{fig:pb186-spec}, and are compared to 
the experimental ones \cite{andreyev2000,data}. 
The mapped IBM-CM predicts three bands, which 
are mainly composed of the spherical, oblate, 
and prolate configurations, respectively. 
Specifically, the wave functions of 
the $0^+_1$, $0^+_2$, and $0^+_3$ states 
and those states built on them 
were shown to be made mainly of the normal ($0p-0h$), 
intruder $4p-4h$ and $2p-2h$ configurations, 
respectively, which are associated with the 
spherical global, prolate, and oblate minima in
the energy surface (see Fig.~\ref{fig:pb186-pes}). 
The two excited bands of the prolate and oblate 
characters also exhibit strong $\Delta I=2$ 
in-band $E2$ transitions.

\subsection{Systematic studies}

A benchmark study in \cite{nomura2012sc} 
showed that the IBM-CM worked reasonably 
in describing overall features 
of the low-lying levels in the neutron-deficient 
Pb isotopes. 
The shape evolution and coexistence in the 
even-even $^{172-204}$Hg were studied 
in \cite{nomura2013hg}, in which 
the Gogny-D1M HFB calculations were performed 
as a starting point, and the $\hat L \cdot \hat L$ 
and cubic (\ref{eq:3b}) terms were 
included in the unperturbed Hamiltonian (\ref{eq:ham1}). 
In the Hg chain, 
in addition to the yrast band of the weakly oblate
deformed configurations, which is more or less stable 
against the neutron number $N$, an intruder band 
of strongly prolate deformed configuration exhibits 
a parabolic dependence on $N$ centered around the middle of the 
neutron major shell $N \approx 104$. 
In Ref.~\cite{nomura2013hg}, 
observables that indicate the oblate-prolate 
shape coexistence were computed, i.e., 
the energy levels, $B(E2)$, quadrupole moments, 
monopole transitions $\rho^2(E0)$, 
and mean-square radii.

%-------- Zr-Sr PES --------------
\begin{figure*}[h]
\centering
\includegraphics[width=\linewidth]{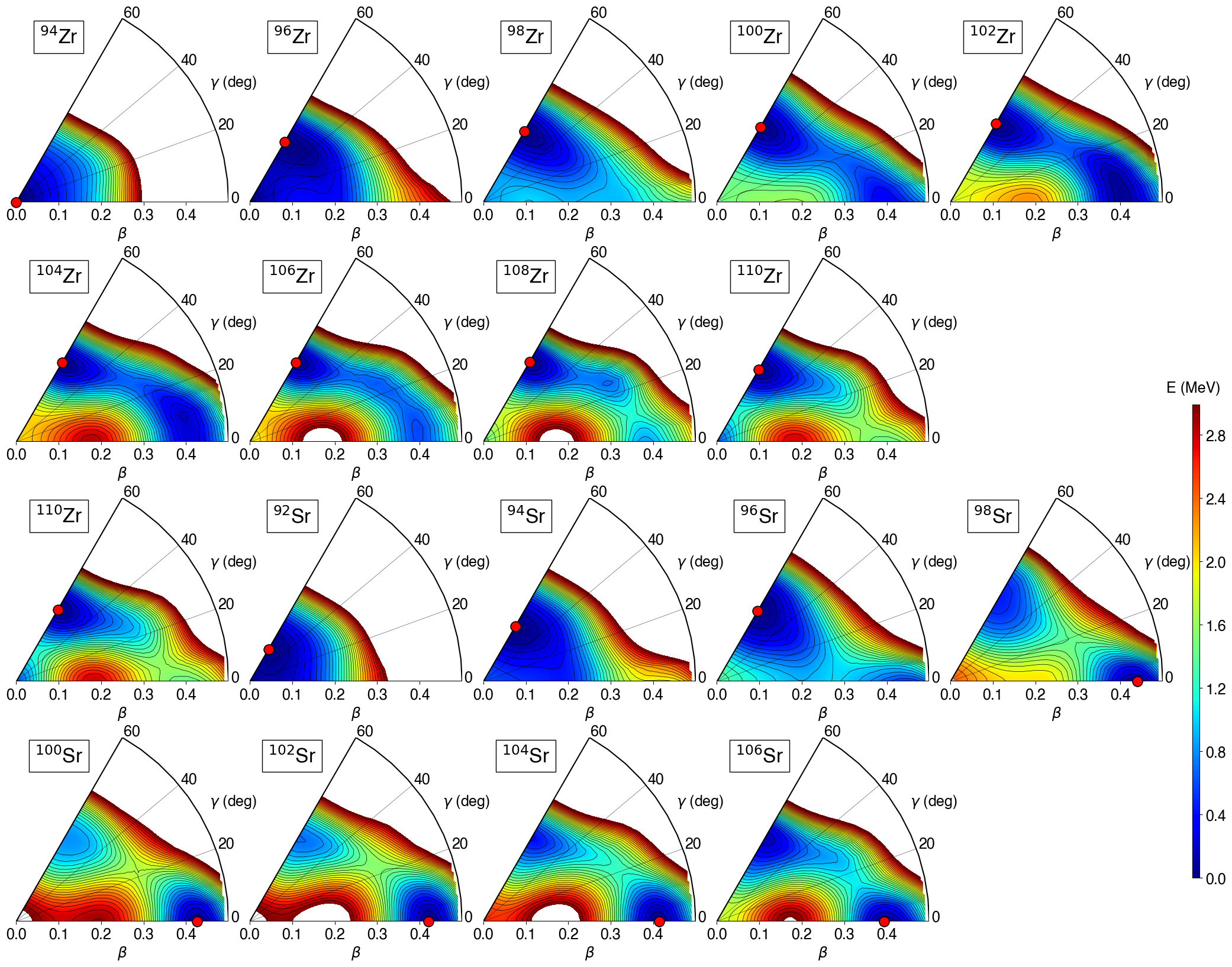}
\caption{
Triaxial quadrupole $(\beta,\gamma)$ PESs
for the even-even nuclei 
$^{94-110}$Zr and $^{92-108}$Sr 
computed by the constrained Gogny-HFB method 
using the D1M interaction.
The energies are 
drawn up to 3 MeV from the global minimum, 
which is indicated by the solid circle.
The energy difference between neighboring contours 
is 100 keV.}
\label{fig:zr-pes1}
\end{figure*}
%---------------------------------

Another interesting but challenging case in which 
shape coexistence is expected to emerge is 
the neutron-rich nuclei in 
the mass $A \approx 100$ region. 
The shape transitions in the Zr isotopes are 
of particular interest. 
Along this isotopic chain 
a rapid nuclear structure change occurs 
at $N \approx 60$, and it is regarded as a manifest 
shape QPT, which is accompanied
by shape coexistence
\cite{togashi2016,kremer2016,gavrielov2019,garciaramos2020,leoni2024}. 
For heavier Zr nuclei with $N \approx 70$,
an emergence of triaxial deformations
has been suggested
\cite{togashi2016,moon2024}.
The underlying mechanism of the 
rich variety of nuclear structure phenomena 
in Zr consists in a subtle interplay 
between the single-particle and collective 
degrees of freedom, and has nowadays been 
a subject of extensive theoretical and 
experimental investigations
(see Ref.~\cite{leoni2024} for a recent review).

%-------- Zr yrast ---------------
\begin{figure}[h]
\centering
\includegraphics[width=\linewidth]{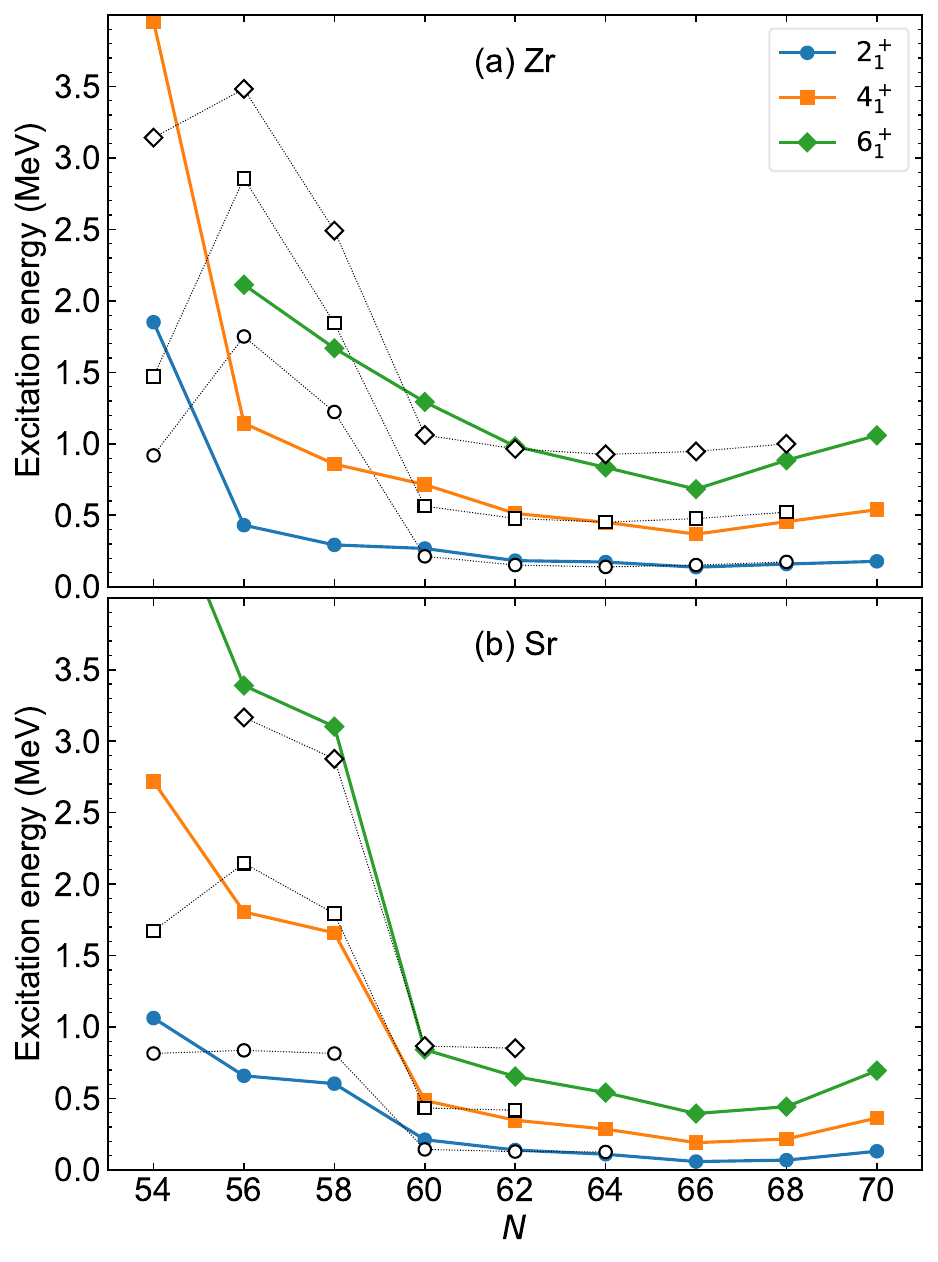}
\caption{
Excitation energies of the $2^+_1$, $4^+_1$, and $6^+_1$
states for $^{94-110}$Zr and $^{92-108}$Sr 
obtained from the Gogny-D1M
mapped IBM-CM (shown as the solid symbols 
connected by solid lines), 
and the corresponding experimental data \cite{data}
(open symbols connected by broken lines).}
\label{fig:zr-gs}
\end{figure}
%---------------------------------

%-------- Zr yrare ---------------
\begin{figure}[h]
\centering
\includegraphics[width=\linewidth]{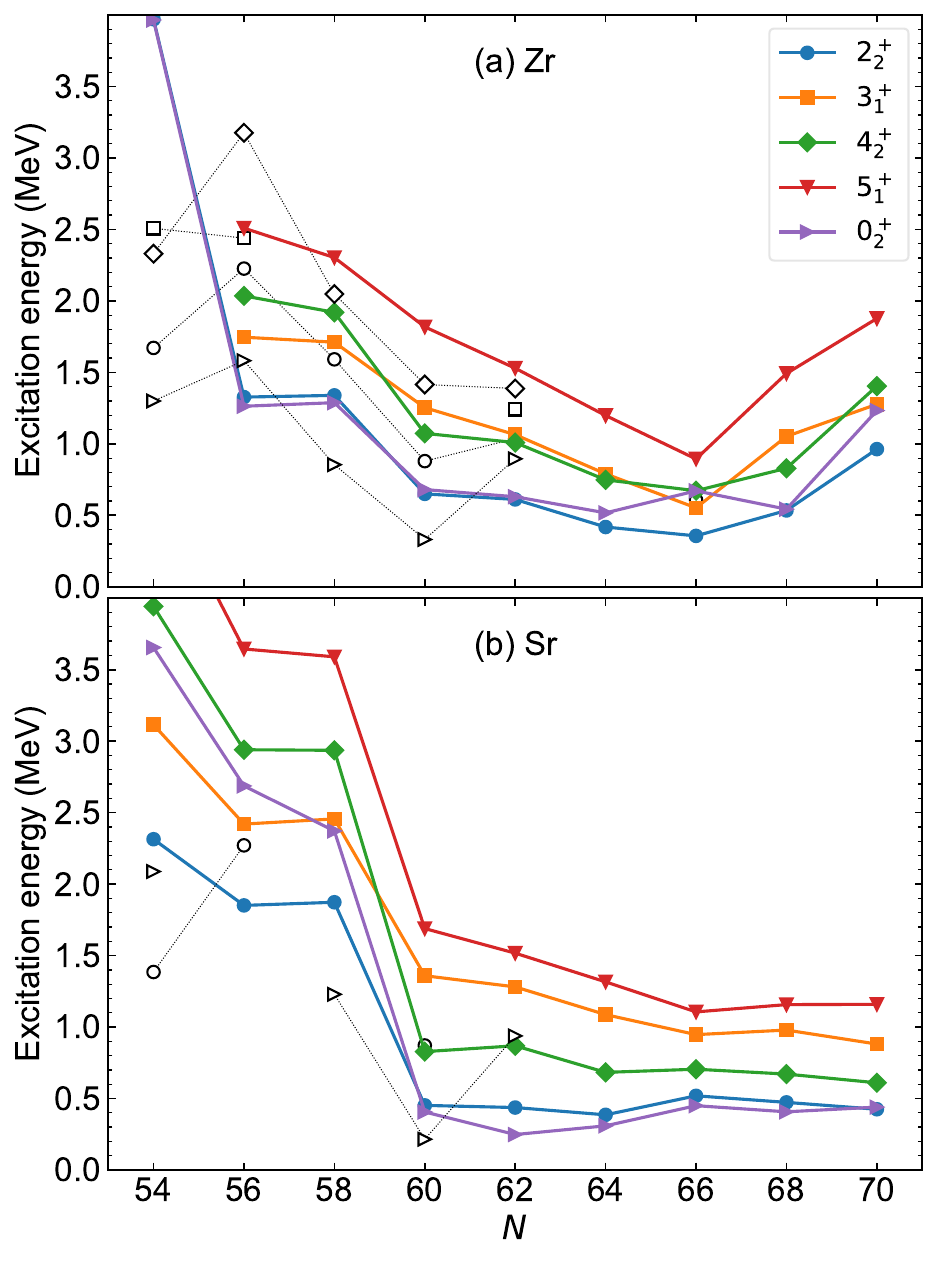}
\caption{Same as the caption to Fig.~\ref{fig:zr-gs}, 
but for the $2^+_2$, $3^+_1$, $4^+_2$, $5^+_1$, 
and $0^+_2$ states.}
\label{fig:zr-qg}
\end{figure}
%---------------------------------

Shape phase transitions 
and coexistence in the even-even 
Sr, Zr, Mo, and Ru nuclei 
with $N=54-70$ were investigated in the 
mapped IBM-CM including 
the proton $0p-0h$, $2p-2h$, and $4p-4h$
excitations across the $Z=40$ sub-shell 
closure \cite{nomura2016zr}, 
for which the microscopic input was provided 
by the constrained HFB calculations 
employing the Gogny D1M EDF \cite{D1M}. 
The Gogny-HFB calculations 
for the isotopes $^{94-110}$Zr 
and $^{92-108}$Sr suggested a variety of shapes 
and shape evolution and, in particular, the appearance 
of more than one mean-field minimum 
in the corresponding triaxial quadrupole 
$(\beta,\gamma)$ PESs in many of these nuclei 
(Fig.~\ref{fig:zr-pes1}): 
a spherical minimum is obtained for 
$^{94}$Zr, and there appears coexistence 
of a weakly oblate global minimum and a spherical 
local minimum in $^{96}$Zr. 
A more pronounced oblate deformation 
is suggested for $^{98}$Zr. 
For $^{100}$Zr, coexistence of an oblate 
global minimum and a strongly deformed 
prolate local minimum is predicted.
Similar structure is suggested 
for heavier nuclei $^{102,104,106}$Zr. 
For $^{108,110}$Zr, the global minimum occurs 
on the oblate side, and several local minima 
in the triaxial, prolate, and spherical configurations 
are suggested. 
The variation of the PES as a function of 
neutron number $N$ in the neighboring Sr isotopes 
appears to be more or less 
similar to that in the Zr isotopic chain.

The mapped IBM-CM results on the energy spectra 
exhibit signatures of the rapid 
shape transitions at $N\approx 60$. 
(see Fig.~\ref{fig:zr-gs} for yrast states, 
and Fig.~\ref{fig:zr-qg} for those 
states in non-yrast bands). 
The variation of the energy levels with $N$ 
also reflects the evolution of the intrinsic
shapes indicated at the mean-field level. 
One notices that the IBM-CM is not 
able to reproduce the excitation energies 
of the yrast states in $^{96}$Zr and $^{98}$Zr. 
This is because the corresponding Gogny-HFB PESs 
exhibit a pronounced oblate deformation 
(Fig.~\ref{fig:zr-pes1}), whereas a spherical 
ground state is expected for these nuclei, 
which are close to the $N=56$ and $Z=40$ 
sub-shell closures. 
An empirical signature of shape coexistence is the 
lowering of the non-yrast band levels
including those of the $0^+_2$ states near the 
transition point $N=60$. 
This trend is reasonably reproduced  
by the mapped IBM-CM. 

\begin{table}[!htb]
\caption{\label{tab:zrsr-frac}
Fractions in percent of the $2mp-2mh$ ($m=0,1,2$) 
configurations in the IBM-CM wave functions 
of the $0^+_1$ and $0^+_2$ states of the $^{96-110}$Zr 
and $^{98-110}$Sr. 
$[0]$, $[1]$, and $[2]$ denote the boson
subspaces associated with a nearly spherical,
an oblate, and a prolate minima, respectively.
Only for $^{96}$Sr, the $[0]$ and $[1]$ spaces 
correspond to the oblate and prolate minima, 
respectively.
}
\centering
\begin{tabular}{lcccccc}
\hline\hline
 & \multicolumn{3}{c}{$0^+_1$} & \multicolumn{3}{c}{$0^+_2$} \\
\cline{2-4}\cline{5-7}
Nucleus & $[0]$ & $[1]$ & $[2]$ & $[0]$ & $[1]$ & $[2]$ \\
%Nucleus & Sph. & Obl. & Prol. & Sph. & Obl. & Prol. \\
\hline
$^{96}$Zr & 1.9 & 98.1 &  & 91.9 & 8.1 &  \\
$^{98}$Zr & 0.7 & 91.2 & 8.0 & 0.0 & 20.3 & 79.7 \\
$^{100}$Zr & 0.5 & 67.0 & 32.4 & 0.5 & 30.6 & 68.9 \\
$^{102}$Zr & 0.1 & 17.0 & 82.9 & 0.4 & 77.3 & 22.3 \\
$^{104}$Zr & 0.0 & 27.3 & 72.7 & 0.1 & 64.2 & 35.7 \\
$^{106}$Zr & 0.1 & 23.1 & 76.8 & 0.1 & 43.7 & 56.2 \\
$^{108}$Zr & 0.3 & 82.0 & 17.7 & 0.0 & 20.6 & 79.4 \\
$^{110}$Zr & 0.6 & 94.4 & 5.1 & 0.4 & 39.9 & 59.7 \\
\hline
$^{96}$Sr & 98.4 & 1.6 &  & 4.0 & 96.0 &  \\
$^{98}$Sr & 2.4 & 60.6 & 37.0 & 1.9 & 37.6 & 60.5 \\
$^{100}$Sr & 10.6 & 28.2 & 61.2 & 41.3 & 24.9 & 33.7 \\
$^{102}$Sr & 0.7 & 33.8 & 65.5 & 2.1 & 66.2 & 31.8 \\
$^{104}$Sr & 0.1 & 5.2 & 94.7 & 1.7 & 93.2 & 5.0 \\
$^{106}$Sr & 0.2 & 9.3 & 90.6 & 2.4 & 88.9 & 8.6 \\
$^{108}$Sr & 2.2 & 86.3 & 11.4 & 0.1 & 19.9 & 80.0 \\
\hline\hline
\end{tabular}
\end{table}

Structures of the IBM-CM wave functions for 
the $0^+$ states 
were analyzed by the decomposition into the 
components corresponding to the configurations 
$[m]$, representing the $2mp-2mh$ ($m=0,1,2$) 
excitations. 
Fractions of these configurations in the 
$0^+_1$ and $0^+_2$ wave functions are summarized 
in Table~\ref{tab:zrsr-frac}. 
The spaces $[0]$, $[1]$, and $[2]$ are 
associated with a nearly spherical, oblate, and 
prolate minima, respectively, for 
the $^{96-110}$Zr and $^{98-108}$Sr. 
For $^{96}$Sr, the $[0]$ and $[1]$ spaces 
correspond to the oblate and prolate minima, 
respectively. 
Also for $^{96}$Zr and $^{96}$Sr up to 
$2p-2h$ configurations were considered, and 
for $^{94}$Zr and $^{92}$Sr only one configuration 
was used, because the PES exhibits a single minimum. 
One can see that 
the oblate $[1]$ and prolate $[2]$ 
configurations are strongly mixed 
in those Zr and Sr nuclei near $N=60$. 

%-------- Zr-Sr E2 ---------------
\begin{figure}[h]
\centering
\includegraphics[width=\linewidth]{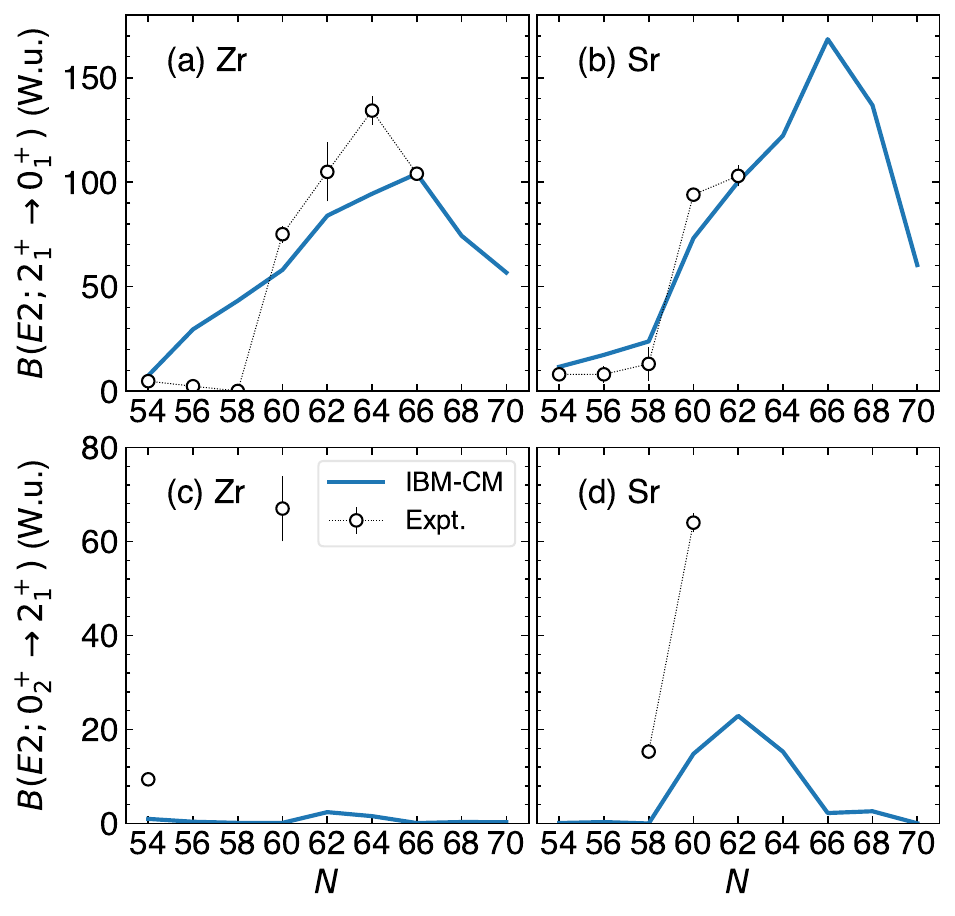}
\caption{Predicted 
and experimental $B(E2;2^+_1 \to 0^+_1)$ and 
$B(E2;0^+_2 \to 2^+_1)$ values for Zr and Sr isotopes. 
The neutron and proton boson effective charges are chosen to 
be equal, and the same values are employed 
for the three configuration spaces. 
These effective charges for the Zr 
and Sr isotopic chains are determined
so as to reproduce the 
experimental $B(E2;2^+_1 \to 0^+_1)$ values 
of the $^{106}$Zr and $^{100}$Sr nuclei, 
respectively. 
}
\label{fig:zr-e2}
\end{figure}
%---------------------------------

Figure~\ref{fig:zr-e2} gives the mapped IBM-CM 
results of the $B(E2;2^+_1 \to 0^+_1)$ and 
$B(E2;0^+_2 \to 2^+_1)$ transition probabilities 
for the Zr and Sr nuclei. 
The increase of the $B(E2;2^+_1 \to 0^+_1)$ from 
$N=56$ to 62 in both isotopic chains is reproduced 
by the calculations, while for the Zr isotopes 
the variation of the theoretical values is 
much more gradual than the experimental ones. 
This reflects the gradual decrease in energy 
of the yrast states from $N=58$ to 60, and appears 
to occur because 
the Gogny-HFB PESs exhibit an unexpectedly large 
deformation for these nuclei. 
The calculated $B(E2;2^+_1 \to 0^+_1)$ values 
for the Sr nuclei are consistent with 
the experimental values. 
The $B(E2;0^+_2 \to 2^+_1)$ transition is 
considered to be a signature 
of shape mixing/coexistence. 
The mapped IBM-CM predicts much lower values 
for this transition in both isotopic chains, 
indicating that the mixing may not be properly 
accounted for by the IBM-CM. 
The results on the $B(E2)$ systematic call for 
further improvements of the theoretical framework 
at the mean-field or/and IBM levels.

The mapped IBM-CM has been applied to the spectroscopic 
calculations on the shape-related phenomena in several 
other mass regions, or by using the different EDF inputs.
Using the HF+BCS method with 
the Skyrme-SLy6 EDF \cite{sly}, spectroscopic 
calculations were carried out for the $^{96,98,100}$Mo 
nuclei \cite{thomas2013,thomas2016},
and suggested coexistence of a nearly spherical and 
a shallow triaxial minima for the nucleus $^{98}$Mo, 
and that considerable contributions 
of the intruder configurations are included 
in both the ground and second excited $0^+$ states.

By means of the mapped IBM-CM employing 
the relativistic and nonrelativistic Gogny EDFs, 
a systematic study was carried out for 
describing the low-energy spectroscopic 
properties of the even-even Kr isotopes from 
the neutron-deficient ($N=34$) to 
the neutron-rich ($N=74$) sides \cite{nomura2017kr}, 
This study suggested many examples for the
prolate-to-oblate shape transitions and shape 
coexistence on both neutron-deficient and 
neutron-rich sides of the isotopic chain. 
In a similar fashion, the mapped IBM-CM 
calculations based on the Gogny D1M EDF 
\cite{nomura2017ge} was performed to investigate 
coexistence of the prolate and oblate shapes 
in the neutron-rich and neutron-deficient 
even-even nuclei $^{66-94}$Ge and $^{68-96}$Se. 
Many of these nuclei were shown to be $\gamma$ 
soft in intrinsic shape, and the coexistence 
of the spherical and $\gamma$ soft and 
of the prolate and oblate shapes were 
indicated.

Both of these studies in Refs.~\cite{nomura2017kr,nomura2017ge}
provided reasonable descriptions of
the observed low-energy 
and low-spin levels, including the $0^+_2$ one, 
and $B(E2)$ and $\rho^2(E0)$ values in Ge, Se, and Kr isotopes,
and also provided predictions
for the neutron-rich nuclei
for which data are scarce or do not exist.
For the IBM configuration mixing, the proton boson 
space was taken to be the $Z=28-50$ major shell, 
and the intruder excitations were assumed to 
occur from the $Z=28$ shell closure. 
It is worth to note that 
the standard IBM, i.e., IBM-2 and IBM-1, 
consists of the neutron-neutron and proton-proton pairs, 
and this assumption is reasonable for 
describing heavy nuclei, 
in which protons and neutrons usually occupy 
different orbits. 
In the neutron-deficient nuclei such as 
these neutron-deficient Ge, Se, and Kr nuclei, however, 
the neutron and proton numbers 
are approximately equal $N\approx Z$, 
in which cases contributions of the 
neutron-proton pairs may not be negligible 
for a more realistic analysis.

Nuclei close to the proton $Z=50$ magic number 
such as those in the Cd and Sn isotopic chains 
were considered classic examples for an anharmonic 
vibrator exhibiting a multi-phonon level structure. 
There have been, however, experimental data suggesting
additional low-spin levels close in energy
to the two- or three-phonon
multiplets \cite{heyde2011,garrett2022,leoni2024},
and the appearance of
these states cannot be explained without 
considerations of the intruder contributions. 
To identify intruder states in even-even Cd nuclei,
the mapped IBM-CM calculation with 
the microscopic input from the HF+BCS method 
based on the Skyrme SLy6 EDF was 
carried out \cite{nomura2018cd} 
to obtain low-lying excitation 
spectra, electric quadrupole and monopole 
transition rates of $^{108-116}$Cd. 
The mapped IBM-CM calculation predicted 
several intruder states in the Cd nuclei, 
as suggested experimentally, and shed light upon 
the interpretation of vibrational spectra 
in the $Z\approx 50$ region.

\section{Higher-order deformations\label{sec:ho}}

The intrinsic shape of most medium-heavy and heavy 
nuclei is characterized by the 
quadrupole deformation, but in specific mass regions 
higher-order deformations play important 
roles at low energy. 
This section reviews extensions 
of the IBM mapping method to include 
the octupole (Sec.~\ref{sec:oct}), 
and hexadecapole (Sec.~\ref{sec:hex}) 
shape degrees of freedom, 
and the dynamical pairing degree 
of freedom (Sec.~\ref{sec:pv}). 
The mapped IBM calculations on the 
octupole deformations and collectivity 
are discussed in a recent review article \cite{nomura2023oct}. 
Section~\ref{sec:oct} in the present review 
therefore intends to outline 
the formalism and a few key results, 
without going into much details.

\subsection{Octupole deformations\label{sec:oct}}

The octupole correlations are expected to be 
enhanced for those nuclei in which coupling occurs 
between the spherical single-particle states differing 
in the quantum numbers by 
$\Delta j = \Delta l = 3\hbar$ \cite{butler1996}. 
These combinations are possible 
at specific nucleon numbers, 
that is, $N$ and/or $Z$ equal to 34, 56, 88, 134, etc. 
Experimental evidence for the static octupole 
shape has been found in a few nuclei 
in light actinides ($^{220}$Rn and $^{224}$Ra \cite{gaffney2013}) 
and lanthanides ($^{144}$Ba \cite{bucher2016} 
and $^{146}$Ba \cite{bucher2017}), which exhibit 
strong electric octupole ($E3$) and dipole ($E1$) 
transitions and low-lying negative-parity bands.

The most relevant ingredients to 
octupole correlations in the IBM are $f$ bosons, 
which reflect collective pairs 
with spin and parity $J^{\pi}=3^-$. 
The corresponding IBM-2 space therefore contains 
$s_\nu$, $d_{\nu}$, $f_{\nu}$, 
$s_\pi$, $d_{\pi}$, and $f_{\pi}$ bosons. 
The $sdf$-IBM-2 was applied in Ref.~\cite{nomura2022oct56}
to the neutron-rich $A \approx 100$ nuclei 
with $N\approx 56$.
In the following, however, 
the $sdf$-IBM-1 is considered,
as the majority of the earlier
mapped $sdf$-IBM calculations employs
this simpler version of the model.

The $sdf$-boson Hamiltonian employed for the 
IBM mapping is of the form
\begin{align}
\label{eq:hamoct}
\hat H_{sdf}
=\epsilon_{d} \hat n_{d} +\epsilon_{f} \hat n_{f} 
  + \kappa_{2} \hat Q \cdot \hat Q
  + \kappa_{3} \hat O \cdot \hat O 
  + \rho\hat L \cdot \hat L
\; .
\end{align}
In the first (second) term 
$\hat n_d=d^\+\cdot\tilde d$ 
($\hat n_f = f^\+ \cdot \tilde f$), 
with $\epsilon_{d}$ ($\epsilon_{f}$) representing 
the single $d$ ($f$) boson energy relative to the 
$s$-boson one. 
Note the expression $\tilde f_{\mu} = (-1)^{3+\mu} f_{-\mu}$. 
The third and fourth terms in (\ref{eq:hamoct}) stand for 
quadrupole-quadrupole and octupole-octupole 
interactions, respectively. 
The quadrupole $\hat Q$ and octupole $\hat O$ 
operators read
\begin{align}
\label{eq:octquad}
& \hat Q = 
(s^\+ \times \tilde d 
+ d^\+ \times \tilde s )^{(2)}
+ \chi (d^\+ \times \tilde d)^{(2)}
+ \chi' (f^\+ \times \tilde f)^{(2)} \; , 
\\
\label{eq:octoct}
& \hat O = 
(s^\+ \times \tilde f 
+ f^\+ \times \tilde s )^{(3)}
+ \chi''(d^\+\times\tilde f +f^\+\times\tilde d)^{(3)} \; ,
\end{align}
with $\chi$, $\chi'$, and $\chi''$ being 
dimensionless parameters. 
The last term 
$\hat L\cdot \hat L$ with the 
parameter $\rho$ in (\ref{eq:hamoct}) 
has also been introduced in order 
to better describe moments of inertia 
of yrast bands of well deformed nuclei. 
The boson angular momentum operator $\hat L$ 
in the $sdf$-IBM takes the form
\begin{align}
 \hat L=\sqrt{10}(d^\+\times\tilde d)^{(1)}-\sqrt{28}(f^\+\times\tilde f)^{(1)} \; .
\end{align}

The SCMF calculations are performed with the 
constraints on the axially-symmetric quadrupole 
$Q_{20}$ and octupole $Q_{30}$ moments, which 
are related to the geometrical deformation parameters 
$\beta_{20}$ and $\beta_{30}$, respectively:
\begin{align}
\label{eq:beta23}
 \beta_{\lambda 0}
=\frac{\sqrt{(2\lambda+1)\pi}}{3AR_0^{\lambda}}
\braket{\hat Q_{\lambda 0}} \; ,
\end{align}
with $\lambda=2$ or 3. 
The constrained SCMF calculations produce the 
energy surface with the $\beta_{2}$ and $\beta_{3}$ 
degrees of freedom 
(in what follows $\beta_{\lambda 0}$ 
is simply expressed by $\beta_{\lambda}$). 
The assumption of the axial symmetry 
is based on the fact that 
the stable octupole shape has empirically been 
suggested to occur in the actinide and rare-earth nuclei 
that are characterized by the axially symmetric deformation. 
As in the case of the quadrupole deformations, 
the boson analog $\pesibmoct$ of the fermionic 
energy surface is obtained as the expectation value of 
the Hamiltonian (\ref{eq:hamoct}) 
\begin{align}
 \pesibmoct
= \bra{\Phi(\beta_2,\beta_3)} 
\hat H_{sdf}
\ket{\Phi(\beta_2,\beta_3)} \; 
\end{align}
in the coherent state $\ket{\Phi(\beta_2,\beta_3)}$, 
which is defined by
\begin{align}
 \ket{\Phi(\beta_2,\beta_3)} 
= \frac{1}{\sqrt{N_{\rm B}!}}
(\lambda^\+)^{N_{\rm B}}
\ket{0}
\end{align}
with
\begin{align}
 \lambda^\+
=
(1+\beta_{2{\rm B}}^2+\beta_{3{\rm B}}^2)^{-1/2}
(s^\+ + \beta_{2{\rm B}} d^\+_0
+ \beta_{3{\rm B}} f^\+_0) \; .
\end{align}
Here the relation (\ref{eq:cbeta}) 
between the bosonic $\beta_{2{\rm B}}$ 
and fermionic $\beta_2$ deformations is assumed, 
and a similar assumption is made for the 
octupole deformations:
\begin{align}
\beta_{2{\rm B}} = C_{2} \beta_2
\quad , \quad
 \beta_{3{\rm B}} = C_{3} \beta_3 \; ,
\end{align}
with $C_2$ and $C_3$ being 
the coefficients of proportionality. 
The parameters 
$\epsilon_d$, $\epsilon_f$, $\kappa_2$, $\kappa_3$, 
$\chi$, $\chi'$, $\chi''$, $C_2$, and $C_3$ 
are determined by the 
SCMF-to-IBM mapping, $\pesdftoct \approx \pesibmoct$, 
which is carried out in the following steps. 
\begin{enumerate}
 \item 
The parameters 
$\epsilon_d$, $\kappa_2$, $\chi$, and $C_2$, 
which concern $s$ and $d$ bosons, 
are determined so as to reproduce the 
SCMF energy curve along the $\beta_2$ deformation 
with $\beta_3=0$, i.e., $E_{\rm SCMF}(\beta_2,0)$. 

\item
The parameter $\rho$ is determined so that 
the cranking moment of inertia in the 
boson system at the equilibrium minimum 
along the $\beta_3=0$ axis should be equal to 
that calculated by the SCMF method, 
using the procedure described in Sec.~\ref{sec:LL}. 

\item
The parameters 
$\epsilon_f$, $\kappa_3$, $\chi'$, $\chi''$ and $C_3$, 
which are related to $f$ bosons, 
are determined so as to reproduce 
the topology of the SCMF PES $\pesdftoct$ 
near the global minimum, such as the location of the 
minimum and softness in the $\beta_3$ direction. 
\end{enumerate}
The mapped $sdf$-IBM Hamiltonian is diagonalized 
in the space consisting of $N_{\rm B} = N_s + N_d + N_{f}$ 
bosons. 
In numerical calculations for realistic nuclei
within the phenomenological $sdf$-IBM,
a truncation has often been made
of the maximum number of $f$ bosons $N_f^{\rm max}$.
Here $N_f$ is taken to vary within the range 
$0 \leqslant N_f \leqslant N_{\rm B}$, that is, 
there is no restriction of the $f$-boson number. 
The large number of $f$ bosons is 
required for the mapping procedure, 
in order to reproduce the steepness of the 
potential valley in the $\beta_3$ 
deformation and the location of 
a non-zero $\beta_3$ deformation in the SCMF PES.

Relevant transition properties to 
the quadrupole-octupole collectivity 
are those of the electric quadrupole ($E2$), 
octupole ($E3$), and dipole ($E1$) modes. 
The $E2$ and $E3$ transition operators are 
written as $\hat T^{(E2)}=e_{\rm B}^{(2)} \hat Q$, 
and $\hat T^{(E3)}=e_{\rm B}^{(3)} \hat O$, 
respectively, with $\hat Q$ and $\hat O$ 
defined in (\ref{eq:octquad}) and (\ref{eq:octoct}). 
$e_{\rm B}^{(2)}$ and $e_{\rm B}^{(3)}$ 
are effective charges. 
The $E1$ operator in the $sdf$-IBM is simply given as 
$\hat T^{(E1)}=e_{\rm B}^{(1)} (d^\+\times\tilde f + f^\+\times\tilde d)^{(1)}$, with $e_{\rm B}^{(1)}$ the effective charge. 
This form of the $E1$ operator may be 
insufficient to describe the observed $E1$ transition 
properties, and dipole or $p$ ($J^\pi=1^-$) boson 
degree of freedom is often required \cite{otsuka1986,otsuka1988,sugita1996}. 
In contrast to $s$, $d$, and $f$ bosons, 
which reflect the collective nucleon pairs, 
the microscopic origin of $p$ bosons 
has not been clarified. In earlier studies, 
this degree of freedom was attributed 
to the giant dipole resonance \cite{sugita1996}, 
or to the spurious center-of-mass motion \cite{engel1987}. 
In either case, since the $E1$ transitions 
are more strongly influenced 
by the single-particle degrees of freedom 
than the $E2$ and $E3$ transitions, 
the validity of the $sdf$-IBM 
in the calculations of the $E1$ 
properties deserves further investigations.

%-------- Th spectra -------------
\begin{figure}[h]
\centering
\includegraphics[width=\linewidth]{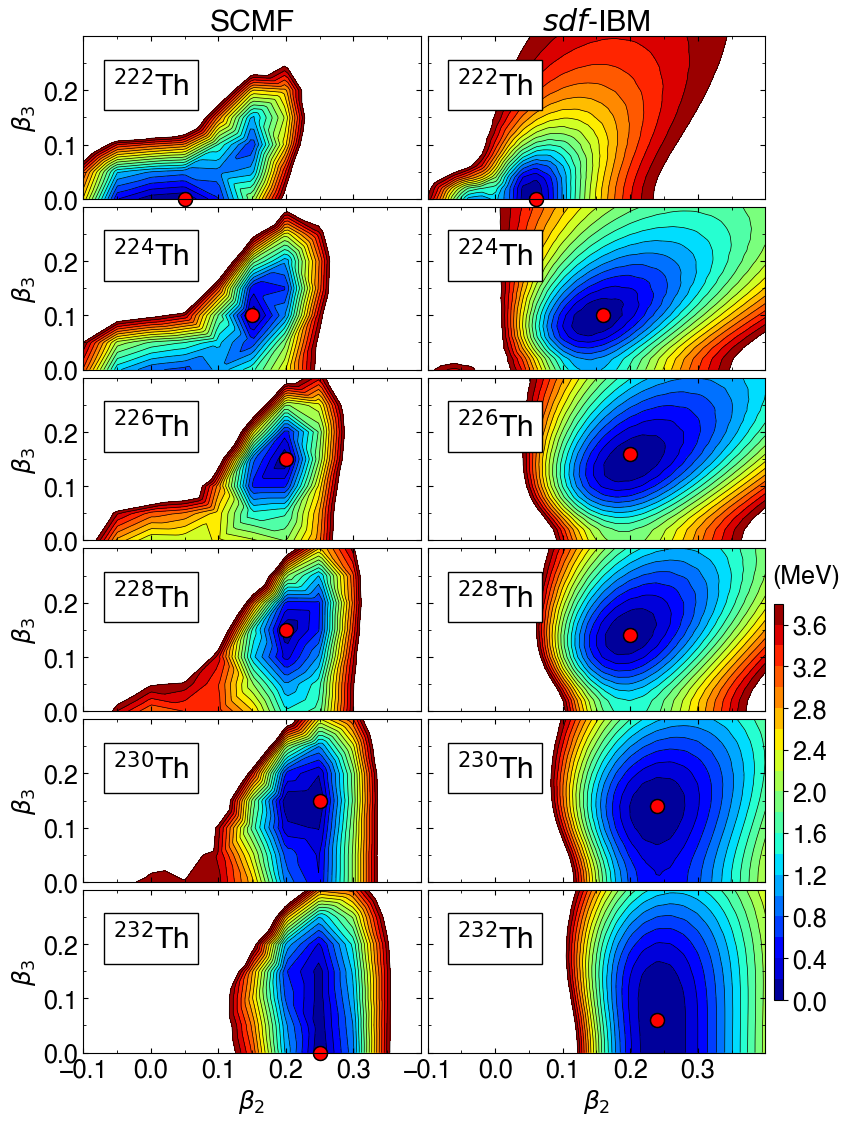}
\caption{
Axially symmetric quadrupole-octupole
$(\beta_2,\beta_3)$ PESs
for $^{220-232}$Th computed 
by the relativistic Hartree-Bogoliubov method
with the DD-PC1 interaction \cite{DDPC1} 
and separable pairing force 
of \cite{tian2009} (left), 
and the corresponding mapped $sdf$-IBM 
PESs (right). 
The global minimum is indicated by the 
solid circle.
}
\label{fig:th-pes}
\end{figure}
%---------------------------------

%-------- Th spectra -------------
\begin{figure}[h]
\centering
\includegraphics[width=.8\linewidth]{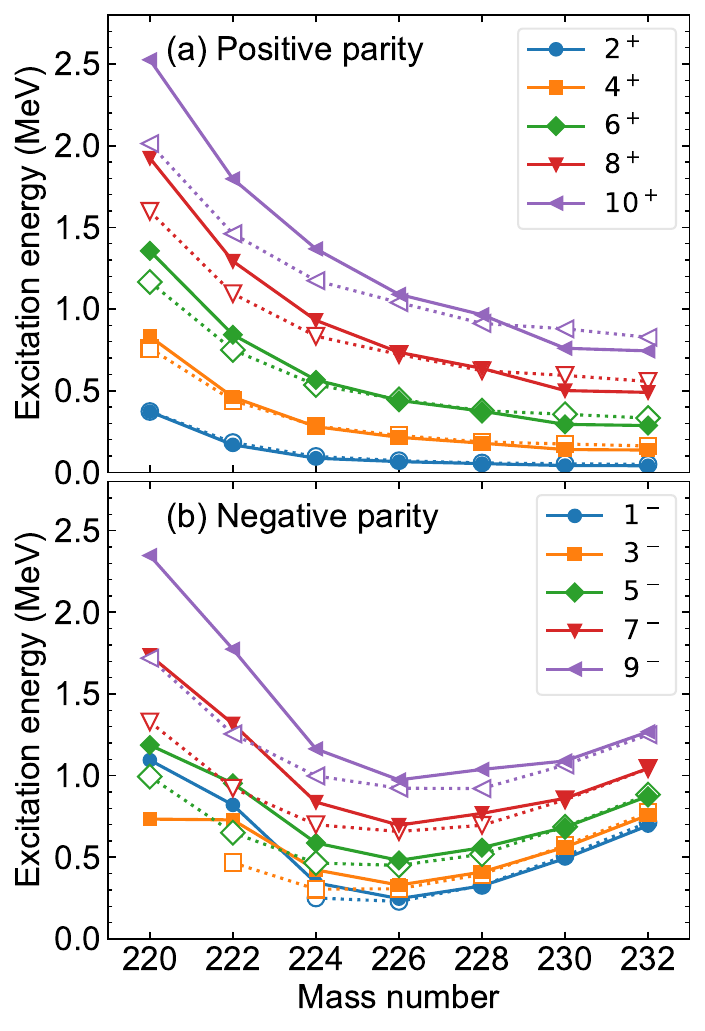}
\caption{
Low-energy positive-parity and negative-parity
spectra for the $^{220-232}$Th isotopes 
obtained from the mapped $sdf$-IBM based on 
the RHB-SCMF model. Experimental data are 
taken from \cite{data}.}
\label{fig:th-level}
\end{figure}
%---------------------------------

The $sdf$-IBM mapping procedure was first developed 
in computing low-lying positive- and negative-parity states 
of the even-even $^{222-232}$Th nuclei \cite{nomura2013oct}. 
The axially symmetric quadrupole-octupole PESs 
for these nuclei, calculated by the RHB method 
with the energy functional DD-PC1 \cite{DDPC1}, 
indicate an octupole-deformed mean-field minimum 
at $\beta_3 \approx 0.15$ for the $^{226}$Th nucleus, 
and a pronounced $\beta_3$ softness in heavier Th nuclei 
(see Fig.~\ref{fig:th-pes}). 
This feature implies a transition from the static 
octupole deformed state near $^{226}$Th and $^{228}$Th 
to the octupole vibrational state 
that is characterized by the $\beta_3$-soft potential 
in $^{230}$Th and $^{232}$Th. 
The calculated odd-spin negative-parity levels 
in the Th isotopes are shown to exhibit a 
parabolic behavior as functions of the neutron number, 
and become lowest in energy at $^{226}$Th, 
forming an approximate alternating parity band 
with the positive-parity even-spin 
ground-state band (see Fig.~\ref{fig:th-level}). 
As a signature of the octupole shape phase transition, 
behaviors of the energy ratios 
$E(I^\pi)/E(2^+_1)$ as functions of the 
increasing spin $I^\pi$ were studied 
in Ref.~\cite{nomura2013oct}. 
These energy ratios exhibit
an odd-even spin staggering that 
emerges in those nuclei heavier than $^{224}$Th. 
The amplitude of the staggering increases for 
heavier Th nuclei, indicating the decoupling 
of the negative-parity band from 
the positive-parity one. 
See Ref.~\cite{nomura2013oct} for detailed discussions. 

The study of Ref.~\cite{nomura2013oct} was further 
extended to a more systematic analysis 
on those Th, Ra, Sm and Ba nuclei 
in which octupole collectivity is expected to be relevant. 
The $sdf$-IBM mapping procedure based on the 
constrained RHB calculations with the 
DD-PC1 EDF provided 
key spectroscopic observables that indicate the 
octupole shape phase transition from the 
stable octupole deformed to octupole-soft regimes.

%-------- Oct sys -------------
\begin{figure*}[h]
\centering
\includegraphics[width=.8\linewidth]{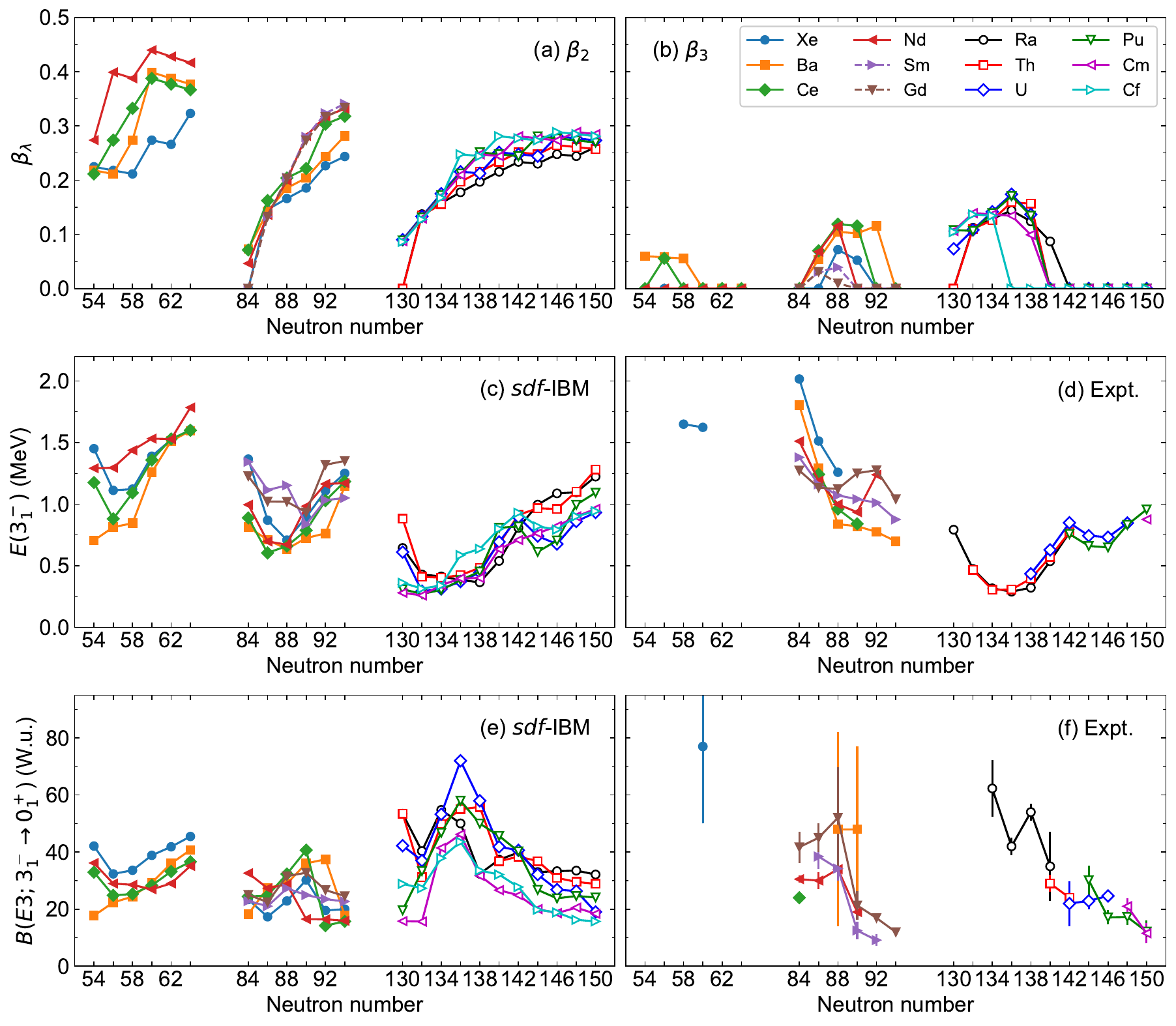}
\caption{
Intrinsic deformations (a) $\beta_2$ and (b) $\beta_3$
predicted by the Gogny-D1M HFB SCMF calculations,
(c) $3^-_1$ energy levels obtained from the
mapped $sdf$-IBM and (d) the corresponding 
experimental data \cite{data},
(e) $B(E3;3^-_1 \to 0^+_1)$ transition
probabilities in W.u. predicted
by the mapped $sdf$-IBM
and (f) the experimental data \cite{wollersheim93,bucher2016,bucher2017,data,kibedi2002,DEANGELIS2002,gaffney2013,butler2020a,pascu2025-prl}
for the Xe, Ba, Ce, and Nd isotopes 
with the neutron numbers $N=54-64$ \cite{nomura2021oct-zn}
and $N=84-94$ \cite{nomura2021oct-ba}, 
rare-earth nuclei Sm and Gd with $N=84-94$ \cite{nomura2015}, 
and actinides Ra, Th, U, Pu, Cm, Cf
with $N=130-150$ \cite{nomura2020oct,nomura2021oct-u}. 
}
\label{fig:oct-sys}
\end{figure*}
%---------------------------------

Systematic investigations of the octupole 
correlation effects on low-lying collective states 
in those nuclei corresponding to 
$(N,Z)\approx$ $(56,56)$, $(88,56)$, $(134,88)$ 
have been made within the mapped $sdf$-IBM that is 
based on the Gogny-HFB calculations 
\cite{nomura2015,nomura2020oct,nomura2021oct-u,nomura2021oct-ba,nomura2021oct-zn}. 
The calculated intrinsic deformations 
$\beta_2$ and $\beta_3$, excitation energies 
for the $3^-_1$ state, and $B(E3; 3^-_1 \to 0^+_1)$ 
transition strengths for these nuclei 
are summarized in Fig.~\ref{fig:oct-sys}.

The Gogny-EDF mapped $sdf$-IBM in 
Ref.~\cite{nomura2015} 
reproduced the observed spectroscopic properties 
of $^{146-156}$Sm and $^{148-158}$Gd nuclei 
as accurately as the RHB-mapped $sdf$-IBM 
of Ref.~\cite{nomura2014}. 
Both the relativistic \cite{nomura2014} 
and nonrelativistic (Gogny) \cite{nomura2015}
EDF-based $sdf$-IBM studies suggested a pronounced 
octupole collectivity in the Sm and Gd isotopes 
near $N\approx 88$. 
In Ref.~\cite{nomura2015}, the nature of the $sdf$-IBM 
wave functions of the lowest 15 $0^+$ states 
of the Sm and Gd nuclei 
was analyzed by calculating the expectation 
values of the operator $\hat n_f$. 
It was found that many of these excited 
$0^+$ states, including the $0^+_2$ one, 
were suggested to be 
of double octupole phonon (or $f$ boson) nature, 
that is, the expectation values 
$\braket{\hat n_f}\approx 2$. 
This shed light upon the origin of a number  
of low-lying $0^+$ states observed 
in the rare-earth nuclei.

Systematic studies
on the even-even Ra, Th, U, Pu, Cm, and Cf nuclei
with $N=130-150$ within the Gogny-D1M mapped $sdf$-IBM
in \cite{nomura2020oct,nomura2021oct-u}
suggested the nuclear structural 
evolution from nearly spherical non-octupole to 
static octupole, and to octupole vibrational states. 
The behaviors of the predicted odd-spin 
negative-parity states 
and enhanced $B(E3;3^-_1 \to 0^+_1)$ transition rates 
suggested that the maximal octupole 
collectivity is reached near the empirical neutron 
``octupole magic number'' $N=134$. 
One can see parabolic behaviors of the calculated $3^-_1$ levels in 
Fig.~\ref{fig:oct-sys}(c), and that the calculated 
$B(E3;3^-_1 \to 0^+_1)$ rates exhibit inverse 
parabolas near $N=134$ in Fig.~\ref{fig:oct-sys}(e). 
These studies also provided spectroscopic predictions 
on those exotic nuclei that are close to 
the neutron $N=126$ major shell closure 
and to the proton dripline.

The Gogny-HFB SCMF calculations 
on the neutron-rich ($86 \leqslant N \leqslant 94$) 
\cite{nomura2021oct-ba} and neutron-deficient 
($54 \leqslant N \leqslant 64$) 
\cite{nomura2021oct-zn} even-even Xe, Ba, Ce, and Nd isotopes
produce a non-zero octupole deformation 
for many of those nuclei that 
correspond to $Z\approx 56$ and $N\approx 88$ 
[see Fig.~\ref{fig:oct-sys}(b)].
The calculated spectroscopic 
properties such as 
the low-lying negative-parity levels 
and $B(E3)$ rates suggest onset of  
octupole collectivity at $N\approx 88$
[Figs.~\ref{fig:oct-sys}(c) and \ref{fig:oct-sys}(e)], 
which is much less pronounced 
than in the actinide region.
On the neutron-deficient side, 
the octupole
ground state is found for a few nuclei  
close to $Z=56$ and $N=56$ 
($^{110}$Ba, $^{112}$Ba, $^{114}$Ba and $^{114}$Ce) 
in the SCMF PES, in which
the potential was shown to be rather soft in 
$\beta_3$ deformation. 
These mean-field results indicate that the 
octupole correlations are relevant in 
the neutron-deficient regions. 
Spectroscopic properties were shown to
exhibit certain signatures
of the octupole correlations, which are 
however much less pronounced than 
for the neutron-rich nuclei.

Furthermore, a framework was developed \cite{nomura2022octcm} 
to incorporate the octupole degrees 
of freedom in the IBM-CM. 
This extension is particularly relevant to those 
cases in which the shape coexistence 
and octupole deformation are likely to occur
simultaneously. 
The $sdf$-IBM-CM Hamiltonian that includes up to the $4p-4h$ 
intruder configurations was determined in such 
a way to reproduce both the 
triaxial quadrupole ($\beta_2,\gamma$) and 
axially symmetric quadrupole and octupole ($\beta_2,\beta_3$)
SCMF PESs calculated within the
constrained RHB method \cite{DIRHB,DIRHBspeedup}.
Illustrative applications were made
to the $^{72}$Ge, $^{74}$Se, 
$^{74}$Kr, and $^{76}$Kr nuclei, which are well known 
cases for the quadrupole shape coexistence, and 
which are also expected to be influenced by 
octupole correlations because they are 
close to the neutron and 
proton octupole ``magic number'' 34. 
For $^{72}$Ge and $^{74}$Se the spherical-oblate 
shape coexistence was suggested
in the RHB-SCMF ($\beta_2,\gamma$) PES.
Triple shape coexistence of 
(nearly) spherical, oblate, and 
prolate shapes in $^{76}$Kr and $^{74}$Kr 
was suggested to occur at the mean-field level. 
In the ($\beta_2,\beta_3$) PES for $^{76}$Kr, 
in particular, a non-zero octupole
deformed minimum at $\beta_2 \approx 0$ 
and $\beta_3 \approx 0.05$ is predicted.
The inclusion of the intruder configuration 
in the $sdf$-IBM was shown to have 
an effect of lowering the 
excited $0^+$ levels significantly to be 
consistent with experiment, and improved 
slightly the description of the negative-parity levels. 
The mapped $sdf$-IBM-CM method provided several 
low-energy positive-parity bands in which 
the (nearly) spherical, 
oblate, and prolate deformed configurations 
are strongly mixed. 
The calculated negative-parity bands were shown 
to be largely composed of the deformed intruder 
configurations.

\subsection{Hexadecapole deformation\label{sec:hex}}

Hexadecapole degrees of freedom represent 
a next-to-leading-order 
effect on low-energy collective states 
of positive parity. 
Experimental evidence for the hexadecapole 
deformations has been found mainly in heavy nuclei
in the rare-earth 
\cite{BM,hendrie1968,erb1972,wollersheim1977,ronningen1977,garrett2005,phillips2010,hartley2020} 
and actinide \cite{bemis1973,zamfir1995} regions, 
but the hexadecapole correlations
have also been shown to be relevant
in light-mass regions \cite{gupta2020,gupta2023}. 
More recently the 
hexadecapole correlations were shown to appear 
in exotic Kr nuclei \cite{spieker2023}. 
Empirical signatures of the hexadecapole collectivity 
include the appearance of the low-energy $K^\pi=4^+$ 
bands and enhanced electric hexadecapole ($E4$) 
transitions. 

%-------- Gd-hex PES -------------
\begin{figure}[h]
\centering
\includegraphics[width=\linewidth]{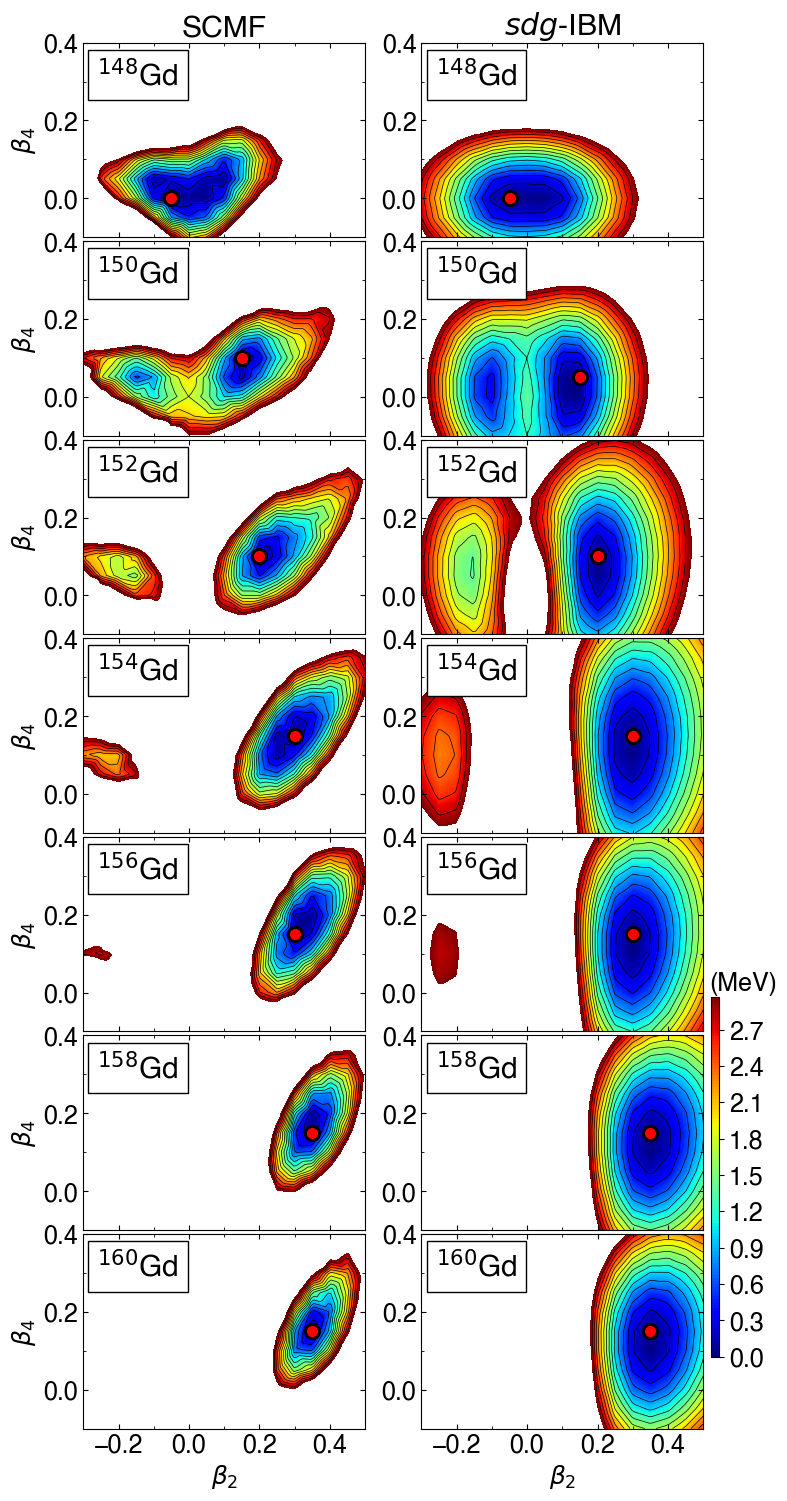}
\caption{
Axially symmetric quadrupole-hexadecapole
$(\beta_2,\beta_4)$ PESs
for $^{148-160}$Gd computed 
by the RHB-SCMF method 
with the DD-PC1 interaction \cite{DDPC1} 
and separable pairing force 
\cite{tian2009} (left), 
and the corresponding mapped $sdg$-IBM 
PESs (right). 
The global minimum is indicated by the 
solid circle.}
\label{fig:gd-hex-pes}
\end{figure}
%---------------------------------

In Ref.~\cite{lotina2024hex-1} hexadecapole $g$ bosons, 
with spin and parity $J^{\pi}=4^+$, were introduced 
in the IBM mapping procedure. 
The simpler $sdg$-IBM-1 can be used,
as in the case of the $sdf$-IBM mapping. 
The following $sdg$-IBM Hamiltonian was 
considered \cite{lotina2024hex-1}. 
\begin{align}
\label{eq:hamhex}
 \hat H_{sdg} = \epsilon_d \hat n_d + \epsilon_g \hat n_g 
+ \kappa \hat Q \cdot \hat Q
+ \kappa (1-\chi^2) \hat Q' \cdot \hat Q' \; ,
\end{align}
where the first term is defined in (\ref{eq:hamoct}), 
and the second term is the $g$-boson number operator
$\hat n_g = g^\+ \cdot \tilde g$. 
The third term represents 
the quadrupole-quadrupole interaction. 
The quadrupole operator, $\hat Q$, reads
 \begin{align}
  \hat Q
=&s^\+\tilde d + d^\+ s
+\chi
\biggl[
\frac{11\sqrt{10}}{28}(d^\+\times\tilde d)^{(2)}
\nonumber\\
&- \frac{9}{7} (d^\+\times\tilde g + g^\+\times\tilde d)^{(2)} 
+ \frac{3\sqrt{55}}{14} (g^\+\times\tilde g)^{(2)}
\biggr] \; .
 \end{align}
This form of the operator corresponds to 
a generator of the $sdg$-SU(3) limit, 
if $\chi=1$ \cite{vanisacker2010,kota1987}, 
and was used in Ref.~\cite{vanisacker2010}. 
The last term in (\ref{eq:hamhex})
represents the hexadecapole-hexadecapole 
interaction, and the hexadecapole 
operator $\hat Q'$ takes a simplified form
\begin{eqnarray}
\hat Q' = s^\+\tilde g + g^\+s \; .
\end{eqnarray}

As in the case of the $sdf$-IBM mapping, 
strength parameters for 
the $sdg$-boson Hamiltonian $\hat H_{sdg}$, 
$\epsilon_d$, $\epsilon_g$, $\kappa$, 
$\chi$ and $C_2$, $C_4$, 
are fixed by mapping the axially symmetric 
quadrupole-hexadecapole SCMF PES onto 
that of the $sdg$-IBM:
\begin{align}
\label{eq:map4}
 \pesdfthex \approx \pesibmhex \; ,
\end{align}
where $\beta_4$ stands for the axial hexadecapole 
deformation, which is related to the 
hexadecapole moment by the formula 
in (\ref{eq:beta23}) with $\lambda=4$. 
In analogy to the $sd$-IBM and $sdf$-IBM mappings, 
the bosonic and fermionic $\beta_{\lambda}$ 
deformations are related by using the formulas
\begin{align}
\label{eq:cbeta4}
 \beta_{2{\rm B}} = C_{2} \beta_2 \, 
\quad , \quad
\beta_{4{\rm B}} = C_{4} \beta_4 \; ,
\end{align}
with $\beta_{4{\rm B}}$ being the boson analog 
of the axial hexadecapole deformation. 
To see effects of $g$ bosons on low-energy 
spectroscopic properties, 
the simple $sd$-IBM-1 mapping was 
also carried out, with $\beta_4=0$ in the 
formula (\ref{eq:map4}).

%-------- Gd-hex level------------
\begin{figure}[h]
\centering
\includegraphics[width=\linewidth]{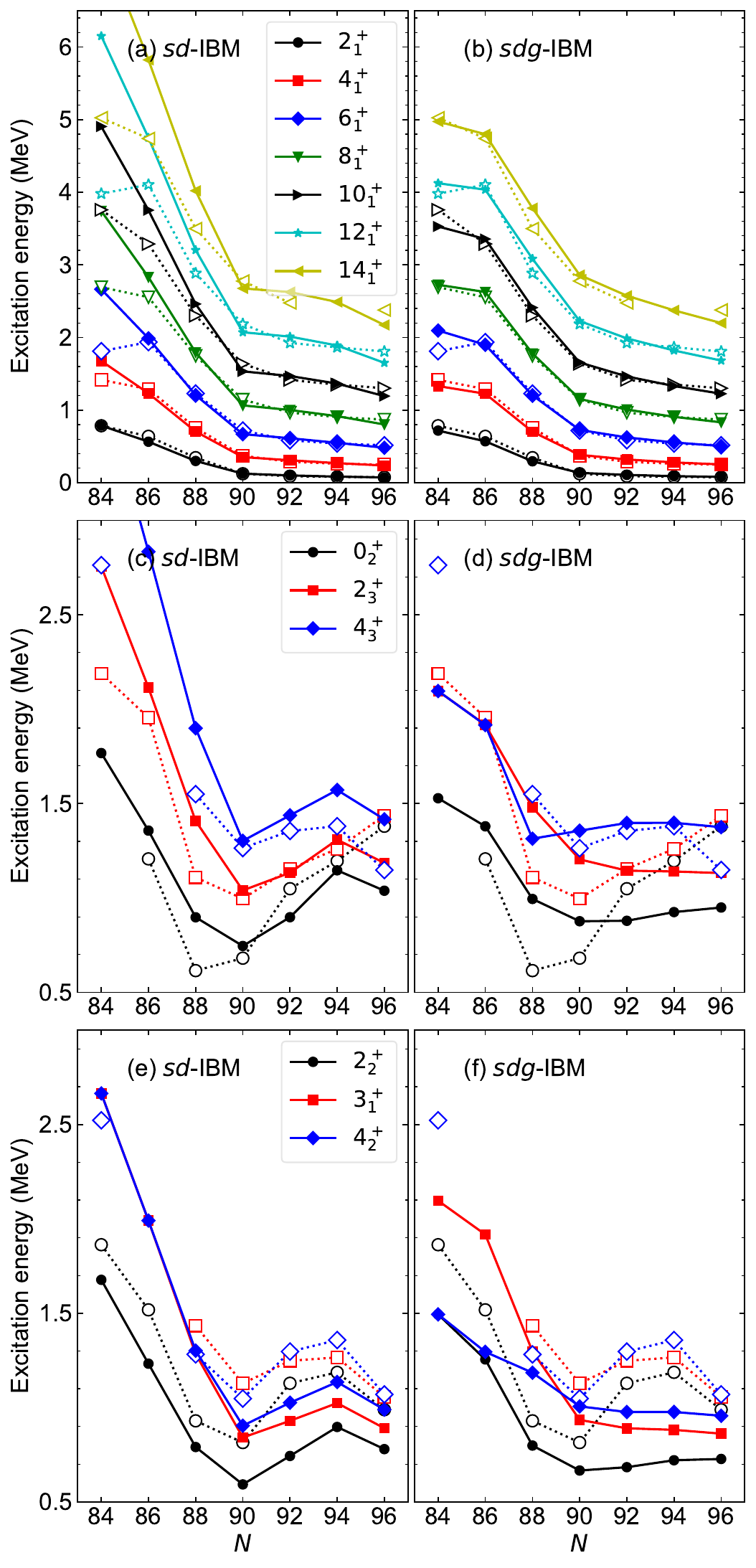}
\caption{
Evolution of low-lying positive-parity levels 
for $^{148-160}$Gd obtained from the 
mapped $sd$-IBM (left) and $sdg$-IBM (right).
The experimental data are adopted from Ref.~\cite{data}.}
\label{fig:gd-hex-level}
\end{figure}
%---------------------------------

The $sdg$-IBM mapping 
procedure was first implemented in 
describing the low-energy 
positive-parity states in the even-even 
$^{148-160}$Gd nuclei, starting from the 
RHB-SCMF calculations using the 
relativistic DD-PC1 EDF. 
The ($\beta_2,\beta_4$)-constrained 
RHB-SCMF PESs, shown in Fig.~\ref{fig:gd-hex-pes}, 
suggest for all the considered Gd, 
but for $^{148}$Gd, a non-zero hexadecapole 
minimum $\beta_4 \neq 0$. 
The corresponding mapped $sdg$-IBM PESs 
are also shown in Fig.~\ref{fig:gd-hex-pes}.

Figure~\ref{fig:gd-hex-level} gives 
the calculated energy levels of the 
ground-state yrast band 
and of those states that are considered 
to belong to the $\beta$-vibrational 
and $\gamma$-vibrational bands 
within the $sd$-IBM and $sdg$-IBM. 
A notable $g$-boson effect is that 
the ground-state-band 
levels with spin $I \geqslant 4^+$ 
for those nearly spherical 
nuclei corresponding to $N \leqslant 86$, 
which are rather close to the neutron 
$N=82$ shell closure, are significantly lowered. 
This is due to the fact that 
large amounts of $g$ boson contributions 
are present in corresponding wave functions with 
higher-spin states of the ground-state band. 
For $^{148}$Gd, in particular, the $g$-boson 
contents, i.e., the expectation 
values $\braket{\hat n_{g}}$ 
in the $4^+_1$, $6^+_1$, 
$8^+_1$, and $10^+_1$ states are 
calculated to be $\braket{\hat n_g} \approx 1$, 
and for the $12^+_1$ and $14^+_1$ states 
$\braket{\hat n_g} \approx 1.5$. 
For transitional ($N \approx 90$) and 
deformed ($N \geqslant 90$) regions, 
the inclusion of $g$ bosons does not make a 
notable contribution to states of the ground-state band. 
The inclusion of $g$ bosons also has influences 
on the $2^+_3$ and $4^+_3$ 
[Figs.~\ref{fig:gd-hex-level}(c) 
and \ref{fig:gd-hex-level}(d)], 
and on the $2^+_2$ and $4^+_2$ states 
[Figs.~\ref{fig:gd-hex-level}(e) 
and \ref{fig:gd-hex-level}(f)]. 
The non-yrast $4^+$ levels 
are significantly lowered.

The study of Ref.~\cite{lotina2024hex-2} was 
extended to a more extensive analysis 
of the collective quadrupole 
and hexadecapole states in the Nd, Sm, Gd, Dy, and Er 
isotopic chains \cite{lotina2024hex-2}. 
In nearly spherical nuclei with $N\leqslant 86$, 
in particular, by including $g$ bosons 
the energy levels of the yrast bands with 
spin $I \geqslant 4^+$ were shown to be lowered 
appreciably, and the observed 
energy ratios of $R_{4/2}<2$ were reproduced 
by the $sdg$-IBM. However, the $g$ boson effects 
on deformed regions were shown to be relatively minor. 
These conclusions are similar to that 
of Ref.~\cite{lotina2024hex-1}, and appear to be 
robust in the rare-earth region.
In fact, Table~\ref{tab:hex-r42} 
summarizes the calculated 
$R_{4/2}$ ratios with the $sdg$-IBM and $sd$-IBM. 
The observed ratios of $R_{4/2}<2$ 
are reproduced by the $sdg$-IBM, in which 
the $4^+_1$ state is largely of $g$-boson 
character, and is predicted to be 
significantly low in energy 
to be close to the $2^+_1$ level.
The fact that a nucleus has
the energy ratio $R_{4/2}<2$ 
reflects, to a good extent, contributions from 
the single-particle excitations, which appear to be 
incorporated effectively in the IBM
by the inclusion of $g$ bosons.

\begin{table}[!htb]
\caption{\label{tab:hex-r42}
Energy ratios $R_{4/2}=E_x(4^+_1)/E_x(2^+_1)$ 
for the Nd, Sm, Gd, Dy, and Er nuclei 
with $N=84$ and 86,
calculated with the mapped $sd$-IBM
and $sdg$-IBM,
and the experimental values \cite{data}.}
\centering
%\begin{ruledtabular}
\begin{tabular}{cccc}
\hline\hline
Nucleus & $sd$-IBM & $sdg$-IBM & Experiment \\
\hline
$^{144}$Nd & 2.11 & 1.78 & 1.89 \\
$^{146}$Nd & 2.25 & 2.05 & 2.02 \\
$^{146}$Sm & 2.12 & 1.83 & 1.85 \\
$^{148}$Sm & 2.20 & 1.98 & 2.14 \\
$^{148}$Gd & 2.13 & 1.86 & 1.81 \\
$^{150}$Gd & 2.18 & 2.15 & 2.02 \\
$^{150}$Dy & 2.15 & 1.71 & 1.81 \\
$^{152}$Dy & 2.21 & 2.16 & 2.05 \\
$^{152}$Er & 2.14 & 1.54 & 1.83 \\
$^{152}$Er & 2.24 & 2.17 & 2.07 \\
\hline\hline
\end{tabular}
%\end{ruledtabular}
\end{table}

%-------- hex E2 -----------------
\begin{figure}[h]
\centering
\includegraphics[width=\linewidth]{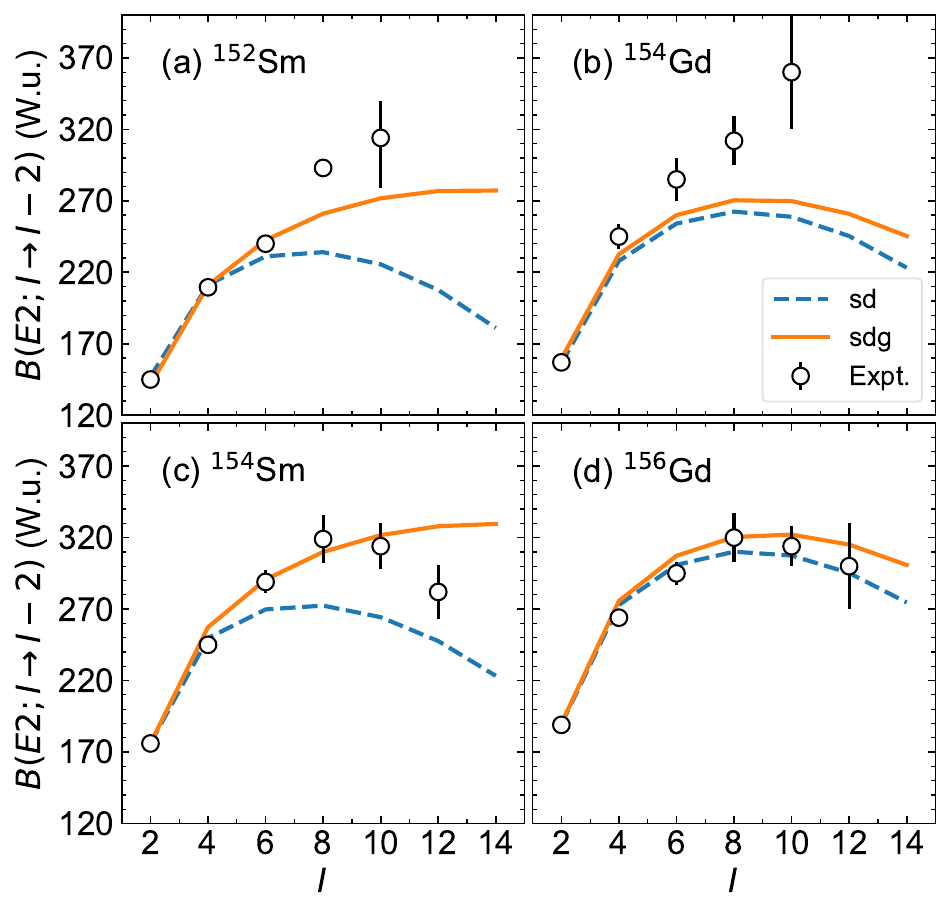}
\caption{
$B(E2)$ transition strengths in W.u.
in the ground-state bands of Sm and Gd 
isotopes with the neutron numbers $N=90$ and 92 
as functions of spin $I$. 
Calculated results from the $sd$-IBM (dashed curves) 
and $sdg$-IBM (solid curves), 
and experimental values (open circles)
\cite{data} are shown.}
\label{fig:hex-e2}
\end{figure}
%---------------------------------

The mapped $sdg$-IBM also gives $\Delta I=1$ 
bands built on $4^+$ states, which exhibit 
stronger $B(E4;4^+_{K=4^+} \to 0^+_1)$ transition 
rates than the $sd$-IBM \cite{lotina2024hex-1}. 
The $E4$ operator in the $sdg$-IBM takes 
the form 
$e_{\rm B}^{(4)}[\hat Q'+(d^\+ \times \tilde d)^{(4)}]$, 
while the corresponding operator 
in the $sd$-IBM reads 
$e_{\rm B}^{(4)}(d^\+ \times \tilde d)^{(4)}$, 
where $e_{\rm B}^{(4)}$ are effective charges. 
For the $^{154}$Gd, for instance, 
the $sdg$-IBM predicted the $K^{\pi}=4^+$ band 
built on the $4^+_3$ state, and the $E4$ 
transition rate of 
$B(E4;4^+_{3} \to 0^+_1)=93$ W.u. 
The $sd$-IBM also produced a $K^{\pi}=4^+$ band 
based on the $4^+_3$ state, but the 
corresponding $E4$ transition 
$B(E4;4^+_{3} \to 0^+_1)=1.3$ W.u.
is much weaker than in the $sdg$-IBM. 
Note that the effective charges for the 
$E4$ operator were chosen so as to fit 
the experimental $B(E4;4^+_1 \to 0^+_1)$ value.

The presence of $g$ bosons affects
the $E2$ transition properties. 
Figure~\ref{fig:hex-e2} depicts the 
$B(E2)$ transition strengths between states 
in the ground-state band as functions of 
spin for those Sm and Gd  
nuclei with $N=90$ and 92. 
The $sdg$-IBM gives greater $B(E2)$ rates than
the $sd$-IBM for those states
with spin higher than 4, and reproduces the
experimental data more accurately.

In a more recent study of 
Ref.~\cite{lotina2025hex}, the mapped 
$sdg$-IBM calculations were carried out 
for Sm and Gd nuclei using 
the constrained HFB method 
with the Gogny-D1S \cite{D1S} EDF. 
It was shown that the Gogny EDF provided the 
axially symmetric ($\beta_2,\beta_4$) PESs which are 
qualitatively quite similar to those obtained 
within the relativistic EDF framework. 
The main conclusions regarding 
the $g$-boson effects on energy spectra 
were also shown to be robust, regardless of 
whether the relativistic or nonrelativistic 
EDF is chosen as the starting point.

\subsection{Dynamical pairing degrees of freedom\label{sec:pv}}

In addition to shape oscillations,
pairing vibrations have been shown
to be relevant to nuclear spectroscopy, 
in particular, to the interpretation of the 
low-energy $0^+$ states and bands built on them, 
and electric monopole transitions
(see Ref.~\cite{nomura2021pv-1}, for
references to empirical evidence and
some earlier theoretical studies).
This degree of freedom has been incorporated
as an additional collective coordinate in
the DFT framework for the descriptions, e.g., of
nuclear fissions
\cite{giuliani2014,zhao2016,schunck2016,rayner2018,bernard2019,rayner2020,zhao2021,rayner2023,zdeb2025},
and neutrinoless 
$\beta\beta$ ($0\nu\beta\beta$) 
decay \cite{vaquero2013}.

A method of deriving 
the pairing plus quadrupole collective 
Hamiltonian by using the results
of the relativistic mean-field plus BCS (RMF+BCS)
calculations based on the PC-PK1 EDF \cite{PCPK1}
was developed \cite{xiang2020}. 
In Ref.~\cite{nomura2020pv}, 
the pairing degree of freedom was introduced 
in the IBM mapping procedure, using 
the pairing-quadrupole constrained 
RMF+BCS method with the PC-PK1 EDF, 
as in \cite{xiang2020}. 
In these two approaches, 
the pairing vibrations were shown 
to play an important role in lowering significantly 
the excited $0^+$ states in the rare-earth nuclei.

In Refs.~\cite{xiang2020,nomura2020pv}, 
only the axial quadrupole deformation 
was considered as the shape degree of freedom. 
An attempt to include simultaneously 
the dynamical pairing and triaxial quadrupole 
degrees of freedom for the spectroscopic studies 
within the EDF framework 
was first made in terms of the 
IBM mapping procedure in \cite{nomura2021pv-1}. 
The present section illustrates 
the method developed in Ref.~\cite{nomura2021pv-1}, 
by taking the $\gamma$-soft 
nucleus $^{128}$Xe as an example, 
and discusses impacts of the pairing and triaxial 
deformations on the corresponding 
spectroscopic properties.

The self-consistent calculations are performed 
within the RMF+BCS method as in \cite{nomura2020pv}. 
Constraints are imposed in terms of  
the triaxial quadrupole deformations (\ref{eq:beta2}), 
and the intrinsic pairing deformation, denoted by $\alpha$. 
The new coordinate $\alpha$ is related to 
the pairing gap $\Delta$, and is defined as 
the expectation value of the monopole 
pair operator $\hat P$
\begin{align}
\label{eq:pvdft}
 \alpha = \braket{\alpha | \hat P | \alpha }
= \sum_{\rho=\nu,\pi} \sum_{k>0} u^{\rho}_{k} v^{\rho}_{k} \; 
\end{align}
in a BCS state $\ket{\alpha}$, which is given by
(without pairing rotation)
\begin{align}
 \ket{\alpha} = \Pi_{k>0} 
(u_k + v_k c^\+_{\bar{k}} c^\+_k) \ket{0} \; ,
\end{align}
and the operator $\hat P$ is defined as
\begin{align}
 \hat P = \frac{1}{2} \sum_{k>0}
(c_k c_{\bar{k}} + c^\+_{\bar{k}}c^\+_k) \; .
\end{align}
The parameters $u_k$ and $v_k$ are, respectively, 
unoccupation and occupation amplitudes 
for the particle in the state $k$. 
$k$ and $\bar{k}$ denote the single-nucleon and 
corresponding times-reversed states, respectively, 
and $c^{(\+)}_{k}$ are the single-nucleon 
annihilation (creation) operator.

%-------- pv 128Xe PES------------
\begin{figure}[h]
\centering
\includegraphics[width=\linewidth]{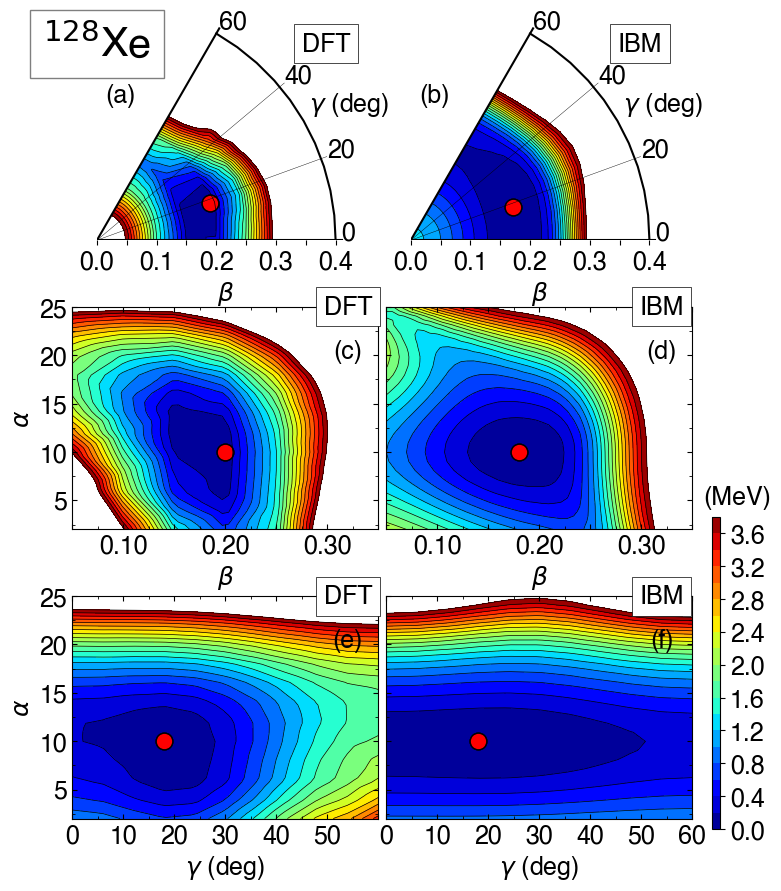}
\caption{
Triaxial quadrupole $(\beta,\gamma)$ - 
pairing ($\alpha$) constrained 
energy surface for $^{128}$Xe, projected onto the 
$(\beta,\gamma)$ [(a) and (b)],
$(\beta,\alpha)$ [(c) and (d)], and 
$(\gamma,\alpha)$ [(e) and (f)] surfaces.
The energy surfaces shown in the left column 
are computed by the RHB-SCMF method with the PC-PK1 EDF 
and separable pairing interaction, and those 
on the right-hand side are 
the corresponding mapped IBM PESs. 
The energy minimum on each surface is indicated 
by the solid circle.}
\label{fig:pv-pes-xe128}
\end{figure}
%---------------------------------

Figure~\ref{fig:pv-pes-xe128} gives 
the $(\beta,\gamma,\alpha)$-constrained SCMF PES, 
which is projected onto the $(\beta,\gamma)$, 
$(\beta,\alpha)$, and $(\gamma,\alpha)$ surfaces, 
for which the $\alpha$, $\gamma$, and $\beta$ values 
are fixed at those corresponding to 
the global minimum in the three-dimensional 
$(\beta,\gamma,\alpha)$ space. 
The $(\beta,\gamma)$ SCMF PES exhibits a pronounced 
$\gamma$-softness with a shallow triaxial minimum 
at $\gamma\approx20^{\circ}$. 
In the $(\beta,\alpha)$ and $(\gamma,\alpha)$ 
SCMF PESs a minimum is obtained at 
a non-zero pairing deformation $\alpha \approx 10$.

The procedure to incorporate 
the pairing vibration 
effects in the IBM is formally similar 
to that in the IBM-CM, described in Sec.~\ref{sec:co}, 
since in both cases several boson spaces that correspond 
to different boson numbers are mixed. 
In the IBM-CM, however, 
the configuration mixing is 
performed between those spaces that 
reflect particle-hole 
excitations from other major shells. 
For the pairing vibration, the configuration 
mixing is assumed to occur 
rather within a given valence space.

To incorporate the pairing vibration effects, 
the number of bosons $N_{\rm B}$ for a given 
nucleus is allowed to vary by 1, that is, 
$N_{\rm B}=N_{\rm B}^{(0)}-1$, 
$N_{\rm B}^{(0)}$, and $N_{\rm B}^{(0)}+1$, 
with $N_{\rm B}^{(0)}$ denoting the boson number 
in the normal space. 
The entire boson model space is represented as
the direct sum of the subspaces 
consisting of the $N_{\rm B}^{(0)}-1$, 
$N_{\rm B}^{(0)}$, and $N_{\rm B}^{(0)}+1$ $sd$ 
bosons:
\begin{align}
\label{eq:spacepv}
[N_{\rm B}^{(0)}-1] \oplus 
[N_{\rm B}^{(0)}] \oplus 
[N_{\rm B}^{(0)}+1] \; .
\end{align}
The corresponding Hamiltonian comprises 
the terms that conserves ($\hat H_{\rm cons}$)
and does not conserve ($\hat H_{\rm noncons}$) 
the boson number:
\begin{align}
\label{eq:hampv1}
\bh = \hat H_{\rm cons} + \hat H_{\rm noncons} \; .
\end{align}
Here the simpler IBM-1 is considered. 
Assuming that the pairing vibrations are 
simulated in terms of $s$ bosons, 
$\hat H_{\rm noncons}$ corresponds to the boson 
analog of the monopole pair transfer operator, 
which is of the form
\begin{align}
\label{eq:hampv2}
 \hat H_{\rm noncons} = \eta \frac{1}{2} (s^\+ + \tilde s) \; ,
\end{align}
with the strength parameter $\eta$. 
The Hamiltonian $\hat H_{\rm cons}$ 
takes the form
\begin{align}
\label{eq:hampv3}
 \hat H_{\rm cons}
&=
\hat{\delta} 
+ \epsilon_{d}\hat{n}_{d} 
+ \kappa \hat{Q}\cdot\hat{Q}
+ \rho\hat{L}\cdot\hat{L}
\nonumber\\
&+\theta\sum_{\lambda=2,4}
((d^\+ d^\+)^{(\lambda)} d^\+)^{(3)}\cdot ((\tilde d\tilde d)^{(\lambda)}\tilde d)^{(3)} \; .
\end{align}
The first term 
\begin{align}
 \hat\delta = \epsilon_{0}\hat{n} = \epsilon_{0}(s^{\+}s+d^{\+}\cdot\tilde d)
\end{align}
represents the total boson number, 
which only affects the absolute value 
of the ground-state energy. 
This term is introduced in order to 
determine relative energies of 
the three unperturbed $0^+$ ground states. 
The second, third, and fourth terms represent 
the $d$-boson number operator, quadrupole-quadrupole 
interaction, and rotational terms, the details of which 
are given in Ref.~\cite{nomura2021pv-2}. 
The last term in (\ref{eq:hampv3}) 
stands for a cubic term in the IBM-1, 
introduced in order to produce a 
triaxial minimum.

In a similar manner to that for the IBM-CM, 
the bosonic PES in the $(\beta,\gamma,\alpha)$ space 
is obtained as an energy expectation value in 
the coherent state
\begin{align}
 \ket{\Phi({\vec\alpha})}=
\ket{\Phi_{N_{\rm B}^{(0)}-1}({\vec\alpha})}\oplus
\ket{\Phi_{N_{\rm B}^{(0)}}({\vec\alpha})}\oplus
\ket{\Phi_{N_{\rm B}^{(0)}+1}({\vec\alpha})} \; .
\end{align}
The coherent state for each unperturbed boson space
$\ket{\Phi_{N_{\rm B}}({\vec\alpha})}$ is defined by
\begin{align}
\ket{\Phi_{N_{\rm B}}(\vec{\alpha})}
\frac{1}{\sqrt{N_{\rm B}!}}(\lambda^{\+})^{N_{\rm B}}\ket{0} \; ,
\end{align}
where 
\begin{align}
 \lambda^{\+}=
\frac{1}{\sqrt{\mathcal{N}}}
\Biggl[
\alpha_{s}s^{\+} + \beta_{\rm B}\cos{\gamma_{\rm B}}d_{0}^{\+}
+\frac{1}{\sqrt{2}}\beta_{\rm B}\sin{\gamma_{\rm B}}(d_{+2}^{\+}+d_{-2}^{\+})
\Biggr] \; ,
\end{align}
with a normalization factor 
${\mathcal{N}}=\alpha_{s}^2+\beta_{\rm B}^{2}$, 
is characterized by 
the three amplitudes 
$\{\beta_{\rm B},\gamma_{\rm B},\alpha_{s}\}$. 
Note that the vector $\vec{\alpha}$ represents 
the set $\{\beta_{\rm B},\gamma_{\rm B},\alpha_{s}\}$ 
and $\ket{0}$ is the boson vacuum, that is, the inert core. 
It was proposed in Ref.~\cite{nomura2020pv} 
that $\alpha_s$ is transformed into a new 
variable $\alpha_{\rm B}$ by the formula
\begin{align}
 \alpha_s = \cosh{(\alpha_{\rm B}-\alpha_{\rm B}^{\rm min})} \; ,
\end{align}
where $\alpha_{\rm B}^{\rm min}$ represents the 
$\alpha$ deformation that gives the global minimum 
in the $(\beta,\gamma,\alpha)$ SCMF PES. 
The new variable $\alpha_{\rm B}$ is treated as the boson 
equivalent of the pairing deformation 
$\alpha$ (\ref{eq:pvdft}) in the fermionic system. 
In addition, the relation between the 
fermionic and bosonic $\alpha$ deformations
\begin{align}
 \alpha_{\rm B} = C_{\alpha} \alpha \; ,
\end{align}
and those for the 
quadrupole triaxial shape deformations 
$\beta_{\rm B} = C_{\beta} \beta$ and 
$\gamma_{\rm B} = \gamma$ (\ref{eq:cbeta}) 
are assumed. 
$C_{\alpha}$ and $C_{\beta}$ are coefficients 
of proportionality.

The IBM energy surface is obtained as the lowest 
eigenvalue of the $3\times 3$ coherent-state matrix 
at each set of the deformations 
$\{\beta_{\rm B},\gamma_{\rm B},\alpha_{\rm B}\}$. 
The procedure to obtain the parameters in the 
Hamiltonian (\ref{eq:hampv1}) is described 
in Ref.~\cite{nomura2021pv-2}. 
The mapped IBM PES for the $^{128}$Xe, projected 
onto the $(\beta,\gamma)$, $(\beta,\alpha)$, 
and $(\gamma,\alpha)$ spaces, 
is shown in Fig.~\ref{fig:pv-pes-xe128}.
One can see that the bosonic PESs in these
deformation spaces are similar to the SCMF
counterparts.
The Hamiltonian, 
defined in Eqs.~(\ref{eq:hampv1})--(\ref{eq:hampv3}), 
with the parameters determined by the mapping procedure 
is diagonalized in the space (\ref{eq:spacepv}).

%-------- pv 128Xe levels ------------
\begin{figure}[h]
\centering
\includegraphics[width=.8\linewidth]{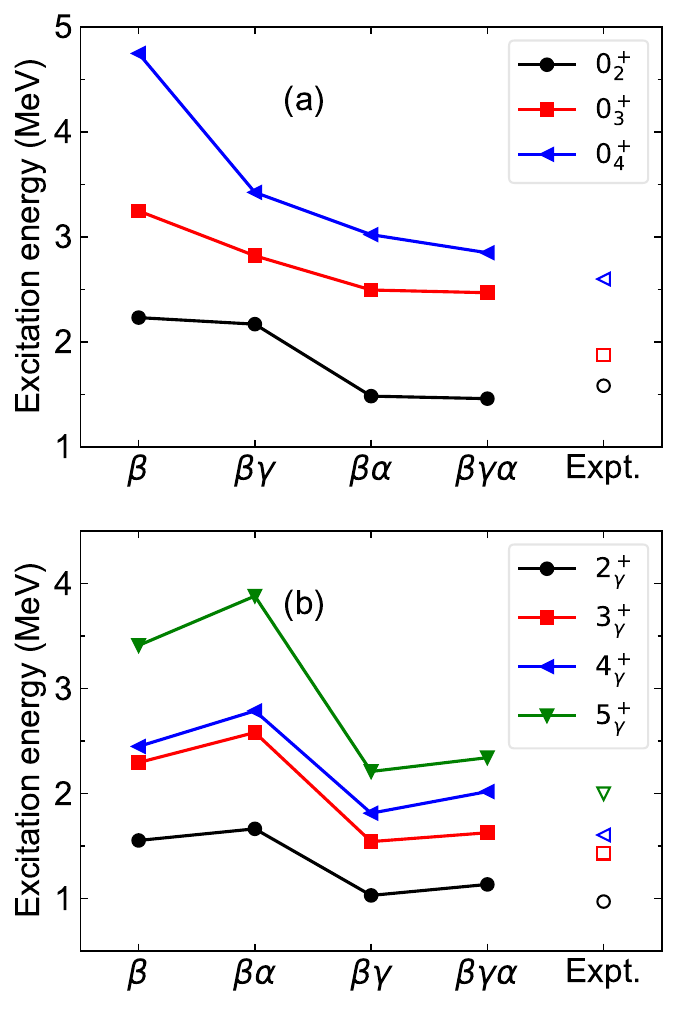}
\caption{
Excitation energies of 
the $0^+_2$, $0^+_3$ and $0^+_4$ states,
and $2^+$, $3^+$, $4^+$ and $5^+$ states
of the $\gamma$-vibrational band, 
resulting 
from the mapped IBM that includes 
the axial quadrupole ($\beta$), 
triaxial quadrupole $(\beta\gamma)$, 
axial quadrupole and pairing ($\beta\alpha$), 
and triaxial quadrupole 
and pairing ($\beta\gamma\alpha$) deformations.}
\label{fig:pv-level}
\end{figure}
%-------------------------------------

Figures~\ref{fig:pv-level}(a) and \ref{fig:pv-level}(b) 
show, respectively, the calculated excitation energies 
of the $0^+_{2}$, $0^+_{3}$ and $0^+_{4}$ states, 
and of the states that are members 
of the $\gamma$-vibrational band.
By including the pairing 
degree of freedom, the $0^+$ energy levels 
are lowered. 
The triaxial deformation, however, does not 
appear to have major influences 
on the excited $0^+$ levels, except for $0^+_{4}$. 
The $\gamma$-band levels are mostly affected 
by the inclusion of the triaxiality, 
which is taken into account by the cubic term 
in the Hamiltonian (\ref{eq:hampv3}).

The mapped $\beta\gamma\alpha$-IBM that is based on 
the relativistic DD-PC1 and PC-PK1 interactions 
was applied to an extensive 
study on the even-even $\gamma$-soft 
nuclei \cite{nomura2021pv-2} in the 
$A\approx 130$ ($^{128}$Xe and $^{130}$Xe), 
and $A\approx 190$ 
($^{188,190,192}$Os and $^{192,194,196}$Pt) regions. 
In the Os-Pt region, in particular, 
the prolate-to-oblate shape phase transitions 
and the $\gamma$-soft shapes at the transitional regions 
are empirically suggested to occur. 
This systematic study 
confirmed the importance of considering 
these degrees of freedom simultaneously for the 
descriptions of $\gamma$-soft nuclei.

\section{Particle-boson coupling\label{sec:odd}}

The microscopic description of 
low-lying states of 
those nuclei with odd number(s) of nucleons 
has been a major theoretical challenge. 
Within the EDF framework, 
the full GCM configuration-mixing calculations 
that restore time-reversal symmetries and 
take into account blocking effects were made for 
a few light odd-mass (i.e., Mg) nuclei
\cite{bally2014,borrajo2016,zhou2024}. 
Theoretical calculations on odd nuclei, 
i.e., odd-mass and odd-odd nuclei, 
are not only important from a nuclear 
structure point of view, but 
are crucial for modeling electroweak
decay processes such as the nuclear $\beta$ decay. 
This section outlines the extension of 
the EDF-mapped IBM to describe the low-energy 
excitations in heavy odd nuclei.

\subsection{Formalism\label{sec:odd-formulas}}

In order to study odd nuclei, 
fermionic degrees of freedom representing 
unpaired nucleons, and their 
couplings to the even-even boson core system 
are considered within the interacting 
boson-fermion model (IBFM) \cite{IBFM,iachello1979}
for odd-mass nuclei, and the interacting 
boson-fermion-fermion model (IBFFM) \cite{brant1984,IBFM} 
for odd-odd nuclei. 
The theoretical framework for the IBFFM, which 
includes two unpaired particles, is more general 
than the IBFM, in which a single fermion 
degree of freedom is considered. 
In the following, therefore, 
the formulations are given for the IBFFM. 
In addition, 
the neutron-proton version of the 
IBFFM, denoted by IBFFM-2, is mainly considered. 
The distinction between neutron 
and proton bosons are especially important for 
calculating single and double $\beta$ decays,
in which both the neutron
and proton degrees of freedom must 
be explicitly considered.

The Hamiltonian for the IBFFM-2 is written 
in general as
\begin{align}
\label{eq:hamffm}
 \hamffm = \bh + \hf^{\nu} + \hf^{\pi} + \hbf^{\nu} + \hbf^{\pi} + \hff \; .
\end{align}
The first term represents the IBM-2 core
Hamiltonian, e.g., of the form (\ref{eq:ham1}).
The second (third) term of 
Eq.~(\ref{eq:hamffm}) represents 
the single-neutron (proton) Hamiltonian 
of the form
\begin{align}
\label{eq:hf}
 \hf^{\rho} = -\sum_{\jr}\epsilon_{\jr}\sqrt{2\jr+1}
  (a_{\jr}^\+\times\tilde a_{\jr})^{(0)}
\equiv
\sum_{\jr}\epsilon_{\jr}\hat{n}_{\jr} \; ,
\end{align}
where $\epsilon_{\jr}$ stands for the 
single-particle energy of the odd neutron 
or proton orbit $\jr$. 
$a_{\jr}^{(\+)}$ represents 
the particle annihilation (creation) operator, 
with $\tilde{a}_{\jr\mr}=(-1)^{\jr -\mr}a_{\jr-\mr}$. 
$\hat{n}_{\jr}$ stands for the number operator 
for the odd nucleon in the orbit $\jr$. 

The $\hbf^{\nu}$ and $\hbf^{\pi}$ 
in (\ref{eq:hamffm}) denote, respectively, 
the interactions
between a single neutron and the even-even boson core, 
and between a single proton and the boson core.
A general form of $\hbf$ is highly complicated, but 
the following simplified form is shown to be adequate 
to describe low-lying states of odd nuclei in 
most realistic calculations \cite{iachello1979,IBFM}. 
\begin{equation}
\label{eq:hbf}
 \hbf^{\rho}
=\Gamma_{\rho}\hat{V}_{\mathrm{dyn}}^{\rho}
+\Lambda_{\rho}\hat{V}_{\mathrm{exc}}^{\rho}
+A_{\rho}\hat{V}_{\mathrm{mon}}^{\rho} \; ,
\end{equation}
where the first, second, and third terms 
are quadrupole dynamical, exchange, 
and monopole interactions, respectively, 
with the strength parameters 
$\Gamma_\rho$, $\Lambda_\rho$, and $A_{\rho}$. 
The three terms in (\ref{eq:hbf}) are 
further simplified using the generalized seniority 
framework \cite{scholten1985,IBFM}: 
\begin{align}
\label{eq:dyn}
&\hat{V}_{\mathrm{dyn}}^{\rho}
=\sum_{\jr\jr'}\gamma_{\jr\jr'}
(a^{\+}_{\jr}\times\tilde{a}_{\jr'})^{(2)}
\cdot\hat{Q}_{\rho'} \; ,
\\
\label{eq:exc}
&\hat{V}^{\rho}_{\mathrm{exc}}
=\left(
s_{\rho'}^\+\times\tilde{d}_{\rho'}
\right)^{(2)}
\cdot
%\left(
\sum_{\jr\jr'\jr''}
\sqrt{\frac{10}{N_{\rho}(2\jr+1)}}
\nonumber\\
&\quad\quad
\times
\beta_{\jr\jr'}\beta_{\jr''\jr}
:\left[
(d_{\rho}^{\+}\times\tilde{a}_{\jr''})^{(\jr)}\times
(a_{\jr'}^{\+}\times\tilde{s}_{\rho})^{(\jr')}
\right]^{(2)}:
\nonumber\\
& \quad\quad
+ (\text{H.c.}) \; ,
\\
\label{eq:mon}
&\hat{V}_{\mathrm{mon}}^{\rho}
=\hat{n}_{d_{\rho}}\hat{n}_{\jr} \; ,
\end{align}
where the notations for the $j$-dependent factors 
\begin{align}
\label{eq:gjj}
&\gamma_{\jr\jr'}=(u_{\jr}u_{\jr'}-v_{\jr}v_{\jr'})q_{\jr\jr'} \\
\label{eq:bjj}
&\beta_{\jr\jr'}=(u_{\jr}v_{\jr'}+v_{\jr}u_{\jr'})q_{\jr\jr'}
\end{align}
are used, with 
$q_{\jr\jr'}=\braket{\ell_{\rho}\frac{1}{2}\jr \| Y^{(2)} \| \ell'_\rho\frac{1}{2}\jr'}$ being the matrix element of the fermion 
quadrupole operator in the single-particle basis. 
$\hat{Q}_{\rho'}$ in (\ref{eq:dyn}) is the 
bosonic quadrupole operator of (\ref{eq:bquad}).  
The notation $:(\cdots):$ in Eq.~(\ref{eq:exc}) 
stands for normal ordering. 
Furthermore, in the seniority scheme 
the unperturbed single-particle 
energy, $\epsilon_{\jr}$, in Eq.~(\ref{eq:hf}) 
is replaced with the quasiparticle 
energy $\tilde\epsilon_{\jr}$. 
The use of the specific forms for the 
boson-fermion interactions in 
(\ref{eq:dyn})--(\ref{eq:mon}) 
is based on the assumption that 
both the dynamical and exchange 
terms are dominated by the interaction between 
unlike particles, i.e., between 
an odd neutron (proton) and proton (neutron) bosons, 
and that, for the monopole term, like-particle 
interaction, i.e., the interaction
between an odd neutron (proton)
and neutron (proton) bosons plays an dominant role \cite{IBFM}.

The residual interaction 
between the unpaired neutron and proton
$\hff$ (\ref{eq:hamffm}) is of the form
\begin{align}
\label{eq:hff}
\hff
=& 4\pi
(
{\vd}
+
{\vssd}
\bm{\sigma}_\nu \cdot \bm{\sigma}_\pi
)
%\delta(\bm{r})
\delta(\bm{r}_{\nu}-r_0)
\delta(\bm{r}_{\pi}-r_0)
\nonumber\\
&
-\frac{1}{\sqrt{3}}
\vsss \bm{\sigma}_\nu \cdot \bm{\sigma}_\pi
\nonumber\\
&
+ \vt
\left[
\frac{3({\bm\sigma}_{\nu}\cdot{\bf r})
({\bm\sigma}_{\pi}\cdot{\bf r})}{r^2}
-{\bm{\sigma}}_{\nu}
\cdot{\bm{\sigma}}_{\pi}
\right] \; .
\end{align}
The first, second, third, and fourth 
terms stand for the delta, spin-spin 
delta, spin-spin, 
and tensor interactions, with 
$\vd$, $\vssd$, $\vsss$ and $\vt$ 
being strength parameters, respectively. 
Note that $\bm{r}=\bm{r}_{\nu}-\bm{r}_{\pi}$ 
is the relative coordinate of the 
neutron and proton, and $r_0=1.2A^{1/3}$ fm. 
The matrix element of $\hff$, 
denoted by $V_{\nu\pi}'$, depends on the 
occupation $v_j$ and unoccupation $u_j$ 
amplitudes \cite{yoshida2013}: 
\begin{align}
\label{eq:vres}
V_{\nu\pi}'
&= (u_{j_\nu'} u_{j_\pi'} u_{j_\nu} u_{j_\nu} + v_{j_\nu'} v_{j_\pi'} v_{j_\nu} v_{j_\nu})
V^{J}_{j_\nu' j_\pi' j_\nu j_\pi}
\nonumber \\
& {} - (u_{j_\nu'}v_{j_\pi'}u_{j_\nu}v_{j_\pi} +
 v_{j_\nu'}u_{j_\pi'}v_{j_\nu}u_{j_\pi}) \nonumber \\
&\times \sum_{J'} (2J'+1)
\left\{ \begin{array}{ccc} {j_\nu'} & {j_\pi} & J' \\ {j_\nu} & {j_\pi'} & J
\end{array} \right\} 
V^{J'}_{j_\nu'j_\pi j_\nu j_\pi'} \; ,
\end{align}
with 
\begin{align}
V^{J}_{j_\nu'j_\pi'j_\nu j_\pi} 
= \langle j_\nu'j_\pi';J| \hff |j_\nu j_\pi;J\rangle
\end{align}
being the matrix element in 
the bases defined in terms of 
the neutron-proton pair 
coupled to the angular momentum $J$. 
The bracket in (\ref{eq:vres}) 
represents the Racah coefficient. 
By following the procedure of 
Ref.~\cite{morrison1981}, 
those terms resulting from contractions are 
neglected in Eq.~(\ref{eq:vres}).

The procedure to determine 
the parameters for the Hamiltonian (\ref{eq:hamffm}) 
was developed in Refs.~\cite{nomura2016odd,nomura2019dodd}, 
and is summarized as follows. 
\begin{enumerate}
 \item 
The parameters of the IBM-2 even-even 
core Hamiltonian are determined by the
PES mapping procedure
described in Sec.~\ref{sec:model}.

\item
The quasiparticle energies $\tilde\epsilon_{\rho}$, 
and occupation probabilities $v^2_{\jr}$ of the 
odd nucleon in orbits $\jr$ are computed by
the SCMF calculations for the spherical configuration. 
For these calculations 
the standard HFB method, i.e., without blocking, 
is employed, but imposing the odd nucleon number 
constraint. 
These quantities are input to 
the single-fermion Hamiltonian $\hf$ (\ref{eq:hf}), 
to the factors $\gamma_{\jr\jr'}$ (\ref{eq:gjj}) 
and $\beta_{\jr\jr'}$ (\ref{eq:bjj}) 
for the boson-fermion interaction $\hbf$, and 
to $V_{\nu\pi}'$ (\ref{eq:vres}).

\item
The coupling constants $\Gamma_{\rho}$, 
$\Lambda_{\rho}$, and $A_{\rho}$ are treated as 
phenomenological parameters, and are determined 
to reproduce, to a reasonable accuracy, 
the experimental data for a few low-lying 
levels of each odd-$N$ (for $\rho=\nu$) and 
odd-$Z$ (for $\rho=\pi$) nucleus, 
separately for positive-parity 
and negative-parity states.

\item
The strength parameters for the neutron-proton 
interaction (\ref{eq:hff}) are determined so as 
to reproduce the observed low-lying 
levels of each odd-odd nucleus.
\end{enumerate}
The IBFFM-2 Hamiltonian determined by the 
above procedure is diagonalized in the 
basis $\ket{L_\nu L_\pi(L);j_{\nu}j_{\pi}(J):I}$, 
where $L_\nu$ ($L_\pi$) and $L$ stand for the 
angular momentum of the neutron (proton) boson 
system and total angular momentum of the 
even-even boson core system, respectively.

The $E2$ and $M1$ transitions are calculated 
using the corresponding operators that depend on 
the $u$ and $v$ amplitudes, 
which are calculated by the SCMF method. 
The $E2$ transition operator $\hat T^{(E2)}$ reads
\begin{align}
 \label{eq:e2ffm}
\hat T^{(E2)}
= \hat T^{(E2)}_\text{B}
+ \hat T^{(E2)}_\text{F} \; ,
\end{align}
where the boson part $\hat T^{(E2)}_\text{B}$ 
is defined in (\ref{eq:e2op}), 
and the fermion part is of the form
\begin{align}
 \label{eq:e2f}
\hat T^{(E2)}_\mathrm{F}=
&-\frac{1}{\sqrt{5}}
\sum_{\rho=\nu,\pi}
\sum_{\jr\jr'}
(u_{\jr}u_{\jr'}-v_{\jr}v_{\jr'})
\nonumber\\
&\times
\left\langle
\ell_\rho\frac{1}{2}\jr 
\bigg\| 
e^\mathrm{F}_\rho r^2 Y^{(2)} 
\bigg\|
\ell_\rho'\frac{1}{2}\jr'
\right\rangle
(a_{\jr}^\dagger\times\tilde a_{\jr'})^{(2)} \; .
\end{align}
$e^\mathrm{F}_\nu$ and
$e^\mathrm{F}_\pi$ are the neutron 
and proton effective charges, respectively. 
The $M1$ transition operator 
$\hat T^{(M1)}$ reads 
\begin{align}
 \label{eq:m1}
\hat T^{(M1)}
=\sqrt{\frac{3}{4\pi}}
&\sum_{\rho=\nu,\pi}
\Biggl[
g_\rho^\mathrm{B}\hat L_\rho
-\frac{1}{\sqrt{3}}
\sum_{\jr\jr'}
(u_{\jr}u_{\jr'}+v_{\jr}v_{\jr'})
\nonumber \\
&\times
\left\langle \jr \| g_l^\rho{\bf \ell}+g_s^\rho{\bf s} 
\| \jr' \right\rangle
(a_{\jr}^\+\times\tilde a_{\jr'})^{(1)}
\Biggr] \; .
\end{align}
$g_\rho^\mathrm{B}$ are $g$-factors for 
the neutron or proton bosons, and
$g_\ell^\nu=0\,\mu_N$ and $g_s^\nu=-3.82\,\mu_N$
($g_\ell^\pi=1.0\,\mu_N$ and $g_s^\pi=5.58\,\mu_N$) 
are free values of the neutron (proton) orbital 
and spin $g$-factors, respectively.

The formulation described above is 
based on the generalized seniority framework, 
and has been used in the IBFM-2 and IBFFM-2
calculations for realistic nuclei
\cite{IBFM}.
In these studies, the $u$ and $v$ factors 
and quasiparticle energies 
are usually obtained by solving a simple  
BCS equation with empirical single-particle energies 
and pairing gap $\Delta=12/\sqrt{A}$, and 
the parameters for the even-even core Hamiltonian 
and coupling constants for the boson-fermion interactions 
are determined in a purely phenomenological way. 
In the mapped IBM-2 framework, 
the even-even boson core Hamiltonian, 
the occupation probabilities and quasiparticle 
energies are determined by the self-consistent 
calculations based on the nuclear EDF. 
While the coupling constants of the 
boson-fermion and residual neutron-proton 
interactions are determined so as to reproduce 
the experimental low-energy spectra, 
the method allows to study 
even-even, odd-mass, and odd-odd nuclei consistently 
by using as microscopic input the results 
of the EDF-SCMF calculations, 
and reduces significantly the number 
of adjustable parameters.

\subsection{Structure of odd-mass nuclei\label{sec:odda}}

The method described in Sec.~\ref{sec:odd-formulas} 
was initially applied to calculating 
the low-energy spectra of Eu nuclei, 
which are coupled systems of an odd proton and 
even-even Sm nuclei as boson cores. 
The single-particle space is the proton 
major oscillator shell $Z=50-82$, which includes 
the $3s_{1/2}$, $2d_{3/2}$, $2d_{5/2}$, $1g_{7/2}$, 
and $1h_{11/2}$ orbits. 
The corresponding even-even Sm nuclei exhibit 
a shape phase transition 
between the nearly spherical vibrational 
and strongly deformed rotational states 
as a function of the neutron number. 
It is thus of particular interest
to study how the nuclear structure
in the neighboring odd-mass systems changes
in the presence of the unpaired particle.

%-------- Eu negative-parity leves ---
\begin{figure}[h]
\centering
\includegraphics[width=\linewidth]{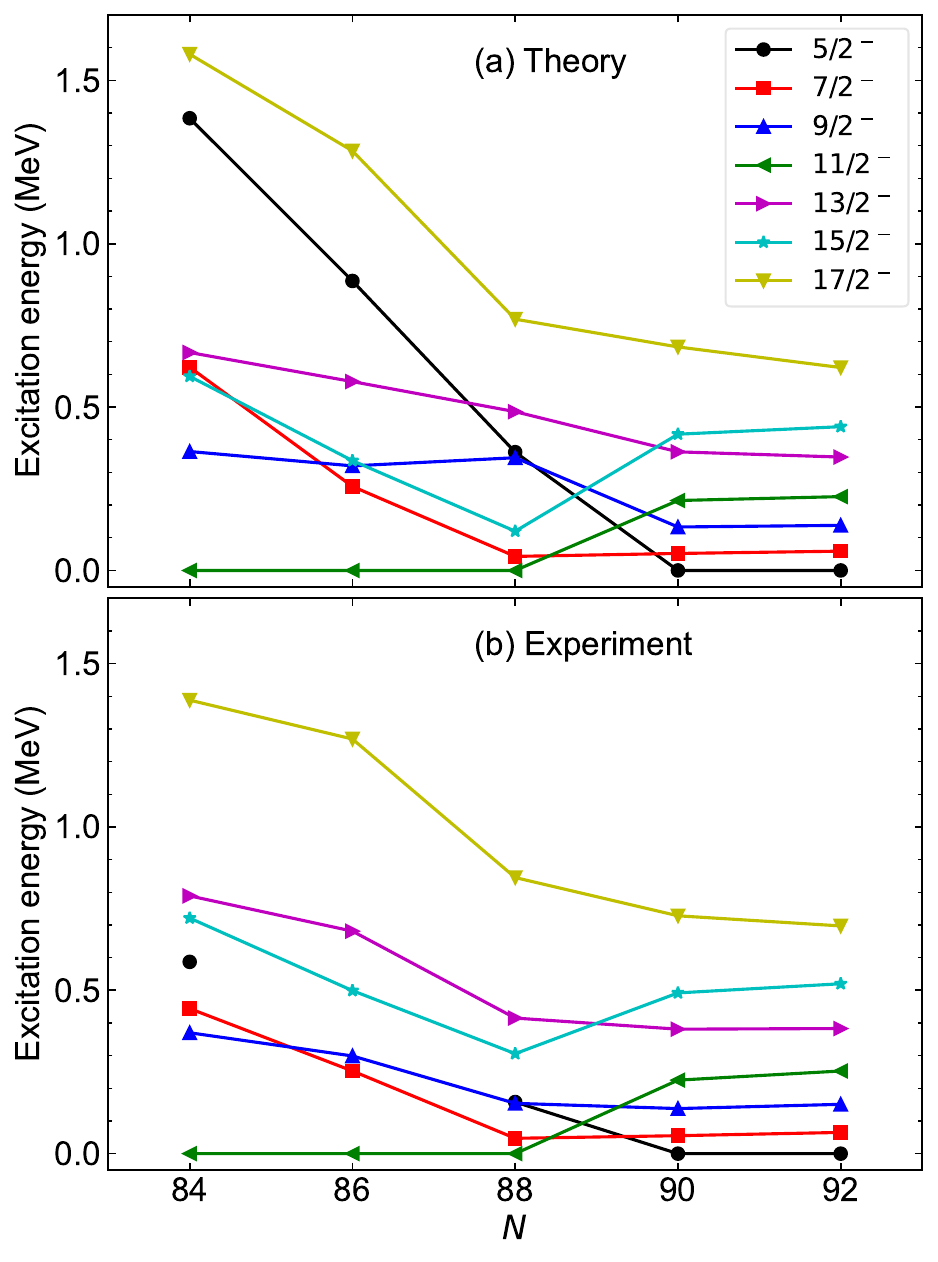}
\caption{
(a) Excitation energies of the
negative-parity yrast states
calculated for the odd-$Z$ nuclei $^{147-155}$Eu
within the IBFM-1 that is based on
the RHB-SCMF method with the DD-PC1 EDF,
and (b) the observed \cite{data} energy levels.
}
\label{fig:eu-level-neg}
\end{figure}
%-------------------------------------

Evolution of the calculated 
\cite{nomura2016odd,nomura2016qpt}
excitation spectra for the negative-parity states 
of the odd-mass nuclei $^{147-155}$Eu is
shown in Fig.~\ref{fig:eu-level-neg},
and is compared with the experimental data \cite{data}. 
Note that, in this example,
the simpler IBFM-1 was employed \cite{nomura2016odd}
as an initial implementation of the
EDF-based IBFM.
The boson-fermion interactions for the IBFM-1 
are similar to those defined in Eq.~(\ref{eq:hbf}), 
but the exchange term (\ref{eq:exc}) 
takes a simpler form 
\begin{eqnarray}
 \sum_{j j' j''}
\Lambda_{jj'j''}
%
%-2\Lambda_0\sqrt{\frac{5}{2j''+1}}
%\beta_{jj''}^{(2)}\beta_{j'j''}^{(2)}
%
:\left[(a_{j}^\+\times\tilde d)^{(j'')} 
\times (d^\+\times\tilde a_{j'})^{(j'')}\right]^{(0)}: \; ,
\end{eqnarray}
where the coefficients $\Lambda_{jj'j''}$ 
are expressed in terms of the $u$ and $v$ 
factors, as in Eqs.~(\ref{eq:exc}) and (\ref{eq:bjj}).
The microscopic inputs were calculated 
by the constrained RHB approach with the 
DD-PC1 EDF and pairing force of Ref.~\cite{tian2009}. 
The negative-parity states are obtained 
from the coupling of the proton $1h_{11/2}$ orbit 
to the even-even $sd$ boson core. 
The behaviors of the negative-parity levels 
reflect a rapid structural change in the 
even-even Sm isotopes. The shape transition 
in the odd-mass systems is characterized by 
a change in the spin of the ground state at 
a particular nucleus. 
In the example shown in Fig.~\ref{fig:eu-level-neg}, 
the spin of the ground state is ${11/2}^-$ for 
those Eu nuclei with $N=84-88$, 
but is ${5/2}^-$ for $N=90$ ($^{153}$Eu) 
and $N=92$ ($^{155}$Eu). 
In the even-even Sm neighbors, the transition 
from spherical to deformed shapes indeed occurs 
near $N=90$. 
Both the calculated and experimental spectra for 
the odd-mass Eu resemble vibrational levels 
for $N \leqslant 86$, for which low-lying 
bands exhibit the $\Delta I=2$ $E2$ transitions 
of the weak coupling limit. 
In the region after the transition point, 
i.e., $N\geqslant 90$, the levels look more like 
rotational bands, which exhibit 
the $\Delta I=1$ pattern of strong-coupling limit. 
The positive-parity levels of the odd-mass Eu isotopes 
are described in terms of the coupling of the 
normal-parity ($sdg$) single-particle orbits to 
the boson core, and exhibit rapid structural evolution 
near $N=90$ characterized by the change 
in the ground-state spin, 
similarly to the negative-parity levels.

In Refs.~\cite{nomura2016odd,nomura2016qpt}, 
evolution of low-lying positive-parity
and negative-parity states
in the odd-mass Sm isotopes was studied,
which are systems of an odd neutron
coupled to the even-even Sm cores.
Signatures of the shape phase transitions in 
the axially symmetric odd-mass Eu and Sm isotopes were 
proposed \cite{nomura2016qpt}, such as
the excitation energies, 
intrinsic ($\beta$) deformation, quadrupole shape invariant, 
and one-neutron separation energies. 
These quantities were shown to exhibit a discontinuity 
at $N\approx 90$, and can be thus 
considered to be order parameters
of the shape phase transitions. 
In Ref.~\cite{nomura2017odd-1}, 
the shape phase transitions 
in odd-mass $\gamma$-soft 
Xe, Cs, Ba, and La nuclei in the 
mass $A\approx 130$ region 
were analyzed. 
The effective $\beta$ and $\gamma$ deformations 
calculated by the $E2$ matrix elements 
resulting from the IBFM-1 demonstrated 
signatures of the transition between the 
$\gamma$-unstable O(6) and nearly spherical 
U(5) limits. 
In order to show the robustness of the EDF-based 
IBFM, the spectroscopic properties of the same 
odd-mass nuclei as those considered in 
Refs.~\cite{nomura2016qpt,nomura2017odd-1} 
were studied by adopting the nonrelativistic 
Gogny EDF \cite{nomura2017odd-2,nomura2017odd-3}. 
It was shown that the EDF-based IBFM 
was able to reproduce the experimental 
low-energy collective properties of these 
odd-mass systems reasonably well, 
regardless of whether the relativistic 
or nonrelativistic (Gogny) EDF is employed 
as a starting point.

Other types of the shape phase transitions 
in odd-mass systems were analyzed within the 
IBFM-2 that is based on the Gogny EDF. 
In Ref.~\cite{nomura2018pt}
the effective triaxial 
deformations $\gamma$ calculated by using 
the $E2$ matrix elements in the IBFM-2 
indicated a prolate-to-oblate transition 
in the odd-mass Pt, Os, and Ir nuclei in 
the mass $A\approx 190$ region. 
The structural evolution in the neutron-rich 
odd-mass nuclei $^{87-95}$Kr was studied 
in Ref.~\cite{nomura2018kr}. 
In the neighboring even-even 
systems $^{86-94}$Kr, gradual shape evolution 
from $\gamma$-soft to a pronounced 
oblate deformation was suggested 
in the energy surface. 
The IBFM-2 calculation produced detailed  
energy levels and electromagnetic properties 
for the neutron-rich $^{95}$Kr,
for which a $\gamma$-ray spectroscopy has
recently been performed at RIKEN \cite{gerst2022}.

The EDF-based IBFM-2 was applied to describe
the low-lying states of the neutron-rich odd-mass
Zr isotopes \cite{nomura2020zr}.
Theoretical descriptions of these systems
are even more challenging than in the case
of the even-even Zr.
The underlying SCMF calculations were 
carried out within the relativistic framework. 
The quadrupole triaxial PESs for the 
even-even $^{94-102}$Zr nuclei exhibit
a transition from triaxial or $\gamma$-soft 
($^{94,96}$Zr) to prolate ($^{98}$Zr),  
and to triaxial ($^{100,102}$Zr) shapes. 
This systematic is qualitatively 
similar to that obtained from 
the HFB calculations based on the
Gogny-D1M EDF (cf. Fig.~\ref{fig:zr-pes1}). 
The calculated low-energy levels of states 
of both parities, and effective 
quadrupole deformations and their fluctuations 
of given states in the odd-mass Zr isotopes 
were shown to exhibit a pronounced discontinuity 
near the transitional nucleus $^{99}$Zr.

An even more challenging case is the calculation 
of low-lying states in odd-mass octupole deformed nuclei. 
The $sdf$-IBFM with the octupole-deformed even-even core, 
and with the boson-fermion interaction including 
the quadrupole and octupole degrees of freedom, 
was developed in \cite{nomura2018oct}, 
and was applied to investigate 
octupole correlation effects on spectroscopic properties in 
neutron-rich $^{141,143,145}$Ba nuclei. 
The corresponding even-even $^{144,146}$Ba are 
experimentally suggested to have a stable 
octupole deformation \cite{bucher2016,bucher2017}. 
A few experimental findings proposed 
octupole bands at about an excitation energy 
of 500 keV in the odd-mass systems also, 
e.g., $^{145}$Ba \cite{rzacaurban2012}. 
In Ref.~\cite{nomura2018oct}, 
the $sdf$-IBM Hamiltonian for the 
even-even Ba core of the form in (\ref{eq:hamoct}) 
was determined by mapping the axially symmetric 
$(\beta_2,\beta_3)$-constrained RHB PES onto 
that of the $sdf$-IBM, using the procedure described in 
Sec.~\ref{sec:oct}. 
The boson-fermion interaction terms of similar forms 
to those in (\ref{eq:dyn})--(\ref{eq:mon}) were derived 
by means of the generalized seniority scheme 
that includes the $f$ boson degree of freedom. 
The additional boson-fermion interaction 
parameters arising from the 
inclusion of $f$ bosons were determined 
so as to reproduce the bands experimentally 
suggested to be of octupole nature. 
Further details are found in \cite{nomura2018oct}. 
The $sdf$-IBFM calculation confirmed presence of 
several octupole bands in the $^{143,145,147}$Ba nuclei 
at intermediate excitation energies, which also
exhibit sizable amounts of the $E3$ transitions
to the ground-state bands. 
The octupole correlations were shown 
\cite{nomura2024oct} to be
relevant in the low-lying structures of the 
neighboring even-even and odd-mass Xe isotopes 
in the same mass region $A \approx 140$, 
although the effects are shown to be
much less significant than in the Ba nuclei.

In recent years, some low-spin 
non-yrast bands of several $\gamma$-soft nuclei,
e.g., $^{105}$Pd \cite{timar2019},
$^{127}$Xe \cite{CHAKRABORTY2020},
$^{133}$La \cite{biswas2019},
$^{135}$Pr \cite{matta2015,sensharma2019,135Pr-Lv},
$^{183}$Au \cite{nandi2020},
$^{187}$Au \cite{sensharma2020,187Au-guo2022},
were suggested to be wobbling bands, which are 
characterized by the dominant $E2$ transitions 
to yrast bands. 
These interpretations are based on the measurements 
of the $E2$ to $M1$ mixing ratios, $\delta(E2/M1)$, 
that were indeed larger in magnitude than 1, 
indicating the $E2$ dominance over $M1$.  
The theoretical descriptions of the proposed wobbling 
bands have been made mostly within 
the particle-rotor model. 
In Ref.~\cite{nomura2022wob}, 
an alternative interpretation of the proposed 
wobbling bands of $^{105}$Pd, $^{127}$Xe,
$^{133}$La, and $^{135}$Pr
was given in terms of the mapped IBM-2 and IBFM-2. 
The predicted ratios $\delta(E2/M1)$ of the 
$\Delta I=1$ transitions between yrast bands 
and those yrare bands that were proposed 
to be wobbling bands are all smaller than 
the corresponding experimental values. 
The small $\delta(E2/M1)$ mixing ratios indicated 
dominant magnetic character of these transitions, 
and are consistent with the updated experimental data 
on $^{135}$Pr \cite{135Pr-Lv} 
and earlier data on $^{105}$Pd \cite{rickey1977}. 
%
%$^{187}$Au \cite{sensharma2020}, and
%$^{183}$Au \cite{nandi2020},

\subsection{Calculations for odd-odd nuclei\label{sec:oo}}

Application of the EDF-based IBFFM-2 
has been made initially to the Au nuclei 
in the mass $A \approx 190$ region \cite{nomura2019dodd}, 
starting from the Gogny-HFB SCMF calculations. 
The even-even core nuclei for these systems 
are Hg isotopes, the low-lying states of which 
are characterized by a pronounced $\gamma$-softness.

An illustrative example is the structures 
of $\gamma$-soft odd-odd nuclei in the mass $A \approx 130$, 
since in this region a number of bands 
were empirically suggested to be 
chiral bands \cite{frauendorf1997}
e.g., in Cs isotopes \cite{starosta2017}. 
The IBFFM-2 was applied to 
a systematic investigation on the odd-odd $^{124-132}$Cs 
isotopes, with the microscopic input provided by 
the Gogny-EDF calculation \cite{nomura2020cs}. 
The IBFFM-2 produced low-spin positive-parity 
and negative-parity levels that were consistent 
with experiment for lighter isotopes $^{124,126,128}$Cs, 
while certain deviation from experiment was 
observed for those heavier nuclei that 
are rather close to the $N=82$ 
shell closure, in particular, $^{132}$Cs. 
The band structures of higher-spin states in the 
Cs isotopes were also analyzed in \cite{nomura2020cs}. 
Within the IBFFM-2 these bands are mainly 
based on the neutron-proton pair configurations 
$[(\nu h_{11/2})^{-1}\otimes \pi h_{11/2}]^{(J)}$. 
Many of the calculated 
$[(\nu h_{11/2})^{-1}\otimes \pi h_{11/2}]^{(J)}$ 
two-quasiparticle bands were identified as 
candidates for chiral doublet bands, 
which exhibit characteristic features 
such as the odd-even spin staggering 
pattern of the $B(M1;I \to I-1)$ transition rates.

\begin{figure}[h]
\centering
\includegraphics[width=\linewidth]{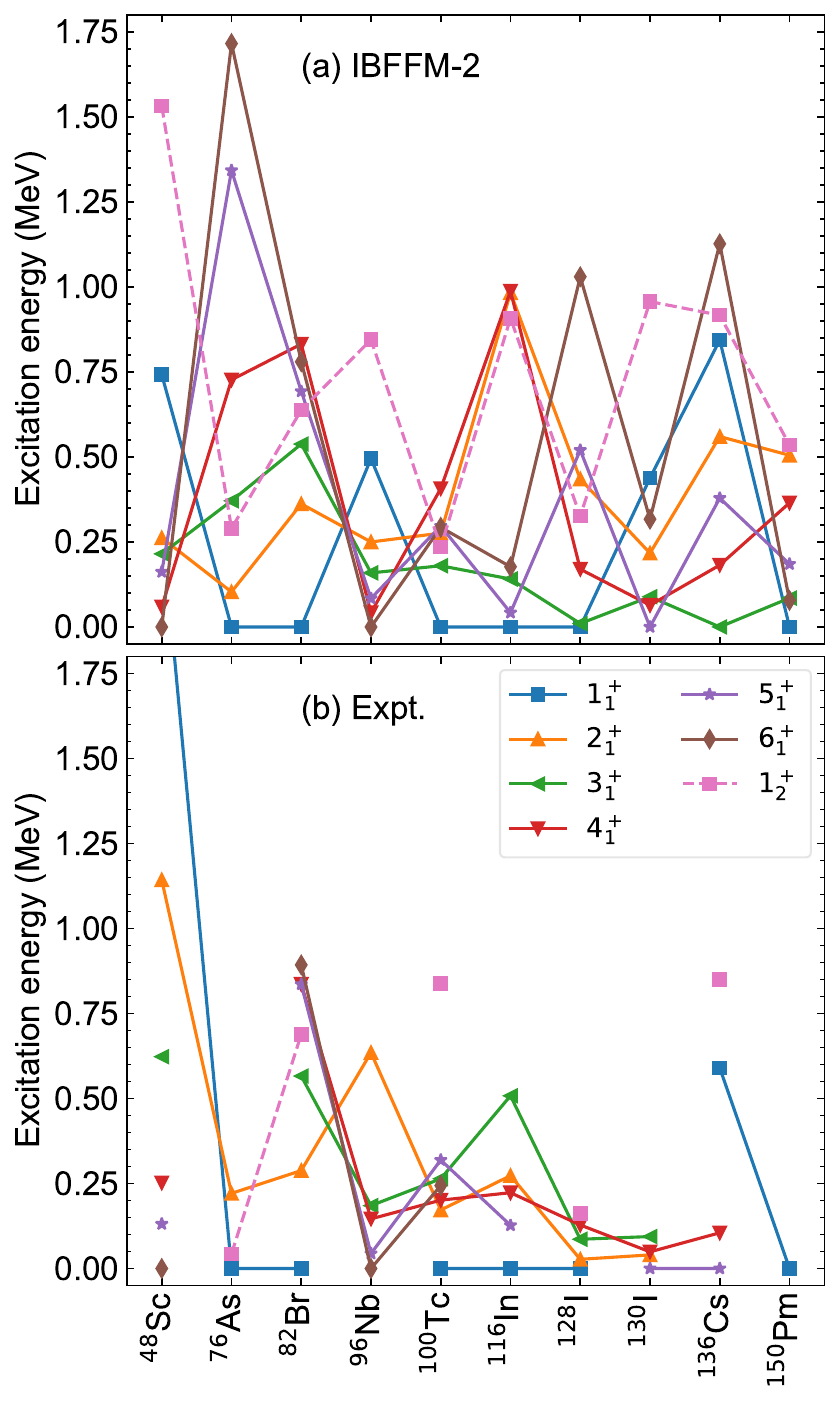}
\caption{Calculated and experimental \cite{data} 
low-energy low-spin 
levels in the odd-odd nuclei that correspond to the 
intermediate states of the $\db$ decays of 
the even-even nuclei $^{48}$Ca, $^{76}$Ge, $^{82}$Se, $^{96}$Zr, 
$^{100}$Mo, $^{116}$Cd, $^{128}$Te, $^{130}$Te, 
$^{136}$Xe, and $^{150}$Nd. 
The RHB-SCMF method using the
DD-PC1 EDF is employed to provide inputs
to the IBM-2 mapping and IBFFM-2 Hamiltonian.}
\label{fig:oo-bb2n}
\end{figure}

The EDF-based IBFFM-2 calculations 
for odd-odd nuclei have been made for 
the predictions on the $\beta$ 
and $\beta\beta$ decays. 
Just to show that the EDF-formulated IBFFM-2 
is able to give a systematic description 
of the odd-odd nuclear structure, 
Fig.~\ref{fig:oo-bb2n} depicts calculated low-energy 
levels for those odd-odd systems that correspond to 
the intermediate states of the probable $\db$ decays. 
The IBFFM-2 calculations are based on 
the RHB-SCMF framework with the 
DD-PC1 EDF and separable 
pairing force \cite{tian2009}. 
The detailed nuclear structure aspects 
including the electromagnetic properties 
are discussed in Refs.~\cite{nomura2022bb,nomura2024bb}.
In most of these calculations, 
the IBFFM-2 parameters are usually chosen 
for each odd-odd nucleus 
so as to reproduce the spin of the ground state 
and the energy levels of the
$1^+_1$ and $1^+_2$ states.

%-------- Odd-odd nuclei in 2vbb -----
%-------------------------------------

\subsection{Applications to $\beta$ and $\beta\beta$ decays\label{sec:beta}}

A formalism for the
$\beta$-decay transition operators 
was developed in the IBFM-2 and IBFFM-2 
using the generalized seniority scheme 
\cite{DELLAGIACOMA1989,IBFM}, 
and was used for phenomenological calculations 
in different mass regions
\cite{DELLAGIACOMA1989,zuffi2003,brant2004,brant2006,yoshida2013,ferretti2020,vsevolodovna2022}. 
Within this formalism,
the boson operators for the 
Gamow-Teller (GT) $\hat{T}^{\rm GT}$ 
and Fermi (F) $\hat{T}^{\rm F}$ transitions
are obtained by the mapping
\begin{align}
\label{eq:ogt}
&
\hat\tau\hat\sigma \longmapsto
\hat{T}^{\rm GT}
=\sum_{j_{\nu}j_{\pi}}
\eta_{j_{\nu}j_{\pi}}^{\mathrm{GT}}
\left(\hat P_{j_{\nu}}\times\hat P_{j_{\pi}}\right)^{(1)} \; , \\
\label{eq:ofe}
&
\hat\tau \longmapsto
\hat{T}^{\rm F}
=\sum_{j_{\nu}j_{\pi}}
\eta_{j_{\nu}j_{\pi}}^{\mathrm{F}}
\left(\hat P_{j_{\nu}}\times\hat P_{j_{\pi}}\right)^{(0)} \; .
\end{align}
$\hat\tau\hat\sigma$ and $\hat\tau$ stand for 
the fermionic GT and 
Fermi transition operators, respectively. 
The coefficients 
$\eta_{j_{\nu}j_{\pi}}^{\mathrm{GT}}$ and 
$\eta_{j_{\nu}j_{\pi}}^{\mathrm{F}}$ are given by
\begin{align}
\label{eq:eta}
\eta_{j_{\nu}j_{\pi}}^{\mathrm{GT}}
&= - \frac{1}{\sqrt{3}}
\left\langle
\ell_{\nu}\frac{1}{2};j_{\nu}
\bigg\|{\bm\sigma}\bigg\|
\ell_{\pi}\frac{1}{2};j_{\pi}
\right\rangle
\delta_{\ell_{\nu}\ell_{\pi}} \\
\eta_{j_{\nu}j_{\pi}}^{\mathrm{F}}
&=-\sqrt{2j_{\nu}+1}
\delta_{j_{\nu}j_{\pi}} \; .
\end{align}
$\hat P_{\jr}$ in Eqs.~(\ref{eq:ogt}) and (\ref{eq:ofe}) 
represents one-particle transfer operator, 
and is given as one of the following 
operators derived from the 
seniority considerations \cite{DELLAGIACOMA1989}. 
\begin{subequations}
 \begin{align}
\label{eq:creation1}
&A^{\+}_{\jr\mr} = \zeta_{\jr} a_{{\jr}\mr}^{\+}
 + \sum_{\jr'} \zeta_{\jr\jr'} s^{\+}_\rho (\tilde{d}_{\rho}\times a_{\jr'}^{\+})^{(\jr)}_{\mr}
\\
\label{eq:creation2}
&B^{\+}_{\jr\mr}
=\theta_{\jr} s^{\+}_\rho\tilde{a}_{\jr\mr}
 + \sum_{\jr'} \theta_{\jr\jr'} (d^{\+}_{\rho}\times\tilde{a}_{\jr'})^{(\jr)}_{\mr} \; ,
\end{align}
and their conjugate operators
\begin{align}
\label{eq:annihilation1}
&\tilde{A}_{\jr\mr}=(-1)^{\jr-\mr}A_{\jr-\mr}\\ 
\label{eq:annihilation2}
&\tilde{B}_{\jr\mr}=(-1)^{\jr-\mr}B_{\jr-\mr} \; .
\end{align}
\end{subequations}
The operators in Eqs.~(\ref{eq:creation1}) 
and (\ref{eq:annihilation1}) 
conserve the boson number, whereas those 
in Eqs.~(\ref{eq:creation2}) and (\ref{eq:annihilation2}) 
do not. 
$\hat{T}^{\rm GT}$ and $\hat{T}^{\rm F}$ are 
formed as a product of two of the operators in 
Eqs.~(\ref{eq:creation1})--(\ref{eq:annihilation2}), 
depending on the particle or hole nature of bosons in 
the even-even IBM-2 core. 
The coefficients $\zeta_{j}$, $\zeta_{jj'}$, 
$\theta_{j}$, and $\theta_{jj'}$ 
are dependent on the $v_{\jr}$ and $u_{\jr}$ 
amplitudes. 
The above formalism for the GT and Fermi 
transition operators has been exploited in the IBM-2 
mapping procedure for studying $\beta$-decay properties 
of odd-mass and odd-odd nuclei. 
The $u$ and $v$ factors, which appear 
in the formulas (\ref{eq:creation1})--(\ref{eq:annihilation2}), 
are computed by the microscopic 
SCMF method, and take the same values as those 
used in the IBFFM-2 calculations for 
the odd-odd nuclei. 

The $ft$ value for the single-$\beta$ decay
is calculated by using the
Fermi, $\mf = \braket{I_f \| \hat T^{\rm F} \| I_i}$, 
and GT, $\mgt = \braket{I_f \| \hat T^{\rm GT} \| I_i}$, 
reduced matrix elements 
for the transitions between the initial 
state $I_i$ of the parent nucleus 
and the final state $I_f$ 
of the daughter nucleus:
\begin{eqnarray}
\label{eq:ft}
 ft = \frac{K}{|\mf|^2+\left(\ga/\gv\right)^2 |\mgt|^2} \; ,
\end{eqnarray}
with the factor $K=6163$ (in seconds), 
$\ga=1.27$ and $\gv=1$ being the 
axial-vector and vector coupling constants, 
respectively.

Within the framework of the mapped IBM-2 that 
is based on the Gogny-D1M EDF SCMF calculations, 
$\beta^-$ decay of odd-mass Xe and Ba, 
and $\beta^+$ decay of odd-mass Cs and La 
with the mass $A \approx 130$
were studied \cite{nomura2020beta-1}. 
A theoretical investigation for the 
$\beta^+$ decay of the even-mass Ba and Xe nuclei 
was done within the IBFFM-2 using the Gogny force 
\cite{nomura2020beta-2}. 
In these studies the $\ft$ values were computed 
by using the IBFM-2 and IBFFM-2 wave functions 
that reasonably describe low-energy 
spectra and electromagnetic properties of the 
parent and daughter nuclei. 
In Ref.~\cite{nomura2022beta-rh}, 
the mapped IBM-2, IBFM-2, and IBFFM-2 
were employed for a systematic 
study on the low-energy spectroscopy 
and $\beta^-$-decay properties 
of the even- and odd-mass Rh nuclei in the 
$N=50-82$ neutron major shell, 
using the Gogny-D1M EDF as the 
microscopic input.

%-------- ft values Zr nuclei --------
\begin{figure}[h]
\centering
\includegraphics[width=\linewidth]{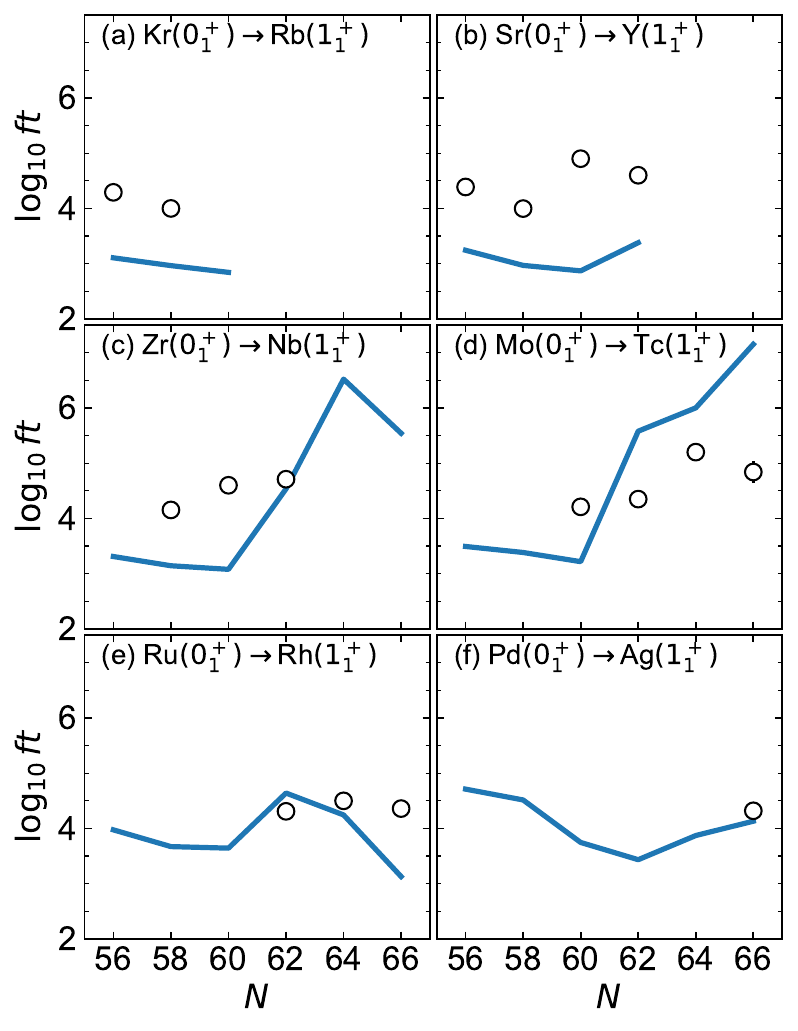}
\caption{
$\ft$ values for the $0^+_1 \to 1^+_1$ 
$\beta^-$ decays of the neutron-rich even-even 
Kr, Sr, Zr, Mo, Ru, and Pd isotopes with 
neutron number $N=56-66$.
Calculated values from the mapped IBM-2 based on 
the RHB-SCMF method are shown 
by solid lines, and the experimental data \cite{data} 
are depicted as open circles.}
\label{fig:ft-zr}
\end{figure}
%-------------------------------------

In Ref.~\cite{nomura2022beta-ge}, 
the $\beta$ decays of the neutron-deficient 
Ge and As nuclei were studied within the 
IBM-2 mapping that is based on the 
RHB-SCMF method. 
This mass region includes the nucleus $^{76}$Ge, 
which is a candidate $\znbb$-decay emitter, and 
the nuclear structure in this region 
is of relevance to the $\db$-decay study. 
In Ref.~\cite{nomura2024beta}, low-energy 
structure and $\beta$-decay properties of 
neutron-rich even-mass nuclei near the neutron number $N=60$ 
from Kr ($Z=36$) to Cd ($Z=48$) 
were investigated using the relativistic EDF. 
Some of the above nuclear systems 
are considered to be the candidate $\znbb$ emitters
and final-state nuclei,
i.e., $^{96}$Zr, $^{96}$Mo, $^{100}$Mo, and $^{100}$Ru.
The calculated low-lying levels for the 
even-even and odd-odd $A\approx 100$ nuclei exhibit a 
phase transitional behavior at $N \approx 60$. 
The resultant $\ft$ values exhibit
a rapid increase from $N=60$ to 62 in Zr and Mo isotopes  
(see Fig.~\ref{fig:ft-zr}), which reflects 
the shape QPTs in the neighboring 
even-even systems.

In the $\beta$-decay studies of 
Refs.~\cite{nomura2020beta-1,nomura2020beta-2,nomura2022beta-ge,nomura2022beta-rh,nomura2024beta}, 
the predicted $\ft$ values for some nuclei
were shown to differ
significantly from the experimental values.
In those nuclei near with $Z\leqslant 42$ 
and $N\leqslant 60$, in particular, 
the calculated $\ft$ values 
for the $0^+_1$ $\to$ $1^+_1$ transitions 
are by a factor of 1-2 
smaller than the experimental values 
[see Figs.~\ref{fig:ft-zr}(a)--\ref{fig:ft-zr}(d)]. 
To identify a major source of this deviation, 
the parameter sensitivity 
of the $\beta$-decay prediction 
within the mapped IBM-2 was analyzed in \cite{homma2024}. 
It was found that the predicted $\ft$ values 
for the $\beta^-$ decay of 
neutron-rich Zr isotopes
depend strongly on the quadrupole-quadrupole 
strength parameter $\kappa$ (\ref{eq:ham1}) 
and on the tensor-type interaction 
strength $\vt$ (\ref{eq:hff}) used for the 
IBFFM-2 Hamiltonian for the odd-odd Nb daughters. 
It was further suggested that, 
in order to reproduce the $\ft(0^+_1 \to 1^+_1)$ 
values, the strength parameter $\kappa$ for the 
daughter nuclei would have to be smaller in magnitude 
than that obtained from the mapping procedure. 
The derived values of the IBM-2 parameters for the 
even-even core nuclei depend largely on the 
topology of the SCMF PESs, which are computed 
by using a given EDF.

The method to compute the GT and Fermi transition 
rates for the even-mass nuclei is readily
applied to the calculations of the nuclear 
matrix elements (NMEs) for the $\beta\beta$ decay.
This decay process is a rare phenomenon,
in which the even-even nucleus
with $(N,Z)$ decays into the one with
$(N\pm2,Z\mp2)$, emitting
two electrons (or positrons) and some light 
particles such as neutrinos. 
The calculations of the two-neutrino 
$\db$ ($\tnbb$) NMEs, in particular, 
require to consider intermediate states in the 
odd-odd nuclei adjacent to the even-even 
parent and daughter nuclei. 
The mapped IBM-2 framework that is based on the 
relativistic EDF was applied \cite{nomura2022bb} 
to predict the $\tnbb$-decay NMEs and half-lives 
of 13 even-even nuclei 
$^{48}$Ca, $^{76}$Ge, $^{82}$Se, $^{96}$Zr, 
$^{100}$Mo, $^{110}$Pd, 
$^{116}$Cd, $^{124}$Sn, $^{128}$Te, $^{130}$Te, 
$^{136}$Xe, $^{150}$Nd, and $^{198}$Pt, 
which are candidates for the $\db$ emitters.

The $\tnbb$-decay NME is given by
\begin{align}
\label{eq:nme}
 M_{2\nu} = \ga^2 \cdot m_e c^2
\left[ M^{\rm GT}_{2\nu} - \left(\frac{\gv}{\ga}\right)^2 M^{\rm F}_{2\nu}\right] \; ,
\end{align}
where
\begin{align}
& M^{\rm GT}_{2\nu}
=\sum_n
\frac{\braket{0^+_f\| \hat T^{\rm GT} \| 1^+_n}
\braket{1^+_n\| \hat T^{\rm GT} \| 0^+_{1,i}}}
{E_n - E_i + (Q_{\beta\beta} + 2m_e c^2)/2}
\\
& M^{\rm F}_{2\nu}
=\sum_n
\frac{\braket{0^+_f\| \hat T^{\rm F} \| 0^+_n}
\braket{0^+_n\| \hat T^{\rm F}  \| 0^+_{1,i}}}
{E_n - E_i + (Q_{\beta\beta} + 2m_e c^2)/2} \; ,
\end{align}
represent, respectively, the GT and Fermi 
matrix elements for the $\tnbb$ decay. 
$E_n$, $E_i$, $Q_{\beta\beta}$, 
and $m_e$ are energies 
of the intermediate and initial state, $\db$-decay 
$Q$-value, and electron mass, respectively. 
In the present framework, 
the nuclear wave functions for the 
initial $\ket{0^+_{1,i}}$ and final $\ket{0^+_{f}}$ 
even-even nuclei are obtained from the 
mapped IBM-2, and the intermediate 
states $\ket{1^+}$ and $\ket{0^+}$ 
of the odd-odd nuclei are calculated within 
the IBFFM-2, which is based on the 
formulation described in Sec.~\ref{sec:odd-formulas}.

%-------- 2vbb NME unquenched --------
\begin{figure}[h]
\centering
\includegraphics[width=\linewidth]{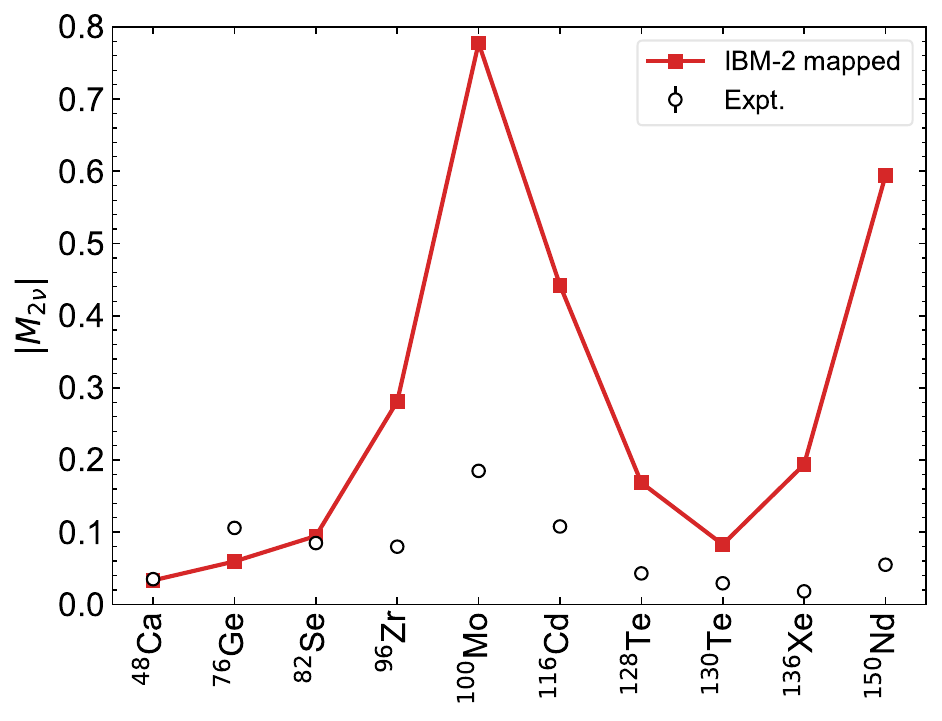}
\caption{
Calculated NMEs of the $\tnbb$ decay 
in the mapped IBM-2 framework that is based 
on the relativistic functional DD-PC1. 
The effective NMEs obtained from the measured 
half-lives \cite{barabash2020} 
are included for comparison.}
\label{fig:nme}
\end{figure}
%-------------------------------------

%-------- eff. NME with other theories --------
\begin{figure}[h]
\centering
\includegraphics[width=\linewidth]{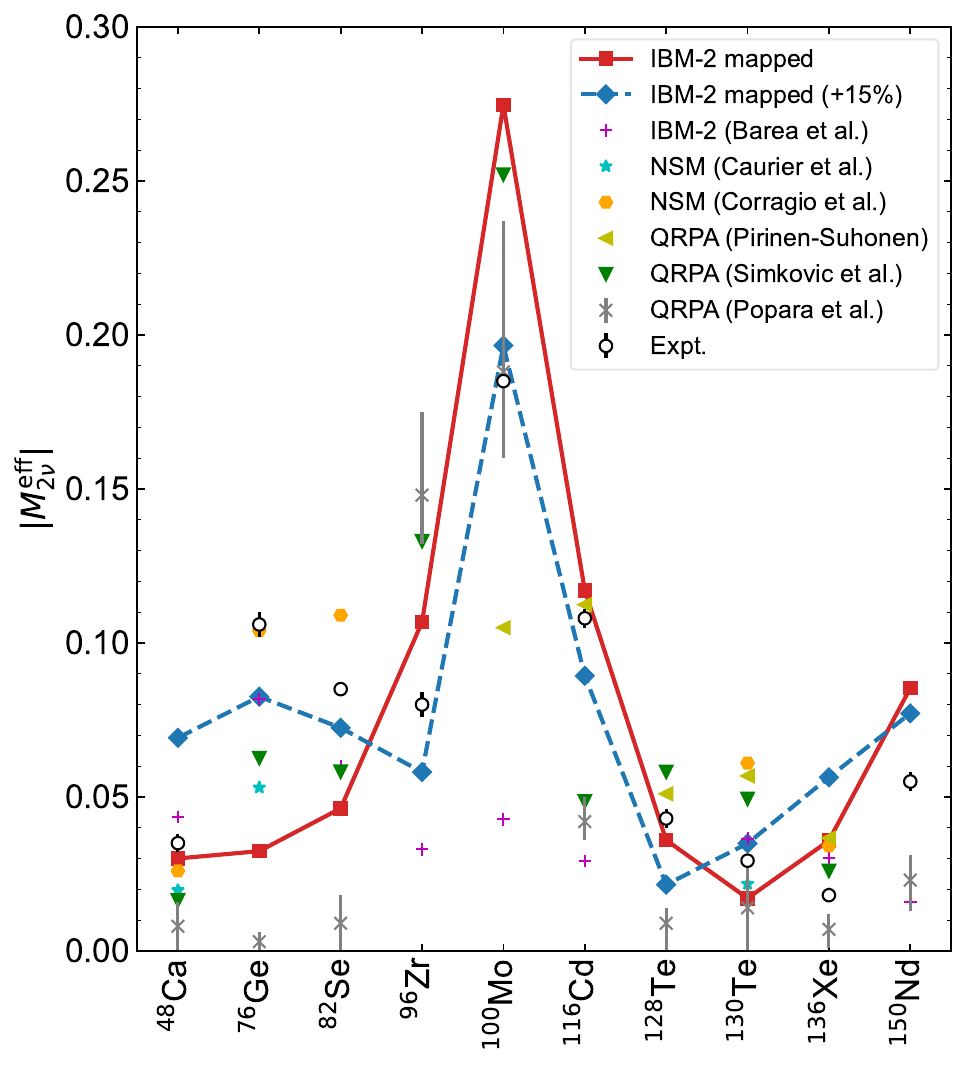}
\caption{
Effective NMEs of $\tnbb$ decay resulting 
from the mapped IBM-2 employing 
the default and 15\% increased 
pairing strengths in the RHB-SCMF 
calculations. 
The experimental 
NMEs \cite{barabash2020}, and those from 
the IBM-2 \cite{barea2015}, 
QRPA \cite{pirinen2015,simkovic2018,popara2022}, 
and Nuclear Shell Model \cite{caurier2007,coraggio2019} 
are shown for comparison. 
The constants for the effective 
$\ga$ factors (\ref{eq:gae})
$(c,d)=(1.86,0.009)$, and $(1.30,0.008)$ 
for the default and increased pairing strengths, 
respectively.}
\label{fig:nme-lit}
\end{figure}
%----------------------------------------------

Figure~\ref{fig:nme} displays the 
calculated $M_{2\nu}$ 
(\ref{eq:nme}) for even-even $\tnbb$ 
emitting isotopes, and the corresponding 
effective NMEs, which are derived from 
the observed half-lives \cite{barabash2020}. 
It is seen that 
the calculation gives the NMEs close to the 
empirical ones for the decays of 
the nuclei $^{48}$Ca, $^{76}$Ge, and $^{82}$Se,
but overestimates the experimental data 
for those nuclei heavier than $A=82$. 
The large NMEs for the heavy nuclei 
indicate a need for quenching the GT transition 
matrix elements, leading to an effective 
$\ga$ factor. 
A mass-number-dependent effective $\ga$ 
factor of the form
\begin{eqnarray}
\label{eq:gae}
 \ga^{\rm eff}=ce^{-dA}
\end{eqnarray}
was proposed in Ref.~\cite{nomura2022bb},
where $c$ and $d$ are numerical constants 
that are fitted to the experimental NMEs. 
The effective NMEs $M_{2\nu}^{\rm eff}$ 
calculated by employing the effective 
$\ga$ factor of (\ref{eq:gae}) are shown 
in Fig.~\ref{fig:nme-lit} (denoted as ``IBM-2 mapped''), 
and are compared with 
those NMEs predicted by the IBM-2 \cite{barea2015}, 
QRPA \cite{pirinen2015,simkovic2018,popara2022}, 
and Nuclear Shell Model \cite{caurier2007,coraggio2019}. 
The calculated values of the NMEs within 
the mapped IBM-2 appear to be more or close 
to these different theoretical predictions.

The fact that the effective $\ga$ factors 
have to be introduced indicates certain 
deficiencies or uncertainties of a given nuclear model. 
The mapped IBM-2 predictions on the $\tnbb$-decay NMEs 
were shown \cite{nomura2022bb,nomura2024bb} 
to be dependent on the parameters 
and assumptions considered, such as the 
choice of the EDF, parameters and forms of the 
IBM-2 and IBFFM-2 Hamiltonians, 
and single-particle energies for the IBFFM-2 Hamiltonian. 
In particular, the effects of 
changing the pairing strength employed 
in the RHB-SCMF calculations 
were investigated in a systematic 
way \cite{nomura2024bb}. 
By increasing the pairing strength by 15\%, 
the SCMF PES was shown to exhibit 
a much less pronounced deformation, that is, 
the potential valley becomes much less steeper, 
than in the case of the calculation with the 
default pairing strength. 
With the less pronounced potential valley, 
the mapping procedure was shown to
result in a quadrupole-quadrupole strength $\kappa$
that is significantly smaller in magnitude
than that obtained with the default pairing.
The calculation with the increased 
pairing strength was shown to give an improved 
description of the non-yrast levels 
in the even-even nuclei, and provide 
larger $\tnbb$-decay NMEs than those 
with the default pairing strength.

Figure~\ref{fig:nme-lit} shows 
the effective NMEs calculated 
with the pairing force that 
is enhanced by 15\% [``IBM-2 mapped ($+15$\%)''], 
which are consistent with the experimental NMEs 
for the $\tnbb$ decays 
$^{76}$Ge $\to$ $^{76}$Se, 
$^{82}$Se $\to$ $^{82}$Kr,  
$^{100}$Mo $\to$ $^{100}$Ru, and  
$^{130}$Te $\to$ $^{130}$Xe. 
However, the calculation with the standard pairing 
strength appears to be adequate to provide 
an overall good description 
of the effective NMEs, and is also
shown \cite{nomura2024bb} to
reproduce the electromagnetic 
properties of the odd-odd nuclei 
better than in the case of the increased 
pairing force.

Many of the ingredients of these 
$\tnbb$-decay NME calculations, including the 
SCMF PESs for even-even nuclei and derived 
IBM-2 Hamiltonian parameters, can be exploited 
for the predictions of the $\znbb$-decay NMEs and 
half-lives. The extension of the calculations 
in Refs.~\cite{nomura2022bb,nomura2024bb} 
to study the $\znbb$ decays 
is in progress \cite{nomura2025-0v}.

\section{Summary and perspective\label{sec:summary}}

The EDF-mapped IBM has been developed, and 
extensively used to study low-energy nuclear structure. 
Assuming that the nuclear surface deformation 
is simulated by bosonic degrees of freedom, 
the parameters of the IBM are determined by mapping 
the PES that is calculated by the SCMF method 
employing a given EDF onto the expectation 
value of the IBM Hamiltonian in the boson intrinsic state. 
The mapped Hamiltonian produces energy spectra and transition 
properties of even-even nuclei. 
For strongly deformed nuclei with axial symmetry, 
the rotational response of the 
nucleonic system is incorporated in the 
boson system by the cranking calculation, 
and turns out to play a crucial role 
in reproducing accurately the rotational bands. 
For studying those nuclear systems 
that are characterized by pronounced 
$\gamma$ softness, an optimal IBM
description should consider three-body boson terms. 
The inclusion of these terms is required 
in order to produce a triaxial minimum 
on the energy surface consistently with the 
self-consistent calculations, and naturally 
explains the energy level systematic of the 
$\gamma$-vibrational bands of medium-heavy 
and heavy nuclei.

These findings indicate that 
the IBM mapping procedure is valid in general 
cases of the quadrupole collectivity, 
that is, nearly spherical vibrational, 
strongly deformed rotational, 
and $\gamma$-soft limits. 
Given the fact that the nuclear EDF 
is fine-tuned to predict accurately 
the intrinsic properties 
of almost all nuclei, it has become 
possible to derive the parameters of 
the IBM Hamiltonian for  
arbitrary nuclei from nucleonic degrees of freedom, 
and to predict low-energy spectroscopic properties 
for those heavy exotic nuclei for which
experimental data do not exist. 
The method has been further extended to 
study phenomena of shape coexistence by the 
inclusion of the intruder states and 
configuration mixing, octupole and hexadecapole 
deformations and collectivity by introducing 
the relevant boson degrees of freedom, 
and low-energy structures of odd nuclei and 
$\beta$ and $\db$ decays by the coupling 
of the single-particle to collective degrees 
of freedom.

There are also several open problems 
for further improvements of the mapped IBM 
predictions for nuclear structure. 
An important next step is to formulate a method 
of deriving the boson-fermion 
interaction strengths only from the microscopic EDF input, 
since in the present version of the framework 
these parameters are determined so as to 
reproduce spectroscopic data. 
The inclusion of additional 
degrees of freedom in the IBM will be essential, 
for instance, dipole $p$ bosons, which 
would be necessary for a quantitative prediction 
of $E1$ properties of those nuclei 
in which octupole correlations play a role. 
In addition, 
the fact that the mapped IBM has not been 
able to reproduce quantitatively some 
observed spectroscopic properties 
in transitional nuclei, 
including the low-lying non-yrast energy levels 
in Sm isotopes (see Fig.~\ref{fig:sm-level}) 
and the yrast levels in neutron-rich Zr near $N=56$ 
(Fig.~\ref{fig:zr-gs}), suggests that certain 
improvements of the method are in order. 
The deficiencies in reproducing these 
spectroscopic data may be attributed to 
the assumptions in the IBM, to the mapping procedure, 
and/or to properties of the underlying EDF 
used as a microscopic input. 
Identifying the source of the deficiencies 
will be a crucial step forward 
in the microscopic descriptions 
of nuclear collective motions within the IBM.

\begin{acknowledgements}
The author is supported by JSPS
KAKENHI Grant No. JP25K07293.
\end{acknowledgements}

\bibliographystyle{unsrt} % unsrt,abbrv
\bibliography{refs}

\end{document}